\documentclass[11pt,a4paper]{article}
\pdfoutput=1
\usepackage{caption}
\usepackage{jheppub}
\usepackage[utf8]{inputenc}
\usepackage{soul,xcolor}
\usepackage{cancel}
\usepackage{makecell}
\usepackage{diagbox}
\usepackage{makecell}
\usepackage[normalem]{ulem}
\newsavebox{\foobox}

\usepackage{multirow}
\usepackage{dcolumn}
\newcommand\Tstrut{\rule{0pt}{2.6ex}}         

\newcommand\redsout{\bgroup\markoverwith{\textcolor{red}{\rule[0.5ex]{2pt}{0.8pt}}}\ULon}   
\usepackage{ulem}
\usepackage{graphicx}
\usepackage{comment}
\usepackage{bm,array}
\usepackage{graphics}
\usepackage{epsf}
\usepackage{color}
\usepackage{amsmath}
\usepackage{amssymb}
\usepackage{latexsym}
\usepackage{slashed}
\usepackage{float}
\usepackage{cases}
\usepackage{multirow}
\usepackage{framed}
\def\be{\begin{equation}}
\def\ee{\end{equation}}
\def\ba{\begin{array}}
\def\ea{\end{array}}

\def\alambda{A_\lambda}
\def\akappa{A_\kappa}
\def\mueff{\mu_\mathrm{eff}}
\def\beff{B_{\mathrm{eff}}}
\def\tanb{\tan\beta}

\def\wpm{W^\pm}

\def\sQ3{\widetilde{Q}_3}
\def\sU3{\widetilde{U}_3}
\def\sD3{\widetilde{D}_3}
\def\hone{h_1}
\def\htwo{h_2}
\def\aone{a_1}
\def\hsm{h_{\rm SM}}

\def\hs{h_{_S}} 
\def\as{a_{_S}}
\def\bino{\widetilde{B}}
\def\wino{\widetilde{W}}
\def\higgsino{\widetilde{H}^0}
\def\higgsinod{\widetilde{H}^0_d}
\def\higgsinou{\widetilde{H}^0_u}
\def\singlino{\widetilde{S}}
\def\ntrli{\chi_{_i}^0}
\def\ntrlj{\chi_{_j}^0}
\def\ntrlk{\chi_{_k}^0}
\def\ntrlone{\chi_{_1}^0}
\def\ntrltwo{\chi_{_2}^0}
\def\ntrlthree{\chi_{_3}^0}
\def\ntrlfour{\chi_{_4}^0}
\def\ntrlfive{\chi_{_5}^0}
\def\ntrlonetwo{\chi_{_{1,2}}^0}
\def\ntrltwothree{\chi_{_{2,3}}^0}
\def\ntrlthreefour{\chi_{_{3,4}}^0}
\def\ntrltwothreefour{\chi_{_{2,3,4}}^0}

\def\charonepm{\chi_{_1}^\pm}
\def\charonemp{\chi_{_1}^\mp}

\def\charonemp{\chi_{_2}^\mp}
\def\mone{M_1}
\def\mtwo{M_2}

\def\mhs{m_{_{h_S}}}
\def\mhssq{m^2_{_{h_S}}}
\def\mas{m_{a_{_S}}}
\def\massq{m^2_{a_{_S}}} 

\def\msinglino{m_{_{\widetilde{S}}}}

\def\mntrli{m_{{_{\chi}}_i^0}}
\def\mntrlj{m_{{_{\chi}}_j^0}}
\def\mntrlk{m_{{_{\chi}}_k^0}}
\def\mntrlone{m_{{_{\chi}}_{_1}^0}}
\def\mntrltwo{m_{{_{\chi}}_{_2}^0}}
\def\mntrlthree{m_{{_{\chi}}_{_3}^0}}

\def\mntrlfive{m_{{_{\chi}}_{_5}^0}}
\def\mntrltwothree{m_{{_{\chi}}_{_{2,3}}^0}}
\def\mntrlthreefour{m_{{_{\chi}}_{_{3,4}}^0}}
\def\mcharone{m_{{_{\chi}}_{_1}^\pm}}
\def\mchartwo{m_{{_{\chi}}_{_2}^\pm}}
\def\mhsm{m_{h_{\mathrm{SM}}}}
\def\mhsmsq{m^2_{h_{\mathrm{SM}}}}

\def\vev{{\it vev}}

\def\vu{v_u}
\def\vd{v_d}

\def\vs{v_{\!_S}}
\def\etmiss{\slashed{E}_T}

\def\nmssmtools{{\tt NMSSMTools}}

\def\micromegas{{\tt micrOMEGAs}}

\newcommand{\beq}{\begin{equation}}
\newcommand{\eeq}{\end{equation}}
\newcommand{\bea}{\begin{eqnarray}}
\newcommand{\eea}{\end{eqnarray}}

\newcommand*\sq{\mathbin{\vcenter{\hbox{\rule{.8ex}{.8ex}}}}} 
\title{A relatively light, highly bino-like dark matter in the $Z_3$-symmetric NMSSM and recent LHC searches}
\author[a,b]{Waleed Abdallah,}
\author[a]{AseshKrishna Datta}
\author[a,c]{and Subhojit Roy}
\affiliation[a]{Harish-Chandra Research Institute, HBNI, Allahabad 211019,
India}
\affiliation[b]{Department of Mathematics, Faculty of Science, Cairo University,
Giza 12613, Egypt}
\affiliation[c]{Regional Centre for Accelerator-based, Particle Physics Harish-Chandra Research Institute, \\ Allahabad 211019, India}
\emailAdd{waleedabdallah@hri.res.in, asesh@hri.res.in, subhojitroy@hri.res.in}
\abstract{A highly bino-like Dark Matter (DM), which is the Lightest 
Supersymmetric Particle (LSP), could be motivated by the stringent upper 
bounds on the DM direct detection rates. This is especially so when its mass
is around or below 100~GeV for which such a bound tends to get most severe. 
Requiring not so large a higgsino mass parameter, that would render the 
scenario reasonably `natural', prompts such a bino-like state to be relatively 
light.
In the Minimal Supersymmetric Standard Model (MSSM), in the absence of 
comparably light  scalars, such an excitation, if it has to be a thermal relic, 
is unable to meet the  stringent experimental upper bound on its abundance 
unless its self-annihilation hits a funnel involving either the $Z$-boson or 
the Standard Model (SM)-like Higgs boson.
We demonstrate that, in such a realistic situation, a highly bino-like DM 
of the popular $Z_3$-symmetric Next-to-Minimal Supersymmetric Standard Model 
(NMSSM) is viable over an extended range of its mass, from our targeted maximum 
in the vicinity of the mass of the top quark down to about 30~GeV.
This is facilitated by the presence of comparably light singlet-like states 
that could serve as funnel (scalars) and/or coannihilating (singlino) states 
even as the bino-like LSP receives a minimal (but optimal) tempering triggered 
by suitably light higgsino states that, in the first place, evade stringent 
lower bounds on their masses that can be derived from the Large Hadron Collider (LHC) 
experiments only in the presence of a lighter singlino-like state.
An involved set of blind spot conditions is derived for the DM direct detection 
rates by considering for the very first time the augmented system of neutralinos 
comprising of the bino, the higgsinos and the singlino which highlights the 
important roles played by the NMSSM parameters `$\lambda$' and $\tanb$ in 
delivering a richer phenomenology.
}
\begin{document}
\begin{flushright}
HRI-RECAPP-2020-10
\end{flushright}
\maketitle
%
\section{Introduction}
\label{sec:Intro}
It is well-known in the context of the Minimal Supersymmetric extension of the 
Standard Model (MSSM) that a pure bino ($\widetilde{B}$) neutralino as the Lightest 
Supersymmetric Particle (LSP; $\ntrlone$) and a Weakly Interacting Massive Particle (WIMP) could be a viable cold Dark Matter (DM) 
candidate, which is a thermal relic, under two specific situations: (i) if it is 
lighter than $\sim 100$~GeV and some sfermions (in particular, the right-handed 
sleptons with large hypercharge coupling; such light squarks being ruled out 
long back) are not much heavier than 100~GeV such that the so-called bulk 
($t$-channel) annihilation of a pair of DM states mediated by such sfermions are optimally 
efficient~\cite{ArkaniHamed:2006mb, Pierce:2013rda, Fukushima:2014yia, Baum:2020gjj} and/or (ii) if there is an efficient 
coannihilation of such a DM state with slepton(s) ($\widetilde{\ell}$) with $\mntrlone \equiv m_{\widetilde{B}} \sim m_{\widetilde{\ell}} \lesssim 
500$~GeV when the sleptons
are degenerate~\cite{Chakraborty:2019wav, Ellis:1999mm}.  However, requiring such a pure bino DM (the lightest neutralino, $\ntrlone$) to be relatively 
light, i.e., $m_{\widetilde{B}}  \sim 100$~GeV, both the possibilities are found to be 
practically ruled out by the Large 
Hadron Collider (LHC) experiments
\cite{Aad:2019vnb, Aad:2019qnd} save for the cases with
light staus with comparable mass~\cite{Aad:2019byo}. Such a situation, keeping up further with the ever-improving lower bounds on the sparticle masses, has prompted studies where
sleptons are assumed to be rather heavy~\cite{Huang:2017kdh, Badziak:2017the, vanBeekveld:2016hbo, Abdughani:2017dqs}. Thus, in the framework of the MSSM, only a 
tempered
bino-dominated LSP, with optimal admixtures of higgsinos~\cite{Feng:2000gh,Baer:2005bu} and/or wino~\cite{Masiero:2004ft,Baer:2005zc,Baer:2005jq,BirkedalHansen:2001is,BirkedalHansen:2002am},
 fits the bill so far as compliance with the experimentally observed upper bound on the DM relic abundance~\cite{Ade:2015xua, Aghanim:2018eyx} is concerned which may be ensured in the presence of resonant 
self-annihilations (funnels) or efficient coannihilation of the DM particles. 

On the other hand, stringent and improved upper bounds on spin-independent (SI)~\cite{Aprile:2018dbl} and spin-dependent (SD)~\cite{Aprile:2019dbj,Amole:2019fdf} DM-nucleus (elastic) scattering cross sections that are routinely arriving from various DM Direct Detection (DMDD) experiments could genuinely motivate a nearly pure bino DM. In fact, such bounds coming from the SI experiments tend to be the most stringent ones when the DM mass is relatively low, around or below 100~GeV~\cite{Aprile:2018dbl}. This could prompt one to suspect a nearly pure bino-like DM hiding in this mass-range. It may further be noted that the excess of
$\gamma$-ray emanating from the galactic center as observed by an indirect DM-detection experiment like the Fermi-LAT experiment~\cite{TheFermi-LAT:2017vmf,Fermi-LAT:2016uux} can also be explained in terms of a (relatively light) bino-dominated DM~\cite{Agrawal:2014oha,Freese:2015ysa,Butter:2016tjc,Achterberg:2017emt}.

In the absence of optimally light sleptons, a scenario that has been discussed in a preceding paragraph, the only ways such a highly bino-like DM of the MSSM could comply with the upper bound
on its relic abundance are the following: (i) presence of an efficient $s$-channel mutual annihilation funnel, e.g., in the form of a resonant $Z$- or a Higgs boson
with $m_{\mathrm{DM}} \approx m_Z/2$ or
$\approx m_{\mathrm{Higgs}}/2$, respectively 
and/or (ii) possibility of coannihilation of the DM neutralino with other electroweakino(s) (the ewinos, i.e., the charginos and the neutralinos) close in mass to the DM state. Furthermore, note that for such a light DM, their otherwise prolific 
annihilations to a $W^+W^-$ pair and to a $t\bar{t}$ pair can both be closed.
Compliance with the reported $2\sigma$ upper bound on its relic abundance thus gets even more challenging.

In any case, in the framework of the MSSM, an efficient enough resonant (funnel) annihilation or a coannihilation process mentioned above necessarily requires some minimal higgsino and/or wino admixture in the otherwise bino-like LSP DM. Note that for funnels involving $Z$-boson and the 
SM-like Higgs boson ($\hsm$), the mass of the annihilating LSP gets almost discretely fixed
to either $m_Z/2$ or $\mhsm/2$, respectively. Given current lower bounds on the masses of the heavier MSSM Higgs states ($\gtrsim 
400$~GeV)~\cite{Aaboud:2017sjh, Sirunyan:2018zut, Bahl:2018zmf, Aad:2020zxo}, those
cannot provide funnels for self-annihilation of the LSP DM with mass
$\leq 200$~GeV which is in 
our focus. On top of that, bounds on SD rates put a stringent bound on the higgsino mass parameter `$\mu$'  ($|\mu| > 270$~GeV) which in turn rules out a possible higgsino-like Next-to-LSP (NLSP) having a mass lighter than that~\cite{Carena:2018nlf}. In addition, recent LHC studies~\cite{Aad:2019qnd,Aaboud:2017leg} also appear to put a bound $\gtrsim 200$~GeV on lighter, nearly degenerate ewinos. These together stand in the way for an LSP with mass $\lesssim 200$~GeV coannihilating with a close by ewino state~\cite{Calibbi:2014lga}.

The situation is different in a crucial way, in the Next-to-MSSM (NMSSM)~\cite{Ellwanger:2009dp}, a popular extension of the MSSM.
In the NMSSM, the so-called $\mu$-term ($\mu \widehat{H}_u \cdot\widehat{H}_d$) in the superpotential is
replaced by a term $\lambda \widehat{S} \widehat{H}_u \cdot\widehat{H}_d $, where $\widehat{S}$ is a chiral superfield and a singlet under the Standard Model (SM) gauge group
and `$\lambda$' is a dimensionless coupling much akin to a Yukawa coupling.
The presence of $\widehat{S}$ gives rise to a singlet scalar field `$S$'. An
effective `$\mu$-term' ($\mueff$) is generated dynamically when `$S$' acquires a vacuum expectation value (\emph{vev}) $\vs$ such that $\mueff=\lambda \vs$
thus offering a possible solution to the infamous $\mu$-problem~\cite{Kim:1983dt} that plagues the MSSM. The~NMSSM Higgs sector possesses two extra singlet scalar states of $CP$-even ($\hs$) and $CP$-odd ($\as$) varieties when compared to the MSSM. In addition, a singlet Majorana fermion, the `singlino', augments the MSSM neutralino sector.

Such a spectrum is known to influence both DM and collider phenomenology in a nontrivial way, in particular, in a $Z_3$-invariant version of the NMSSM with its free (input) parameters being arbitrary at the weak scale (the so-called phenomenological NMSSM or the pNMSSM). Scenarios with light higgsinos and a singlino, in which either of them could turn out to be
the LSP (DM), and accompanied by one or more light singlet-like scalars, have attracted
much attention since the early days of the NMSSM scenario.
This has surely been driven by the existence of nontrivial couplings among
the singlet-like states (the singlino and the singlet scalars) and the doublet-like
ones (the MSSM-like Higgs bosons and the higgsinos) that characterize the NMSSM.

In comparison, dedicated, general studies on a possible bino-dominated DM in the NMSSM are lacking. This is presumably since the NMSSM does not suggest any direct 
tree-level coupling among the bino and the new singlet states that it offers thus leaving the pertinent setup, to a good extent, appearing like one with only the bino and the higgsinos (and the wino) of the MSSM~\cite{Beskidt:2017xsd}.   
However, for a highly bino-like LSP DM, which is in the focus of the present 
work, the much important phenomenon of DM annihilation could find a crucial, additional boost in the presence of possibly light singlet-like scalar states
\cite{Belanger:2005kh, Gunion:2005rw} (in particular, even
lighter than the $Z$-boson) which might act as funnels over an extended
range of DM masses, even  below $m_Z/2$. Some later studies~\cite{Das:2010ww, Vasquez:2010ru, Cumberbatch:2011jp, Han:2014nba}
concentrated on a light bino-like DM ($\mntrlone \lesssim 40$~GeV, down to a few hundreds of MeV) that funnels through such light scalar(s) and were mostly motivated by the then recent reported excesses at experiments like
DAMA, CoGeNT and CDMS-II and the exclusion limits from XENON and CDMS-Si experiments which all seemed to be compatible with such excesses. 
Furthermore, the mentioned $\gamma$-ray excess as observed by the Fermi-LAT
experiment also seems to fit better with a not so heavy bino-like DM of the NMSSM when compared to the MSSM, in the presence of light singlet scalars~\cite{Cheung:2014lqa,Guo:2014gra,Gherghetta:2015ysa,Bi:2015qva,Cao:2015loa,Shang:2018dja}.
Phenomenological
studies of efficient search-strategies of such light scalars (irrespective of the nature of the LSP) have remained to be an active
enterprise~\cite{Dermisek:2005ar, Christensen:2013dra, Cerdeno:2013qta, Cao:2013gba, King:2014xwa,Dutta:2014hma, Bomark:2015fga, Ellwanger:2015uaz, Conte:2016zjp, Guchait:2016pes,  Baum:2017gbj, Guchait:2017ztk, Ellwanger:2017skc, Baum:2019uzg, Barducci:2019xkq, Guchait:2020wqn} while their actual searches constitute a dedicated and a vibrant programme at the LHC
\cite{Aaboud:2018fvk, Aaboud:2018gmx, Aaboud:2018iil, Aaboud:2018esj, Aad:2020rtv, Sirunyan:2018mbx, Sirunyan:2018mot, Sirunyan:2020eum}.

During recent years, the phenomenology of a bino-dominated neutralino DM in the pNMSSM has been discussed as parts of some general studies, in different contexts in refs.~\cite{Baum:2017enm, Cao:2018rix, Domingo:2018ykx}. For example, the analysis in 
ref.~\cite{Baum:2017enm} focuses on the region of parameter space that does not 
bank on large radiative corrections to find the mass of the SM Higgs boson
in the proximity of its observed value of 125~GeV
(and hence can be called a `natural' scenario) 
 and, at the same time, the Higgs sector is attributed with 
the so-called `alignment-without-decoupling'~\cite{Carena:2015moc, Baum:2017gbj, Baum:2020vfl}. Compliance with these two
conditions together, for the doublet Higgs states, requires
$0.5 \lesssim \lambda \lesssim 0.7$ and $\tanb \lesssim 5$.
The study rules out a bino-like DM lighter than the 
top quark.  Ref.~\cite{Cao:2018rix}, on the other hand, allows for relatively 
light sfermions that help achieve the right relic abundance via their coannihilation 
with the neutralino DM. The work infers that a bino-like DM in a `natural' pNMSSM setup is strongly disfavored by results from various DM and the LHC experiments.
Ref.~\cite{Domingo:2018ykx} instead performs a study with the then latest LHC results (with expressly less prejudice towards the DM-related issues)
to shed light on the fate of a bino-dominated DM within the pNMSSM where the wino- and the higgsino-like states are mostly on the lighter side ($\lesssim 250$~GeV) while the singlino-like one could be as light as $\lesssim 70$~GeV. Thus, in that study, all such states could play the role of the NLSP and the LSP can have significant wino and higgsino admixtures.

In this backdrop, in the present work, we study in detail the viability of a highly bino-like DM (with bino content $> 95$\%) of the pNMSSM with mass below $200$~GeV (and, in particular, below $100$~GeV) vis-a-vis the theoretical and the current experimental constraints. The wino is taken to be much heavier.\footnote{A heavy wino straightaway aids compliance with the stringent lower bounds on the masses of the wino-like states as reported by the LHC experiments. Furthermore, this precludes their presence in the cascades of lighter
ewinos thus rendering the scenario much less cluttered for our present analysis of a bino-higgsino-singlino system.} Thus, we allow for up to 5\% tempering (a somewhat `minimal' tempering) of the bino-like LSP DM by the higgsinos and the singlino. We then open ourselves up to a broader region of the NMSSM parameter space, mainly by allowing for smaller values of `$\lambda$' and a larger range of $\tan\beta$, when compared to ref.~\cite{Baum:2017enm}. Furthermore, we consider the sleptons to be much heavier when compared to ref.~\cite{Cao:2018rix} such that bulk annihilation and coannihilation of the DM involving sleptons can no more be instrumental.
The squarks and, in particular, the top squarks are assumed to be rather heavy
(beyond the reach of the LHC). The latter choice might result in some compromise on `naturalness' (or `finetuning')~\cite{Ellis:1986yg, Barbieri:1987fn}. The situation could be ameliorated by restricting $|\mueff|$ to smaller values ($\lesssim 500$~GeV) as has been advocated in the recent past~\cite{Baer:2012up, Baer:2013ava, Mustafayev:2014lqa, Baer:2015rja}. This opens up the interesting possibility of relatively light higgsinos~\cite{Jeong:2014xaa} just heavier than a singlino-like NLSP, the rich phenomenology of which we would discuss in detail. An MSSM-like scenario with a lighter higgsino-like NLSP (while the LSP is still bino-like) is also
viable and we would briefly comment on it in the~pNMSSM context.

However, a relatively small $|\mueff|$ is restricted by both DMDD experiments
\cite{Carena:2018nlf} 
and from the LHC constraints on the invisible decay width of $\hsm$~\cite{Arbey:2012na}.\footnote{Naively, since bino-higgsino mixing is driven by the gauge coupling $g_{1} \, (\sim 0.35)$, a bino LSP could allow a somewhat lower value of $|\mueff|$ without getting into conflict with these constraints when compared to the singlino LSP whose mixing with higgsinos is governed by `$\lambda$' which is generally larger than $g_1$ in the phenomenologically interesting region of the parameter space.}
On top of that, $|\mueff| \lesssim 500$~GeV seems to be getting into a soaring tension with the null results from the direct searches of ewinos at the LHC in the modes
$pp \to \charonepm \charonemp, \, \charonepm \ntrltwo$
followed by their exclusive decays to SM gauge and Higgs bosons~\cite{Sirunyan:2017lae, Sirunyan:2018ubx, Aaboud:2018jiw, Aaboud:2018sua, Aaboud:2018ngk, Aad:2019vnb, Sirunyan:2019iwo, Aad:2019vvf, Aad:2019vvi, Aad:2020qnn, ATLAS:2020ckz}, viz., 
\be
\begin{aligned}
\charonepm \to \wpm \ntrlone, \quad \ntrltwo &\to Z \ntrlone
 \quad &(WZ\mathrm{-mode}) \, , \\
 \ntrltwo &\to \hsm \, \ntrlone
 \quad &(WH\mathrm{-mode}) \, .
\end{aligned}
\label{eqn:wz-wh}
\ee
In the MSSM context, some recent studies
\cite{Calibbi:2014lga, Chakraborti:2014gea, Chakraborti:2015mra, Athron:2018vxy, Datta:2018lup}
have taken a critical look into the robustness of such constraints when these states are higgsino-like. Their production rates at the LHC are known to be smaller than the wino-like states of the same mass. 
This would result in a commensurate degradation of the experimental reach for such states. 
On top of that, a further degradation in sensitivity is unavoidable if the cascades
of such states follow patterns that are different from
the simplistic ones assumed by the LHC experiments. This is indeed possible in the NMSSM.
We thus focus on the region of the parameter space where the concerned ewinos
undergo cascades that involve NMSSM excitations (e.g., the singlet scalars or the singlino) that either compete with or dominate over the MSSM-like ones (involving
the $Z$-boson and $\hsm$).
These then result in relaxed bounds on the masses of such higgsino-like states
($\sim |\mueff|$)~\cite{Ellwanger:2018zxt, Domingo:2018ykx, Abdallah:2019znp} thus making a smaller $|\mueff|$ still viable.

As we will see, the interplay of DM and collider phenomenology for such a DM state is  a highly involved phenomenon. Contrary to an expectation naively guided
by what happens in the MSSM, the pNMSSM scenario indeed allows for
a highly bino-like DM over a broad range of its mass below $200$~GeV and over a significant region of the NMSSM parameter space, even when the sfermions are decoupled. In the process, a crucial role in the DM sector is played by larger values of $\tanb$ ($\geq 5$). On the one hand, this enhances the interaction between a singlet scalar, $\hs$ and/or $\as$ (predominantly,
$\as$)
and a bottom quark pair thus facilitating DM annihilation over an extended
range of its mass dictated by the mass of the singlet scalar state, i.e.,
$\mhs,\mas \approx 2 \mntrlone$.
On the other hand, as we will see, for smaller DM masses only such larger values of $\tanb$ could optimally tame the SI scattering rate of the DM 
by arranging for a so-called `blind spot' in the collective contribution of $CP$-even scalars 
to the same.
The other crucial parameter is `$\lambda$'. A moderate to large value of `$\lambda$' ($0.2 < \lambda  < 0.7$) is found to play a deciding
role in relaxing the LHC bounds on $|\mueff|$ optimally in the presence of a singlino-like NLSP so that a `minimal' but an adequate tempering of a bino LSP with higgsino 
(and hence singlino) admixture stands feasible down to
smaller DM masses.

The paper is organized as follows. In section~\ref{sec:scenario} we outline the
theoretical setup by summarizing the salient features of the NMSSM superpotential and its scalar and ewino sectors. To set the stage, the interplay of the scalars and the ewinos, in terms of correlations among their masses and interactions, is then discussed in necessary details, with a bino-like LSP in mind. Section~\ref{sec:results} presents the main results of this work by
referring to the phenomenology of DM relic abundance, its Direct Detection (DD) rates and some interesting collider aspects of such a scenario. Their collective implications for the future DMDD and the LHC experiments are also indicated. 
We present a few benchmark scenarios that capture the salient aspects
of the involved phenomenology.
Further, a brief observation is made on the issue of `naturalness' in the light of the findings of this paper. In section~\ref{sec:conclusions} we conclude.
A three-part appendix is presented at the end by (i) including the general form
of the scalar-neutralino-neutralino interactions in a convenient basis and to be exploited later, (ii) presenting for the very first time 
relevant results pertaining to the SI and the SD blind spots calculated in the bino-higgsino-singlino basis and finally (iii) discussing $\lambda$-dependence of the relevant
ewino partial widths in terms of the involved couplings that would help in understanding the deeper mechanism of how a small $|\mueff|$ could evade pertinent LHC bounds.
%
\section{The theoretical scenario}
\label{sec:scenario}
The superpotential
of the $Z_3$-symmetric NMSSM is given by~\cite{Ellwanger:2009dp}
\beq
{\cal W}= {\cal W}_\mathrm{MSSM}|_{\mu=0} + \lambda \widehat{S}
\widehat{H}_u \cdot \widehat{H}_d
        + {\kappa \over 3} \widehat{S}^3 \, ,
\label{eqn:superpot}
\eeq
where ${\cal W}_\mathrm{MSSM}|_{\mu=0}$ is the MSSM superpotential without the
higgsino mass term (the so-called $\mu$-term), $\widehat{H}_u, \widehat{H}_d$
are the $SU(2)$ Higgs doublet superfields of the MSSM while $\widehat{S}$ is the gauge
singlet superfield of the NMSSM. `$\lambda$' and `$\kappa$' are dimensionless coupling constants for the interaction among the doublet and the singlet
fields and the self-interaction of the singlet field, respectively. As the
singlet scalar field, `$S$' develops a vacuum expectation value~(\vev)~$\vs$, the
second term in eq.~(\ref{eqn:superpot}) gives rise to an effective $\mu$-term
given by $\mueff=\lambda \vs$ thus offering a solution to the baffling $\mu$-problem 
\cite{Kim:1983dt}. The corresponding soft supersymmetry (SUSY)-breaking Lagrangian is given by
\beq
-\mathcal{L}^{\rm soft}= -\mathcal{L_{\rm MSSM}^{\rm soft}}|_{B\mu=0}+ m_{S}^2
|S|^2 + (
\lambda A_{\lambda} S H_u\cdot H_d
+ \frac{\kappa}{3}  A_{\kappa} S^3 + {\rm h.c.}) \,,
\label{eqn:lagrangian}
\eeq
where the first term on the right-hand side gives the soft part of the MSSM Lagrangian but for
a vanishing soft term ($B\mu$) of the MSSM containing the higgsino mass
parameter `$\mu$' while $m_S$ stands for the soft SUSY-breaking mass
of the singlet scalar field, `$S$' and $\alambda$ and $\akappa$ are the trilinear
soft terms of the NMSSM with the dimensions of mass. The singlet
superfield~$\widehat{S}$ also yields a fermionic partner $\widetilde{S}$ known
as the singlino. Together, these constitute the defining aspects of the scalar (Higgs)
and the chargino/neutralino (ewino) sectors of the NMSSM that directly concern the present work. Hence
we briefly discuss the structure and the salient features of these 
two sectors in the subsequent subsections.
%
\subsection{The scalar (Higgs) sector}
\label{subsec:higgs}
%
The Lagrangian of eq.~(\ref{eqn:lagrangian}) contains the following soft
Lagrangian for the NMSSM Higgs sector that now includes soft masses and
couplings specific to the NMSSM, over and above the relevant MSSM terms:
\beq\label{2.5e}
-{\cal L}^\mathrm{soft} \supset
m_{H_u}^2 |H_u|^2 + m_{H_d}^2 | H_d |^2 
+ m_{S}^2 |S|^2
+\left(\lambda A_\lambda S H_u \cdot H_d + \frac{\kappa}{3}  A_\kappa
S^3 
+ \mathrm{h.c.}\right) .
\eeq
Perturbative calculations would involve expansion of the Lagrangian about the
real \vev's of the neutral Higgs fields ($H_d^0$, $H_u^0$) and the neutral
gauge-singlet scalar field ($S$), i.e.,  $v_d$, $v_u$ and $\vs$. Thus,
\beq\label{2.10e}
H_d^0 = v_d + \frac{H_{dR} + iH_{dI}}{\sqrt{2}} , \quad
H_u^0 = v_u + \frac{H_{uR} + iH_{uI}}{\sqrt{2}} , \quad
S = \vs + \frac{S_R + iS_I}{\sqrt{2}},
\eeq
where `$R$' and `$I$' stand for the $CP$-even and the $CP$-odd states. The symmetric squared mass matrix for the $CP$-even scalars in the basis
$H_{jR}=\{H_{dR}, H_{uR}, S_R\}$ is given by~\cite{Ellwanger:2009dp}
\beq
{\small{
{\cal M}_S^2 =
\left( \begin{array}{ccc}
  g^2 \vd^2 + \mueff \beff \,\tanb
& \:\; (2\lambda^2 - g^2) \vu \vd - \mueff \beff
& \:\; \lambda (2 \mueff \, \vd - (\beff + \kappa \vs) \vu) \\[0.2cm]
 \ldots
& \:\; g^2 \vu^2 + \mueff \beff /\tanb
& \:\; \lambda (2 \mueff \, \vu - (\beff + \kappa \vs)v_d) \\[0.2cm]
 \ldots
 &  \ldots
& \:\:\, \lambda \alambda  \frac{\vu \vd}{\vs} + \kappa \vs (\akappa + 4\kappa \vs)
\label{eqn:cp-even-matrix} 
\end{array} \right),}
}
\eeq
where $B_{\mathrm{eff}}=\alambda+\kappa\vs$, $g^2=(g_1^2+g_2^2)/2$, $g_1$ and $g_2$ are the $U(1)_Y$ and 
$SU(2)_L$ gauge couplings, respectively, and $\vu=v\sin\beta$, $\vd=v\cos\beta$ such that
$v^2=\vu^2+\vd^2 \approx (174$~GeV)$^2$ and $\tan\beta=\vu/ \vd$.
The $CP$-even (scalar, $h_i$) mass eigenstates are then given by
\bea
h_i &=& S_{ij} H_{jR}, \qquad  
\mathrm{with} \quad {i,j=1,2,3}\, ,
\label{eqn:cp-even-scalar-physical-states}
\eea
where the matrix `$S$' diagonalizes ${\cal M}_{S}^2$.

For the $CP$-even scalars, it is, however, more convenient to work in a rotated basis ($\hat{h}, \widehat{H}, \hat{s}$)~\cite{Miller:2003ay, Badziak:2015exr} where $\hat{h} = H_{dR} \cos{\beta} + H_{uR} \sin{\beta}$, $\widehat{H} = H_{dR} \sin{\beta} - H_{uR} \cos{\beta}$ and $\hat{s} = S_{R}$.
This is since $\hat{h} \, (\widehat{H})$ now mimics the SM Higgs (heavy
MSSM (doublet)-like $CP$-even Higgs) field. The physical $CP$-even scalar states are then denoted by 
\bea
h_i &=& E_{h_i \hat{h}} \hat{h} + E_{h_i \widehat{H}} \widehat{H} + E_{h_i \hat{s}} \hat{s} \,,
\label{eqn:hiEhhat}
\eea
where $E_{ab}$ is the matrix that diagonalizes the mass-squared
matrix for the $CP$-even scalars in the rotated basis. To be more specific, $E_{ab}$ stands for the admixture of the `$b$'-th scalar state in the `$a$'-th physical state. The symmetric squared mass matrix for the $CP$-odd scalars in the basis
$H_{jI}=\{H_{dI}, H_{uI}, S_I\}$ is given by~\cite{Ellwanger:2009dp}
\beq
{\small{
{\cal M'}_{P}^2 =
\left( \begin{array}{ccc}
  \mueff \beff \,\tanb
& \mueff \beff
& \lambda \vu (\alambda - 2\kappa \vs) \\[0.2cm]
  \ldots
& \mueff \beff /\tanb
& \lambda \vd (\alambda - 2\kappa \vs) \\[0.2cm]
  \ldots
&  \ldots
& \lambda (\beff + 3 \kappa \vs) \frac{\vu \vd}{\vs} - 3 \kappa \akappa \vs
\label{eqn:cp-odd-matrixp}
\end{array} \right).}
}
\eeq
For ${\cal M'}_{P}^2$,  a similar rotation, as used in the $CP$-even case, of the doublet basis states $H_{dI}$ and $H_{uI}$ would project out the massless Nambu-Goldstone mode that can be dropped. The orthogonal
state $A=\cos\beta H_{uI} + \sin\beta H_{dI}$ is identified as the MSSM (doublet)-like
$CP$-odd (pseudoscalar) Higgs boson and in the basis $\{A, S_I \}$ the mass-squared matrix for
the $CP$-odd scalar is reduced to
\beq\label{eqn:cp-odd-matrix}
{\small
{\cal M}_P^2 =
\left( \begin{array}{cc}
    m_A^2
~&~ \lambda (\alambda - 2\kappa \vs)\, v \\[0.2cm]
    \lambda (\alambda - 2\kappa \vs)\, v
~&~ \lambda (\alambda + 4\kappa \vs)\frac{v_u v_d}{\vs} -3\kappa \akappa \, \vs  
\end{array} \right),
}
\eeq
where, akin to that in the MSSM, $m_A^2= 2 \mueff \beff / \sin2\beta$. The $CP$-odd (pseudoscalar, $a_k$) mass eigenstates are given by
\bea
a_k = {\cal O}_{kA} A + {\cal O}_{kS_I} S_I, 
\qquad \mathrm{with} \quad {k=1,2},
\label{eqn:cp-odd-scalar-physical-states}
\eea
where the matrix `${\cal O}$' diagonalizes ${\cal M}_{P}^2$.
The elements of 
the matrix `${\cal O}$' and the matrix `$P$' that diagonalizes ${\cal M'}_{P}^2$ are related by~\cite{Ellwanger:2009dp}
\beq
 P_{i1} = \sin\beta \, {\cal O}_{iA} \,, \qquad 
 P_{i2} = \cos\beta\,  {\cal O}_{iA}\,, \qquad
 P_{i3} = {\cal O}_{iS_I} \,.
\eeq
The $CP$-odd mass eigenstates in terms of the basis $H_{jI}=\{H_{dI}, H_{uI}, S_I\}$ are then given by
\bea
a_k = P_{kj} H_{jI}, 
\quad \mathrm{with} \quad {k=1,2} \;\; \mathrm{and} \;\; j=1,2,3 \, .
\label{eqn:cp-odd-scalar-weak-states}
\eea
At tree-level, modulo possible mixing with the doublet states $H_{dR}$ 
and $H_{uR}$ (the so-called singlet-doublet mixing), the squared mass of the 
singlet-like $CP$-even physical state is given by the (3,3) component of
${\cal M}_S^2$, i.e.,
\beq
\mhssq \approx {\cal{M}}^2_{S,33} =  
 \lambda \alambda  \frac{v_u v_d}{\vs} + \kappa \vs (\akappa + 4\kappa \vs) \,.
\label{eqn:cp-even-mass}
\eeq
Furthermore, while the mass of the heavier of the two MSSM (doublet)-like
$CP$-even neutral physical states, $H$, is given by $m_H \approx m_A$ ($ m_A \gg m_Z$), the squared mass of the
lighter one mimicking $\hsm$, at one-loop level, is given by
\cite{Ellwanger:2011sk}
\beq
\mhsm^2 = m_Z^2 \cos^2 2\beta + \lambda^2 v^2 \sin^2 2\beta + \Delta_\mathrm{mix}+ 
\Delta_{\mathrm{rad.\,corrs.}} \, .
\label{eqn:hsmmass}
\eeq  
In the above expression, the first and the second terms stand for the tree-level MSSM and NMSSM contributions, respectively, while $\Delta_\mathrm{mix}$ stems from a possible singlet-doublet mixing and, in the weak mixing limit (as is now favored by the LHC data), is given by~\cite{Ellwanger:2011sk}
\beq
\label{eqn:hmix}
\Delta_{\rm mix} = \frac{4 \lambda^2 \vs^2 v^2 
(\lambda -\kappa \sin 2\beta)^2} {\tilde{m}_h^2-m_{ss}^2} \, ,
\eeq
where $\tilde{m}_h^2 = \mhsm^2 - \Delta_{\rm mix}$ and 
$m_{ss}^2 = \kappa \vs(A_{\kappa}+4 \kappa \vs)$. 
The term $\Delta_{\mathrm{rad.\,corrs.}}$ gives the MSSM-like radiative  corrections to
$\mhsm^2$ of which the dominant one at one-loop level, with the top quark and the top squarks (stops) in the loops, is given by
\cite{Haber:1996fp, Djouadi:2005gj}
\beq
\Delta_{\mathrm{rad.\,corrs.}}^{{\rm 1-loop}}\simeq \frac{3 m_t^4}{4 \pi^2 v^2 \sin^2 \beta}
\left[2 \log \frac{M_S}{m_t}
+ \frac{X_t^2}{ M_S^2} \left(1- \frac{X_t^2}{12 M_S^2} \right) \right],
\label{eqn:radcorr}
\eeq
where $m_t$ stands for the mass of the SM top quark,
$M_S=\sqrt{m_{\widetilde{t}_1} m_{\widetilde{t}_2}}$,
$m_{\widetilde{t}_1 (\widetilde{t}_2)}$ being the mass of the lighter (heavier) 
physical top squark state and $X_t=A_t-\mueff \cot\beta$, $A_t$ being the soft trilinear coupling for the top sector.
On the other hand, the squared
mass of the singlet $CP$-odd scalar, up to some mixing with its doublet cousin,
is given by
\beq
\massq \approx {\cal{M}}^2_{P,22} =  
\lambda (\alambda + 4\kappa \vs)\frac{v_u v_d}{\vs} -3\kappa \akappa \, \vs \, .
\label{eqn:cp-odd-mass}
\eeq
Eqs.~(\ref{eqn:cp-even-mass}), (\ref{eqn:hsmmass}) and (\ref{eqn:cp-odd-mass})
indicate nontrivial dependencies of the masses of the singlet scalars
and the SM-like Higgs boson on various input 
parameters of the $Z_3$-symmetric pNMSSM. For example, as is well-known,
$\hsm$ could be either heavier or lighter than $\hs$ and the mass of the
latter is not always strongly correlated with the mass of its $CP$-odd counterpart,
$\as$, although both are singlet states having a common origin. As we
will find, such correlations (or, for that matter, their absence)
could crucially govern their collective phenomenology.
\subsection{The ewino sector}
\label{subsec:ewinos}
As in the MSSM, the ewino sector of the NMSSM is comprised of the neutralino
and the chargino sectors with the exception that the former is now augmented
with a singlino ($\widetilde{S}$) state, the singlet fermionic degree of
freedom that owes its origin to the gauge-singlet superfield appearing in the 
NMSSM  superpotential indicated in eq.~(\ref{eqn:superpot}). As a result,
the symmetric neutralino mass matrix has got a dimensionality of $(5\times 5)$ 
and, in the basis $\psi^0=\{\widetilde{B},~\widetilde{W}^0, ~\widetilde{H}_d^0,
~\widetilde{H}_u^0, ~\widetilde{S}\}$, is given by~\cite{Ellwanger:2009dp}
\beq
\label{eqn:mneut}
{\cal M}_0 =
\left( \begin{array}{ccccc}
\mone & 0 & -\dfrac{g_1 \vd}{\sqrt{2}} & \dfrac{g_1 \vu}{\sqrt{2}} & 0 \\[0.4cm]
\ldots & \mtwo & \dfrac{g_2 \vd}{\sqrt{2}} & -\dfrac{g_2 \vu}{\sqrt{2}} & 0 \\
\ldots & \ldots & 0 & -\mueff & -\lambda \vu \\
\ldots & \ldots & \ldots & 0 & -\lambda \vd \\
\ldots & \ldots & \ldots & \ldots & 2 \kappa \vs
\end{array} \right) \,,
\eeq
where $\mone$ and $\mtwo$ are the soft SUSY-breaking masses for the $U(1)_Y$
and the $SU(2)_L$ gauginos, i.e., the bino and the wino, respectively,  while
the same for the singlino, appearing in the (5,5) element, is given by
$\msinglino = 2\kappa \vs$.
The neutralino mass-eigenstates ($\ntrli$, $i=1, 2,\ldots,5$, in order of
increasing mass as `$i$' increases), in terms of the weak eigenstates
($\psi_j^0$, with $j=1, 2,  \ldots, 5$), are given by
\beq
\ntrli = N_{ij} \psi_j^0 \,,
\label{eqn:diagN2}
\eeq
where `$N$' is the $(5 \times 5)$ matrix that diagonalizes the neutralino
mass-matrix in eq.~(\ref{eqn:mneut}).\footnote{We would finally work with a
neutralino system where the wino is decoupled. For the resulting $(4 \times 4)$ system, various useful relations among $N_{ij}$'s are given explicitly in appendix
\ref{appsec:blindspot}.} 

On the other hand, the ($2 \times 2$) chargino mass matrix of the
NMSSM is same as that of the MSSM but for $\mu \to \mueff$ and, in the bases
$\psi^+ = \{ -i \widetilde{W}^+, \, \widetilde{H}_u^+ \}$ and
$\psi^- = \{ -i \widetilde{W}^-, \, \widetilde{H}_d^- \}$, is given by~\cite{Ellwanger:2009dp}
\beq
{\cal M}_C = \left( \begin{array}{cc}
                    \mtwo   & \quad  g_2 \vu \\
                 g_2 \vd  & \quad \mueff 
             \end{array} \right) .
\eeq
The asymmetric matrix ${\cal M}_C$ can be diagonalized by two ($2 \times 2$) 
unitary matrices `$U$' and `$V$':
\beq
U^* {\cal M}_C V^\dagger = \mathrm{diag} (\mcharone , \mchartwo) \; ; \quad
\mathrm{with} \;\;  \mcharone < \mchartwo  \,.
\label{eqn:uvmatrix}
\eeq

In this work, we focus on a scenario where the LSP DM is a highly bino-like neutralino. 
As noted in the Introduction, to make sure that our setup remains reasonably `natural', we focus on relatively low values of $|\mueff|$, subject to its 
current lower bounds from the experiments. A priori, this yields two relatively 
light higgsino-like neutralinos along with a higgsino-like lighter chargino with
masses $\sim |\mueff|$. Also, we would assume a decoupled wino, i.e., a large~$\mtwo$. This would keep the analysis simpler without having to forego any of
its major features by prohibiting wino(s) in the cascades of heavier ewinos and, 
at the same time, by aiding compliance with stringent bounds on the former
by rendering their direct production rate small enough at the LHC.\footnote{The combined cross section for the associated production of wino-like charginos and a wino-like neutralino at the LHC energies is roughly twice as large the combined one for higgsino-like charginos and higgsino-like neutralinos of which there
are two nearly degenerate states~\cite{Ellwanger:2018zxt}.}

The NMSSM-specific effects in the ewino sector get to be rather interesting when 
the soft singlino  mass $\msinglino =2 \kappa \vs$ becomes comparable to both $\mone$ 
and  $\mueff$. It may be gleaned from eq.~(\ref{eqn:mneut}) that the mixings
of the singlino with the higgsinos are governed directly by~`$\lambda$' while
its mixing with the bino is absent. At the lowest order, such a mixing only occurs through the higgsino portal, though the same can get enhanced for
$|\mone| \lesssim |\msinglino| \sim |\mueff|$. Furthermore, depending on whether 
$|\msinglino| < |\mueff|$ or $|\msinglino| > |\mueff|$, the NLSP would be singlino- or higgsino-dominated. Note that, in the latter case, the neutralino sector reduces to a 
typical bino-higgsino system of the MSSM when the singlino also gets decoupled.
In any case, the actual masses and contents of various neutralinos depend much
on the extent of mixing among various states and could be critically important 
for both DM and collider~phenomenologies.

The setup under consideration thus has one light chargino which is
higgsino-like and four relatively light neutralinos with the lightest one of them
(the LSP) being bino-like while the others are singlino- and higgsino-like.
If the higgsino-like 
neutralinos are lighter than the singlino-like one (while the wino is decoupled), phenomenology at the collider
(LHC) would mostly resemble an MSSM scenario with a higgsino-like NLSP except under a situation when the mass-split between the higgsinos and the LSP is rather
small~\cite{Dutta:2014hma}. However, 
the phenomenology of the DM sector might still be somewhat sensitive to a not so
heavy singlino-like state given even a small admixture that it could lend to the 
bino-like LSP neutralino. On the other hand, a singlino-like NLSP might
affect the collider phenomenology as well since higgsinos, which could be
copiously produced at the colliders, can now cascade to the former. This implies
that the stringent lower bounds on the masses of such ewinos that could be placed by different LHC analyses within the framework of simplified SUSY scenarios might get significantly relaxed thus warranting thorough reanalyses that admit such
possibilities.
\subsection{The interplay of the scalars and the ewinos}
\label{subsec:interplay}
%
The interplay of the relatively light scalars and the ewinos is known to be central 
to the DM and the collider phenomenologies alike in the $Z_3$-symmetric pNMSSM 
scenario. In ref.~\cite{Abdallah:2019znp} we highlighted some of the
crucial aspects of such an interplay in the context of a scenario with a 
singlino-dominated LSP. The study presented here has an unmistakable parallel to that but with 
a flipped mass-hierarchy between the singlino- and the bino-like states. 
Nonetheless, there are crucial quantitative differences which may not be unexpected.

As is common to a scenario with a single SUSY DM candidate, the one under 
consideration is also plagued with a tension between the observed values of
relic abundance and the upper bound on the DMDD cross section. Similar to what 
happens in the MSSM, a pair of pure bino LSPs cannot funnel via $Z$- or the Higgs 
boson as those do not couple to the latter two states. Furthermore, its efficient 
coannihilations with a sfermion and/or an ewino are also not possible if we
assume the latter ones to be much heavier than the LSP. Thus, for a light enough 
bino-like LSP (say $\mntrlone \lesssim 200$~GeV), the only way it could become a 
viable DM candidate is if it acquires some higgsino admixture and if $\mntrlone 
\approx \mhsm/2$ or $\mntrlone \approx m_Z/2$.\footnote{The $Z$-funnel region in 
the MSSM, in general, is nearing complete exclusion due to various recent experimental constraints on the DM sector and those coming from searches for 
the ewinos at the LHC~\cite{Pozzo:2018anw, Barman:2020zpz}.} This is true in the NMSSM as well but now 
with a further possibility that the SM-like Higgs funnel could get efficient if, in 
addition, there is some singlino admixture in the LSP. Notably, there could be 
additional funnels available thanks to relatively light singlet-like scalars of 
the NMSSM which can be either heavier than $\hsm$ or lighter than the $Z$-boson or could even have a mass in between $m_Z$ and $\mhsm$.
However, the tension shows up in the form that if efficient scalar (Higgs) 
funnels are to be banked on for compliance with the observed DM relic abundance, 
the SI cross section tends to breach the upper bound on it. This is since 
the same interactions ($h_k$-$\ntrli$-$\ntrlj$) govern these two phenomena. On the 
other hand, a similar problem arises with the SD cross section when the
$Z$-funnel is in action both of which are governed by the 
$Z$-$\ntrli$-$\ntrlj$ couplings. 

Ramifications of such a scenario at the collider experiments primarily depend on
the altered branching ratios of the heavier ewinos to the lighter ones (that 
include the LSP) in the presence of light singlet-like scalars and/or a singlino-like neutralino. As pointed
out in the Introduction, a multitude of LHC constraints in the form of bounds
on the masses of the ewinos, derived assuming so-called `simplified models'
within the framework of the phenomenological MSSM (pMSSM), could now get
significantly relaxed. In particular, such relaxations for the higgsino-like
ewinos could open up the parameter space with relatively low values of $|\mueff|$
and hence with improved `naturalness'. In our present scenario in which all sparticles
but for the bino- and the singlino-dominated neutralinos, the higgsino-dominated 
neutralinos and chargino and the low-lying doublet- and singlet-like scalars
(both of $CP$-even and $CP$-odd variants) are taken to be too massive and hence
decoupled, the relevant decay branching ratios are mostly governed by the 
interactions of the heavier neutralinos to their lighter cousins and the scalar
states ($h_k/a_k$-$\ntrli$-$\ntrlj$) and the $Z$-boson ($Z$-$\ntrli$-$\ntrlj$). These are the
very same interactions that we have discussed in the previous paragraph in the context of DM 
phenomenology. Furthermore, interactions involving the lighter chargino, the neutralinos lighter than the same and the $W^\pm$-boson, $W^\mp$-$\charonepm$-$\ntrli$, could also be rather instrumental. In fact, situations exist when the lighter chargino, whose decay branching ratio
to the $W^\pm$-boson and the LSP is usually considered to be 100\% by the LHC experiments, could  have a reasonable 
branching ratio to a neutralino other than the LSP and a $W^\pm$-boson. This could potentially relax
the published experimental bounds further.

Clearly, the following few issues would primarily set the contour of the interplay between the scalar (Higgs) and the ewino sectors in the current context: 
(i) masses of the singlet-like scalars, (ii) the same masses and the observed masses of
the $Z$-boson and the SM-like Higgs boson in relation to those of the bino-like LSP and the other 
ewinos, (iii) mass-splittings among the ewinos themselves including the LSP  and
(iv) compositions of the LSP and the other ewinos. All of these have intricate 
dependencies on various input parameters of the pNMSSM. In the subsequent
subsections, we highlight the mass-correlations that exist among the singlet-like states and the interactions among the scalars, the ewinos (that include both singlet and doublet states) and the SM gauge boson.
%
\subsubsection{Correlations among masses of the scalars and the ewinos}
\label{subsubsec:masses}
Correlations among masses of the Higgs-like states and the neutralinos
(higgsino- and singlino-like) have recently been studied in some detail in
ref.~\cite{Baum:2019uzg} in the so-called alignment (without decoupling)
limit~\cite{Carena:2015moc, Baum:2017gbj, Baum:2020vfl} which allows for relatively light non-SM Higgs
states ($CP$-even/odd, singlet/doublet), i.e., with masses $\lesssim 1$~TeV
and hence are likely to be within the reach of the LHC. However, the said work
studies such correlations in a `natural' setup for which MSSM-like radiative
corrections to the SM-like Higgs mass have to be, at best, moderate while the
tree-level NMSSM contribution ($\sim \lambda^2 v^2 \sin^2 2\beta$) to the same is
maximized by choosing relatively large values of `$\lambda$' ($>0.5$) and 
by restraining $\tanb$ to low values ($1 < \tanb < 5$). In contrast, as pointed
out in the Introduction, we open our scans to regions with smaller values
of `$\lambda$' as well. Furthermore, we look for relatively light singlet-like scalar
states ($\lesssim 500$~GeV) while the non-SM doublet scalars (MSSM-like heavier Higgs states) are taken to be indeed
heavy ($>1$~TeV) and hence decoupled. These requirements, in contrast to the
scenario discussed in ref.~\cite{Baum:2019uzg}, would render the
mass-correlation among the corresponding heavy and light eigenstates relatively
milder. Under the circumstances, the correlations among the masses of the
light singlet states (scalars and the singlino) assume importance and worth
a closer look.

The masses of the $CP$-even singlet and the doublet (SM Higgs boson) and that of the
$CP$-odd singlet scalars as described by eqs.~(\ref{eqn:cp-even-mass}), 
(\ref{eqn:hsmmass}) and (\ref{eqn:cp-odd-mass}) along with the mass of the singlino 
appearing as the (5,5) element of the neutralino mass matrix of eq.~(\ref{eqn:mneut}) present us with a rich set of observables whose values are 
intricately correlated. Such correlations have an important bearing on the 
phenomenology and their understanding is paramount for its systematic study.
Generally, there are two crucial implications of such correlations.
First, the masses of these singlet-like scalars and the singlino-dominated state 
control the DM physics in a rather involved manner by influencing
both DM relic abundance and DMDD rates. Second, the
mass-hierarchy among these states would dictate their decay patterns and hence 
what to expect at a collider like the LHC.

As can be seen from eqs.~(\ref{eqn:cp-even-mass}) and (\ref{eqn:cp-odd-mass}),
the masses of the singlet-like scalars, at the lowest order, have rather
involved dependencies on as many as six input parameters like $\mueff$, $\tanb$, $\lambda$,
$\kappa$, $\alambda$ and $\akappa$ while masses of the neutralinos have
relatively simpler (well-known) dependencies on the first four of them plus that
on the two soft masses $\mone$ and $\mtwo$. Thus, it is not quite unexpected that the masses of
the singlet scalars and the singlino would be correlated. In fact, in the
so-called decoupling limit with $\lambda, \kappa \rightarrow 0$ or for a sizable
$\tanb$ and small `$\lambda$', `$\kappa$' and $\alambda$ for which mixing among the
singlet and doublet scalars can be ignored, the following sum-rule holds
\cite{Das:2012rr, Ellwanger:2018zxt}:
\begin{equation}
\label{eqn:sumrule}
{\cal M}_{0,55}^2 \simeq {\cal M}_{S,33}^2 + {1 \over 3} {\cal M}_{P,22}^2
\hskip 10pt \Rightarrow \hskip 10pt 
\msinglino^2 \simeq \mhssq + {1 \over 3} m_{a_{_S}}^2 \, ,
\end{equation}
where $\mhs$ ($m_{a_{_S}}$) is the mass of the singlet-like $CP$-even (odd) scalar excitation.

\begin{figure}[t]
\centering
\hspace*{-0.4cm}
\includegraphics[height=0.27\textheight, width=0.53\columnwidth ,
clip]{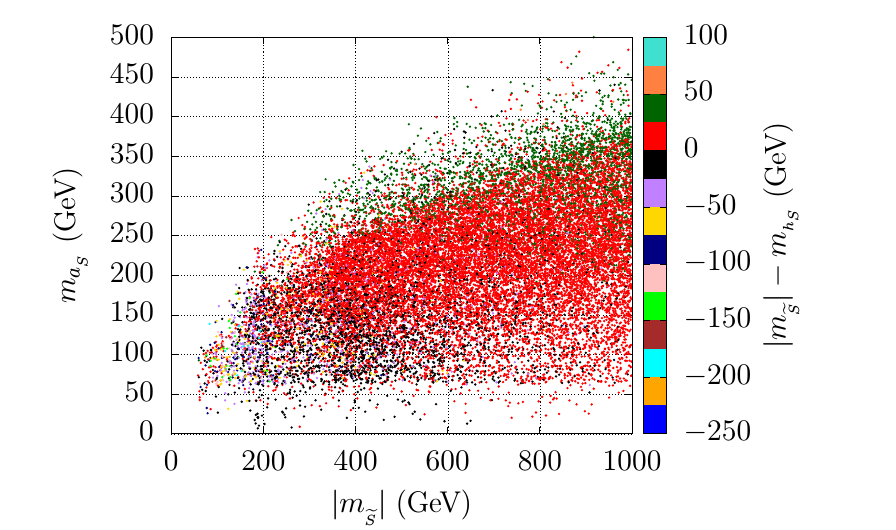}~~
\hskip -18pt
\includegraphics[height=0.27\textheight, width=0.53\columnwidth ,
clip]{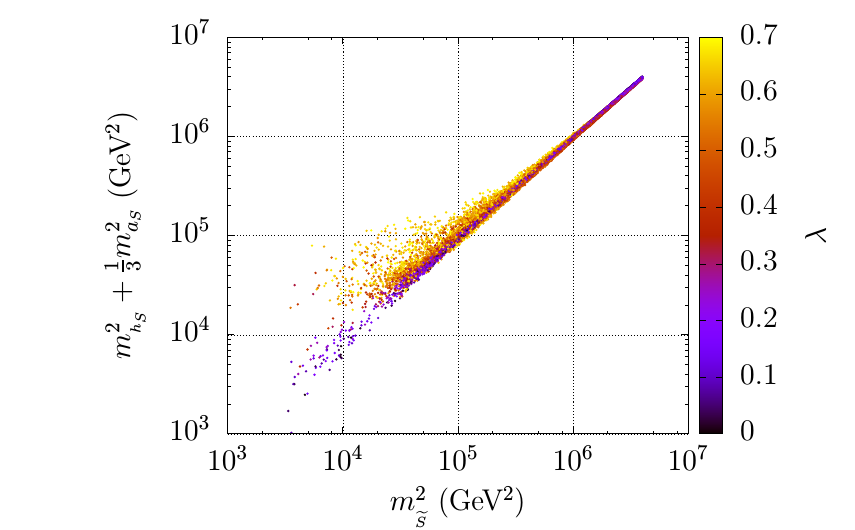}
\caption{Scatter plots
showing the mass-correlation among $|\msinglino|$, $\mas$ and $\mhs$ 
in the $|\msinglino|$--$\mas$ plane
with the (signed) split between $|\msinglino|$ and $\mhs$,
$\Delta_{\msinglino, \mhs} (=|\msinglino|-\mhs)$ being indicated via the palette (left)
and illustrating the validity of (and departure from) the sum-rule of eq.~(\ref{eqn:sumrule}) in the plane formed by the quantities on its left- and
right-hand sides while the palette indicates the magnitude of `$\lambda$' (right). All 
points satisfy various important constraints coming from the DM sector,
colliders and low energy experiments (see section~\ref{sec:results} for
details).
} 
\label{fig:mass-correlation}
\end{figure}

Scatter plots in figure~\ref{fig:mass-correlation} present the
correlations among the masses of the singlet-like $CP$-even and $CP$-odd scalars and the 
singlino-like neutralino (left panel) and illustrate to what 
extent these masses conform to the sum-rule presented in eq.~(\ref{eqn:sumrule}) (right panel). These result from a scan, using the package \nmssmtools-v5.4.1~\cite{Ellwanger:2004xm, Ellwanger:2005dv, Das:2011dg}, over various pNMSSM
parameters the ranges of which are indicated in table~\ref{tab:ranges}.

It can be found from the left plot of figure~\ref{fig:mass-correlation} that 
for larger values of $|\msinglino|$, $\mhs \sim |\msinglino|$.
This (up to some higher order corrections as incorporated in \nmssmtools), already follows from the lowest order expression of $\mhssq$ in eq.~(\ref{eqn:cp-even-mass}) given that the magnitude of $\akappa$ (in the second 
term) never gets large in our scan ($|\akappa|<100$~GeV) and that the first
term being proportional to $\lambda^2$ (as $\vs = \mueff/\lambda$) may be ignored when `$\lambda$' is on the smaller side. Also, for any given value of
$|\msinglino|$ there is a maximum possible value of $\mas$ determined by the
largest value of $|\akappa|$ in use as mentioned above. In fact, the maximum
$\mas$ attainable gets to be increasingly smaller when compared to the 
corresponding value of $|\msinglino|$. This can be understood 
from eq.~(\ref{eqn:cp-odd-mass}) when `$\lambda$' is small (note that this is the limit in which the sum-rule of eq.~(\ref{eqn:sumrule}) is obtained), for which the first term in the same could be ignored and hence
$\massq \approx |-3 \kappa \akappa \vs| = |-{3 \over 2} \akappa \msinglino|$. We have checked that the sharper edge in the distribution (which divides the
plot into a dense red region at the bottom and a sparse region in green above) 
indeed satisfies this relation. In fact, the latter region corresponds to larger `$\lambda$'
values.
\begin{table}[t!]
\begin{center}
\begin{tabular}{|c|c|c|c|c|c|c|c|c|}
\hline
\makecell{Varying \\ parameters}  & $\lambda$ & $|\kappa|$ & $\tanb$& \makecell{$|\mueff|$ \\ (TeV)}&  \makecell{$|\alambda|$ \\ (TeV)} &
\makecell{$|\akappa|$ \\ (GeV)}
 & \makecell{$|\mone|$ \\ (GeV)} & \makecell{$A_t$ \\ (TeV)} \\
\hline
Ranges  & 0.001--0.7 & $\leq$ 0.7 &
1--65 & $\leq 1$ & $\leq$ 10 & $\leq 100$ & $\leq$ 200 & 0--10 \\
\hline
\end{tabular}
\caption{Ranges of various model  parameters employed for
scanning the $Z_3$-symmetric pNMSSM parameter space. All parameters are defined
at the scale
$Q^2 = (2m_{\widetilde{Q}}^2 + m_{\widetilde{U}}^2+m_{\widetilde{D}}^2)/4$,
except for $\tanb$ which is defined at $m_Z$. Masses of 
the gluino and all sfermions from the first two generations (third generation) are kept fixed at 5~TeV (5.5~TeV). The wino mass, $\mtwo$, is kept fixed at 2.5~TeV.}
\label{tab:ranges}
\end{center}
\vspace{-0.5cm}
\end{table}

The situation is, however, different for smaller values of $|\msinglino|$
($\lesssim 250$~GeV) when the first and the third terms in eq.~(\ref{eqn:cp-odd-mass}), which are initially ignored, could turn out to be
comparable in magnitude with the second term. This allows for
$\mas > |\msinglino|$ for such low values of $|\msinglino|$. A little loosely connected patch at the bottom, 
left corner is the only region populated to a certain extent by points for
which $\mhs < \mhsm$ where even the singlet-like $CP$-odd scalar could be heavier than 
its $CP$-even counterpart. We have checked that this region extends over
$60~\mathrm{GeV} \lesssim |\msinglino| \lesssim 120~\mathrm{GeV}$ and
$\frac{\mhsm}{2} \lesssim \mas \lesssim 150~\mathrm{GeV}$. Over this region $\mhs$ tends to 
drop as $\mas$ increases for a given $|\msinglino|$ with
$\mhs > |\msinglino| (\mhs < |\msinglino|)$  at smaller (larger)~$\mas$.

One crucial information from this plot is that one can have a light enough
$\as$ while $\hs$ and the singlino-like neutralino could be much heavier.
An important implication of such a spectrum is that the DM annihilation could just
take advantage of an $\as$-funnel which would need only some higgsino admixture
in the LSP while the SI cross section does not receive much contribution from a heavy enough $\hs$. Note that the bottom part of this 
plot is sparsely populated. This is since $\mhsm > 2\mas$ over this region 
which allows for the decay $\hsm \to \as \as$ and hence attracts significant 
constraint from the LHC experiments~\cite{Aaboud:2018fvk, Aaboud:2018gmx, Aaboud:2018iil, Aaboud:2018esj, Aad:2020rtv, Sirunyan:2018mbx, Sirunyan:2018mot, Sirunyan:2020eum}.

The right plot of figure~\ref{fig:mass-correlation} presents the 
distribution of the scatter points in a plane with the left-hand (right-hand) 
side of eq.~(\ref{eqn:sumrule}) along the abscissa (ordinate) while the 
palette-color indicates the values of `$\lambda$'. That the distribution 
along the diagonal tends to have a lower magnitude of `$\lambda$' clearly 
corroborates the sum-rule holding rather robustly in this regime. Again, 
as expected, one can notice a departure from the sum-rule
for larger values of `$\lambda$'.

Switching to the MSSM ewino sector, it is noted that the mass of the highly bino-like LSP, which is in the focus of the present work, is practically determined by the chosen value of $\mone$ ($\mntrlone \approx \mone$) except in the region of the pNMSSM parameter space where it might receive appreciable admixture(s) of the singlino and/or higgsino components. Note that unlike in the scenario with a singlino-like LSP DM, for which finding a suitable annihilation funnel involving singlet scalar(s) is restricted by their correlated masses, $\mone$ can be varied independently of the parameters defining the singlet sector of the pNMSSM. This offers an added freedom to a bino-like LSP DM in finding singlet scalar(s) as the funnel state(s). Notwithstanding, the existing mass-correlation in the singlet sector could still be phenomenologically important when the singlino happens to be the NLSP.
Furthermore, the absolute values of the masses of the two higgsino-like neutralinos would mostly be
$\approx |\mueff|$ except in situations when these have non-negligible mixings with the singlino, in which case the mass of the singlino-like neutralino would also depart from the value it takes as  a pure state, i.e., $\msinglino$. In any case, the lighter chargino would always be an almost pure higgsino state given that we set $\mtwo$ at a much larger value (2.5 TeV). Thus, the heavier chargino and one of the heaviest neutralinos would be wino-like and would decouple. This renders the scenario crucially simpler thus allowing for a systematic, analytical study without having to compromise any of its major~features.

It should be noted here that in a scenario with a higgsino-like NLSP,
masses of two such nearly degenerate neutralinos have to comply with stringent bounds that can be derived from 
the LHC experiments. Otherwise, the phenomenology of the pNMSSM ewino sector would resemble that of the MSSM (but for the presence of possibly light
singlet-like scalars) when the singlino-like state is much heavier. However, in a scenario with a singlino-like NLSP, the 
NLSP itself would not attract such bounds and hence can be much lighter. In 
addition, its presence could potentially relax the existing mass-bounds on 
the heavier higgsino-like ewinos~\cite{Ellwanger:2018zxt, Abdallah:2019znp}. 
For, the latter might now cascade to new final states by decaying first to the NLSP 
neutralino thus weakening the experimental sensitivities to those final 
states from which the bounds have been actually~extracted.
%
\subsubsection{Interactions among the scalars, the SM gauge bosons and the ewinos}
\label{subsubsec:interactions}
Much of phenomenology, both in the DM sector and in the collider front, is
dictated by how the ewinos, some of which may well be relatively light, interact with the 
relatively light scalar (Higgs) states and the SM gauge bosons. Observed relic 
abundance requires an efficient enough DM annihilation. The presence of a suitable 
funnel-state in the form of the above scalar excitations and/or the $Z$-boson may not prove sufficient unless the involved interactions have the required minimal strengths. However, 
the same interactions often control the DM-nucleus scattering cross sections\
that are experimentally highly constrained by the DMDD
experiments. The tension is quite generic for 
scenarios with a single DM candidate. On the other hand, collider phenomenology 
involving these ewinos depends crucially on their mass-hierarchies,
as highlighted in section~\ref{subsubsec:masses}, provided involved 
interactions remain favorable.

As compared to the MSSM, in the NMSSM the interactions of neutralinos with 
scalar (Higgs) states ($h_i$ and $a_i$) are additionally dictated by terms 
proportional to `$\lambda$' and `$\kappa$' in the superpotential of eq.~(\ref{eqn:superpot}). Thus, in the weak basis, while the gaugino-higgsino-doublet 
Higgs interactions are the same as in the MSSM, a higgsino now have a
$\lambda$-driven interaction with the singlino and a doublet Higgs boson or 
with another higgsino and a singlet scalar state. The generic 
coupling of the $CP$-even and $CP$-odd scalars, $\Phi_i \owns h_i, a_i$, with an 
arbitrary pair of neutralinos $\ntrlj$ and $\ntrlk$, can be cast in the following compact 
form by using eqs.~(\ref{eqn:hinjnk1}) and (\ref{eqn:ainjnk1})~\cite{Ellwanger:2009dp}:
\bea
g_{_{\Phi_i \ntrlj \ntrlk}}
 &=& 
\bigg[ {\lambda \over \sqrt{2}} \, ( M_{i 1} N_{j4} N_{k5} + M_{i 2} N_{j3}N_{k5} + M_{i 3} N_{j3}N_{k4}) - {\kappa \over \sqrt{2}} \, M_{i3} N_{j5} N_{k5} \nonumber \\
&+& \frac{1}{2} (-1)^n  (g_1 N_{j1} - g_2 N_{j2}) \, (M_{i1} N_{k3}  - M_{i2} N_{k4})
\bigg]
 + \bigg[ j \leftrightarrow k \bigg] \, ,
\label{eqn:generic-sinjnk}
\eea
where the matrix elements $M_{il} \owns S_{il}, iP_{il}$ with $S_{il}$ and $P_{il}$ are as defined in eqs.~(\ref{eqn:cp-even-scalar-physical-states}) and (\ref{eqn:cp-odd-scalar-weak-states}) for the $CP$-even 
and $CP$-odd sectors, respectively and $n=0 \, (1)$ for $\Phi \equiv h_i \, (a_i)$.  More useful forms of this compact expression, in terms of convenient basis states for the $CP$-even and $CP$-odd scalars as discussed in section~\ref{subsec:higgs}, are provided in eqs.~(\ref{eqn:hinjnk-reduced-without-approximation}) and
(\ref{eqn:ainjnk-reduced-without-approximation}). 
Given that reference to eq.~(\ref{eqn:generic-sinjnk}) (or, for that matter, to eqs.~(\ref{eqn:hinjnk-reduced-without-approximation}) and (\ref{eqn:ainjnk-reduced-without-approximation})) in our subsequent discussions would be unavoidable, a little description of the
same would be in order. In table~\ref{tab:interactions} we present such a schematic description.
\begin{table}[t]
\begin{center}
\bgroup
\def\arraystretch{1.4}
\begin{tabular}{|@{\hspace{0.15cm}}c@{\hspace{0.15cm}}|@{\hspace{0.15cm}}c@{\hspace{0.15cm}}|@{\hspace{0.15cm}}c@{\hspace{0.15cm}}|@{\hspace{0.15cm}}l@{\hspace{0.15cm}}|}
\hline
Terms (eq.~(\ref{eqn:generic-sinjnk})) & Interaction & Explicit & \hskip 60pt Role(s) played  \\
  (ordered) & states & dependencies & \hskip 18pt (for a bino-dominated LSP DM) \\
\hline
 & & & {\small Moderate (important) for $\lambda < (>) \, g_1$} for \\
& & &
 {\small DM annihilation ($j$=$k$=1) via scalar funnels} \\
 1. $M_{i1} N_{j4} N_{k5}$ & $H_d \higgsinou \singlino$ & & and DM scattering ($j$=$k$=1) off nuclei. \\  
& &   & 
 {\small May assist DM coannihilation ($j(k)$=1} \\
 2. $M_{i2} N_{j3} N_{k5}$ & $H_u \higgsinod \singlino$ & $\lambda$& with {\small $k(j)$=2) and neutralino cascade decays} \\
&  & & {\small ($j \neq k$) for $\widetilde{H}$- or $\singlino$-like NLSP.
       Important} if \\
 3. $M_{i3} N_{j3} N_{k4}$ & $S \higgsinod \higgsinou$ & & {\small $\mone$, $\mueff$ and
        $\msinglino$ are optimally 
  close by.} \\
  & & & {\small The third coupling is the most crucial one} \\
  & & & {\small for the $\as$-funnel.} \\
\hline
  & & &  {\small Instrumental for assisted coannihilation } \\
4. $M_{i3} N_{j5} N_{k5}$ & $S \singlino \singlino$ & $\kappa$ &{\small involving the singlino-like NLSPs} \\
   & & & {\small and the $\as$-funnel.} \\
\hline
5. $M_{i1} N_{j1} N_{k3}$ & $H_d \bino \higgsinod$ & $g_1$ & 
   {\small Generally dominant, in particular when} \\
6. $M_{i2} N_{j1} N_{k4}$ & $H_u \bino \higgsinou$ & (MSSM-like) & 
 {\small $\mone$ and $\mueff$ are close by and $\lambda \ll g_1$.} \\
\hline
7. $M_{i1} N_{j2} N_{k3}$ & $H_d \,\wino^0 \higgsinod$ & $g_2$ & 
   {\small Same as above but for $\mone \to \mtwo$ \& $g_1 \to g_2$.} \\
8. $M_{i2} N_{j2} N_{k4}$ & $H_u\, \wino^0 \higgsinou$ & (MSSM-like) & 
   {\small Dropped in this work ($\mtwo$ made very heavy).} \\
\hline
\end{tabular}
\egroup
\caption{Components of the neutralino-neutralino-scalar (Higgs) coupling, $g_{_{\Phi_i \ntrlj \ntrlk}}$ of eq.~(\ref{eqn:generic-sinjnk}), the participating interaction states present as admixtures in the neutralinos and the scalars,
the model parameter on which each of those explicitly depend on and a brief
description of salient roles the components play in reference to a bino-dominated LSP.}
\label{tab:interactions}
\end{center}
\vspace{-0.5cm}
\end{table}

It should, however, be noted that while the explicit dependencies of different components of the coupling on various parameters, as indicated in table~\ref{tab:interactions}, are already obvious from eq.~(\ref{eqn:generic-sinjnk}), the very same parameters (along with some
others) also play important roles in determining admixtures of various participating interaction states within the involved neutralinos
and the scalars. This way, the actual dependencies of the coupling $g_{_{\Phi_i \ntrlj \ntrlk}}$ on such parameters are of a rather involved nature carrying much phenomenological implication.
These interactions play critical roles in DM annihilation via various scalar funnels thus affecting the relic abundance. The same interactions also control the
appearance of the all-important blind spot
\cite{Cheung:2012qy, Cheung:2014lqa, Badziak:2015exr, Badziak:2016qwg, Badziak:2017uto, Cao:2018rix} and determine the SI cross section. The other generic interaction which is important for our study is that among a pair
of neutralinos and the $Z$-boson. In contrast to gauge interactions of a doublet-like Higgs state with a pair of neutralinos where the former only sees the gaugino admixture in one neutralino and the higgsino admixture in the other (entries 5 to 8 in table~\ref{tab:interactions}), the interaction of the $Z$-boson with such a pair is only sensitive to the higgsino admixtures in both the neutralinos and is given~by 
\beq
g_{_{Z \ntrlj \ntrlk}} =\frac{g_2}{2 \cos\theta_W}(N_{j3} N_{k3} - N_{j4} N_{k4}) \,,
\label{eqn:znjnk}
\eeq
where $\theta_W$ is the weak mixing (Weinberg) angle. Thus, for a LSP pair, i.e., for $j$=$k$=1, the strength of the interaction goes as $g_{_{Z \ntrlone \ntrlone}} \sim N_{13}^2 - N_{14}^2$. This interaction
dictates the cross section for DM annihilation via $Z$-funnel as well as
that for DM-nucleon scattering in the SD case where it could as well hit a blind spot  when $|N_{13}|\approx |N_{14}|$. This happens trivially for $\tanb=1$ both in the MSSM and in the NMSSM. In addition, in the NMSSM such a situation can arise for a bino-like LSP in the presence of a comparably light singlino-like state (see eq.~(\ref{eqn:sigma-SD-1})). On the other hand, for $j \neq k$ eqs.~(\ref{eqn:generic-sinjnk}) and (\ref{eqn:znjnk}) control the rate of coannihilation of the LSP neutralino with the NLSP neutralino ($j(k)=1$ with $k(j)=2$). These also control
some of the crucial cascade decays of the heavier neutralinos to their
lighter cousins along with scalars or the $Z$-boson.
Similarly, the interaction of $W^\pm$-boson with a chargino and a neutralino is given in eq.~(\ref{eqn:gamma-cha-nj-w}).
In the present work, this interaction controls decays of the lighter chargino to 
any neutralino lighter than it and a $W^\pm$-boson. As pointed out in the Introduction, these decays could 
potentially render the simplistic assumption made by the recent
LHC analyses on the branching ratio of the lighter chargino invalid which, 
in turn, is expected to result in relaxation of the mass-bounds on the ewinos derived thereof.

It is thus clear that the interactions that are instrumental for the DM 
phenomenology depend crucially on the higgsino and the singlino admixtures of
the LSP which is taken to be bino-like in our present study. As we will 
see later, these admixtures also affect phenomenology at the colliders in a
nontrivial manner. It is further noted that such admixtures have rather 
intricate variations over the NMSSM parameter space. It follows from eq.~(\ref{eqn:mneut}) that variations with $\tanb$ and `$\lambda$' might be of particular
importance. In addition, the signs on~various mass parameters appearing in eq.~(\ref{eqn:mneut}) also have crucial bearings on these admixtures.

\begin{figure}[t]
\begin{center}
\vspace{-0.5cm}
\hspace{-0.2cm}\includegraphics[width=0.24\linewidth]{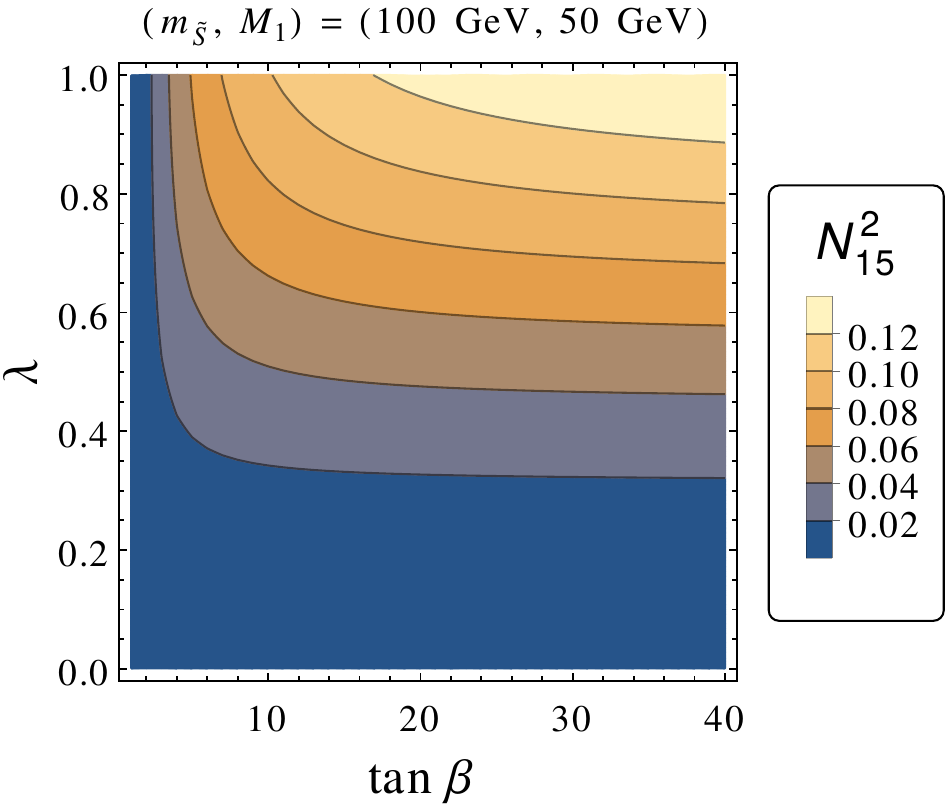}~~
\includegraphics[width=0.24\linewidth]{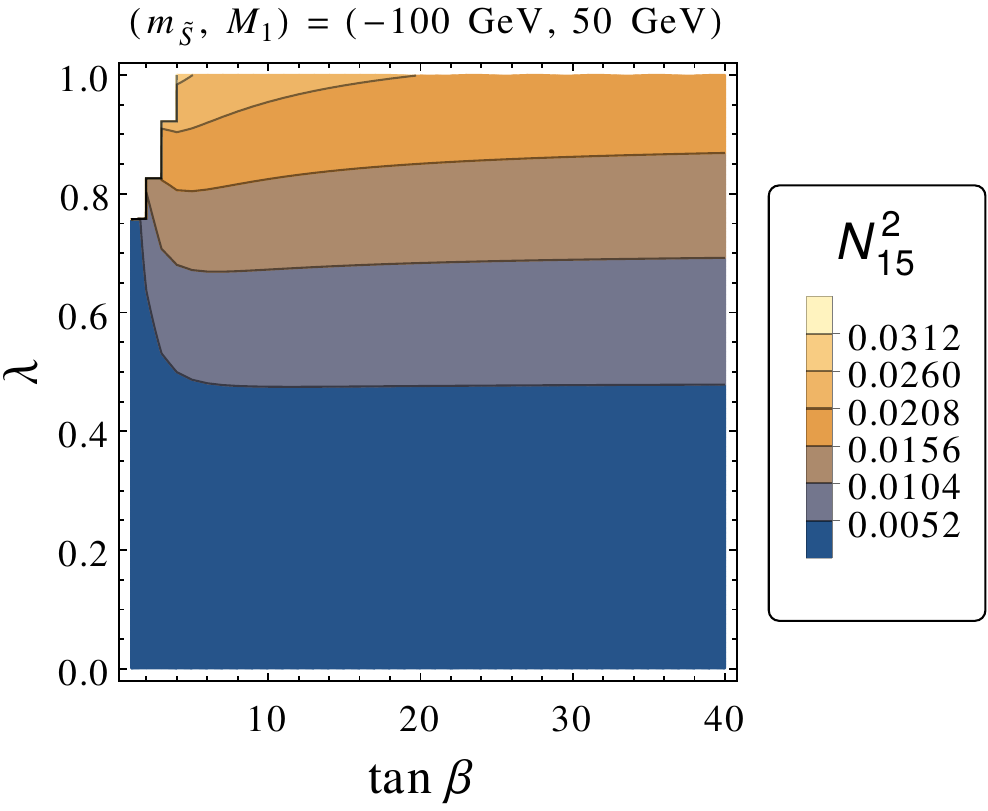}~~
\includegraphics[width=0.24\linewidth]{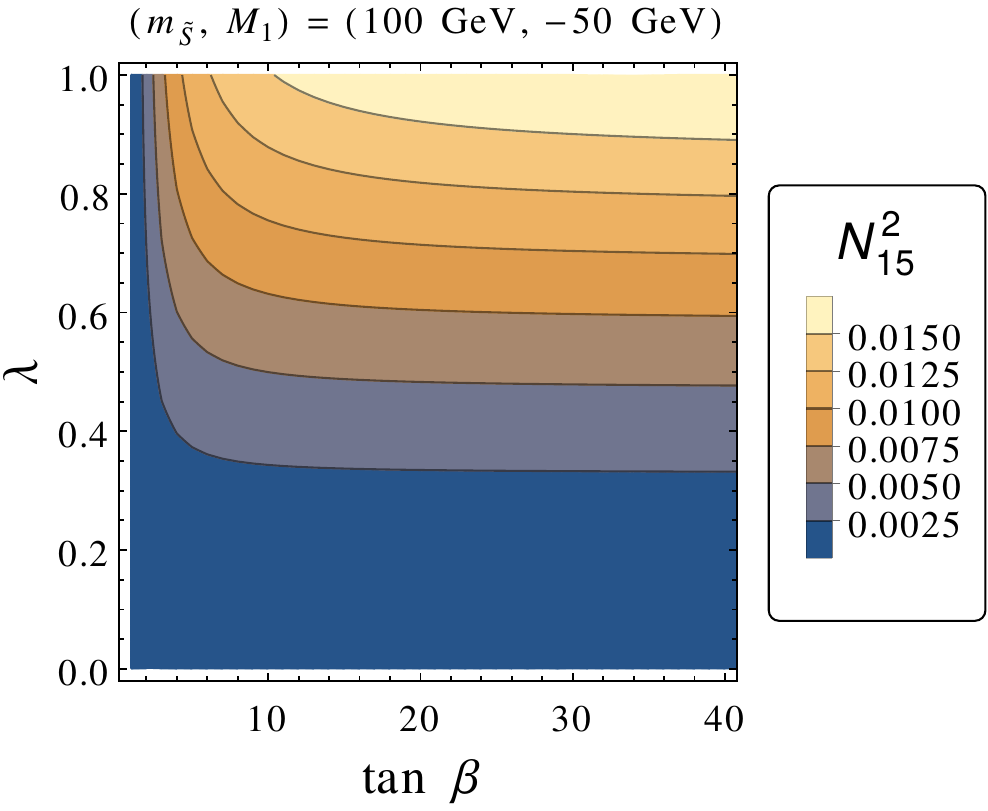}~~
\includegraphics[width=0.24\linewidth]{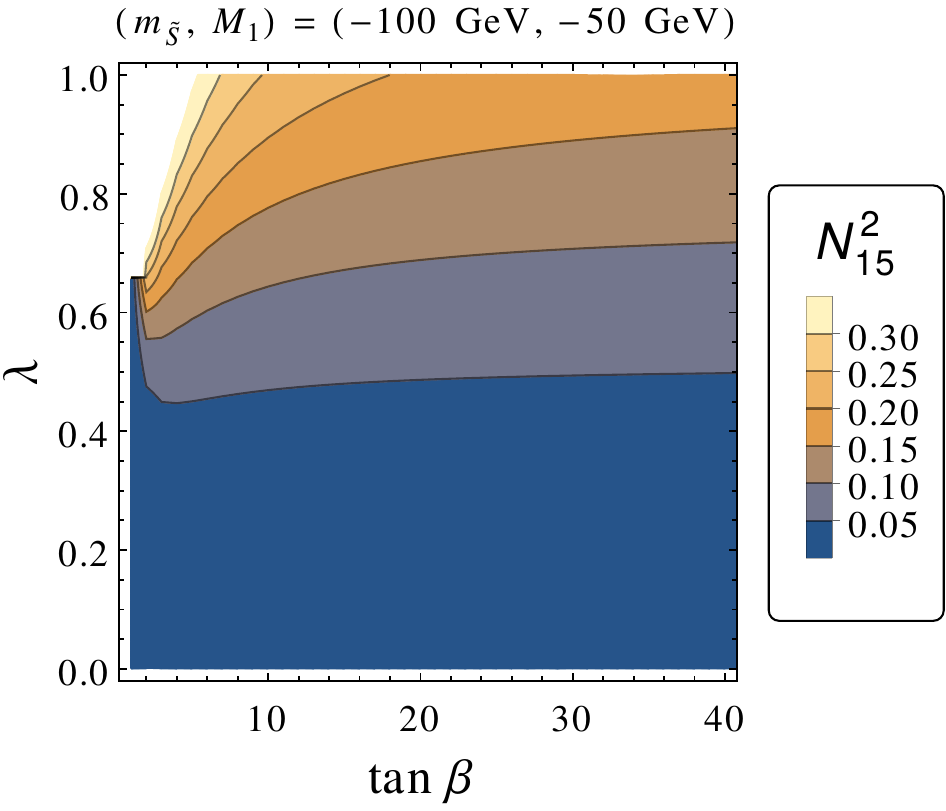}\\
\vskip 10pt
\hspace{-0.2cm}\includegraphics[width=0.24\linewidth]{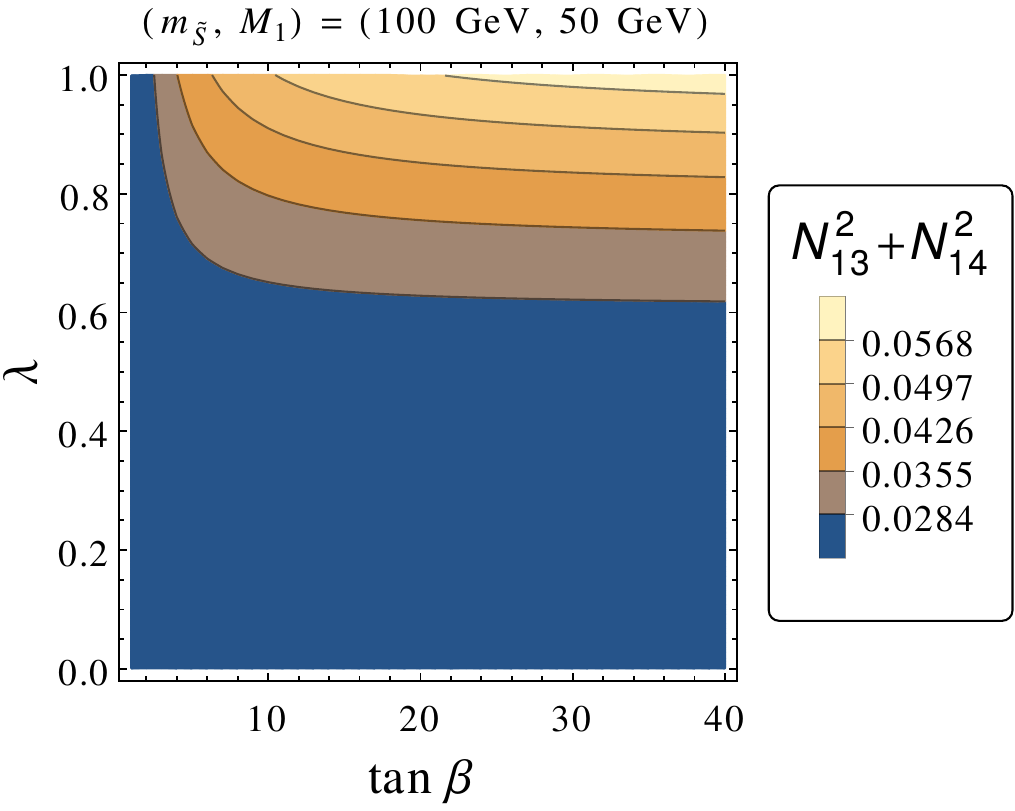}~~
\includegraphics[width=0.24\linewidth]{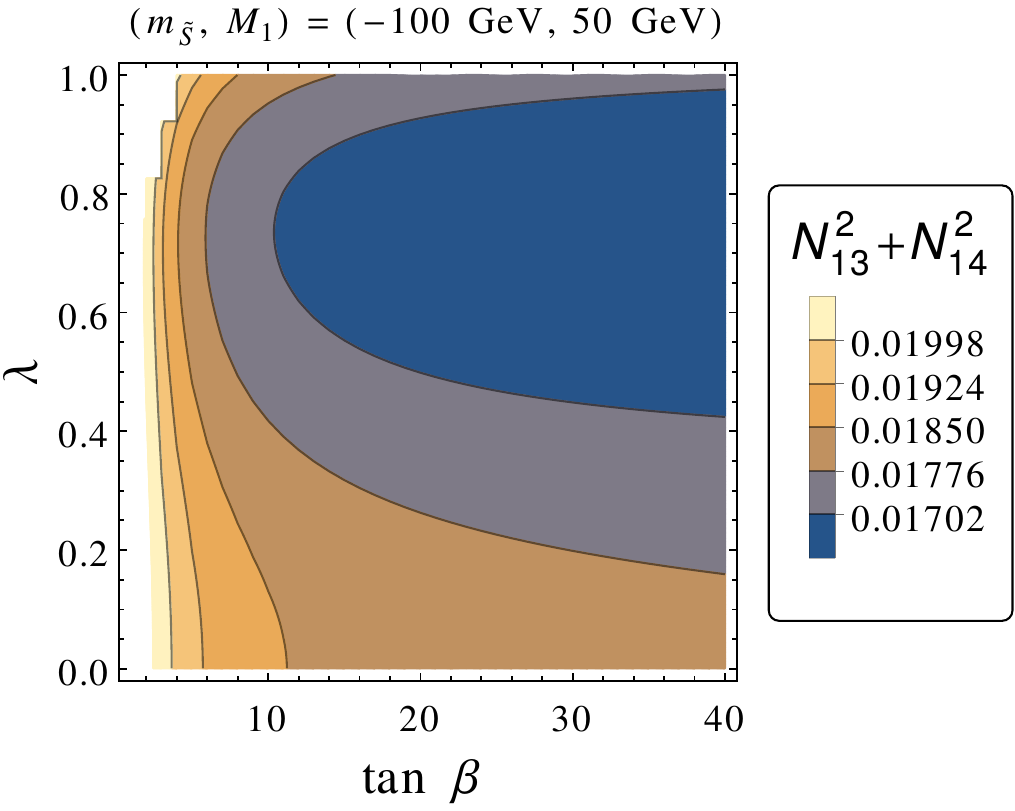}~~
\includegraphics[width=0.24\linewidth]{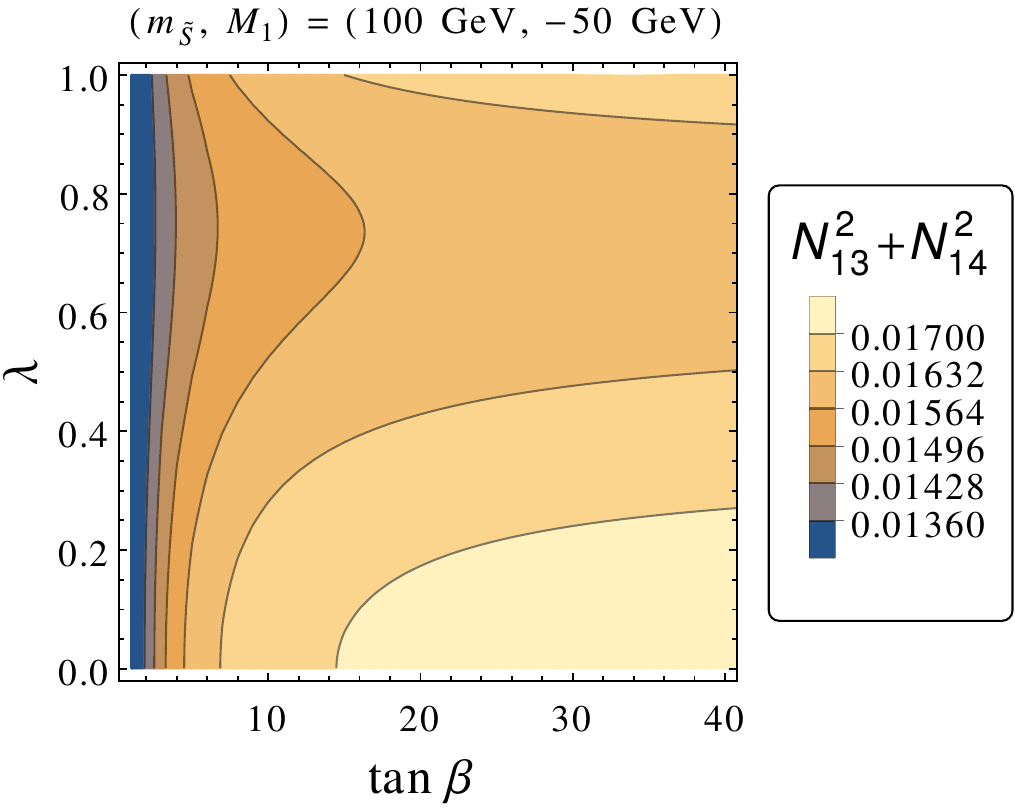}~~
\includegraphics[width=0.24\linewidth]{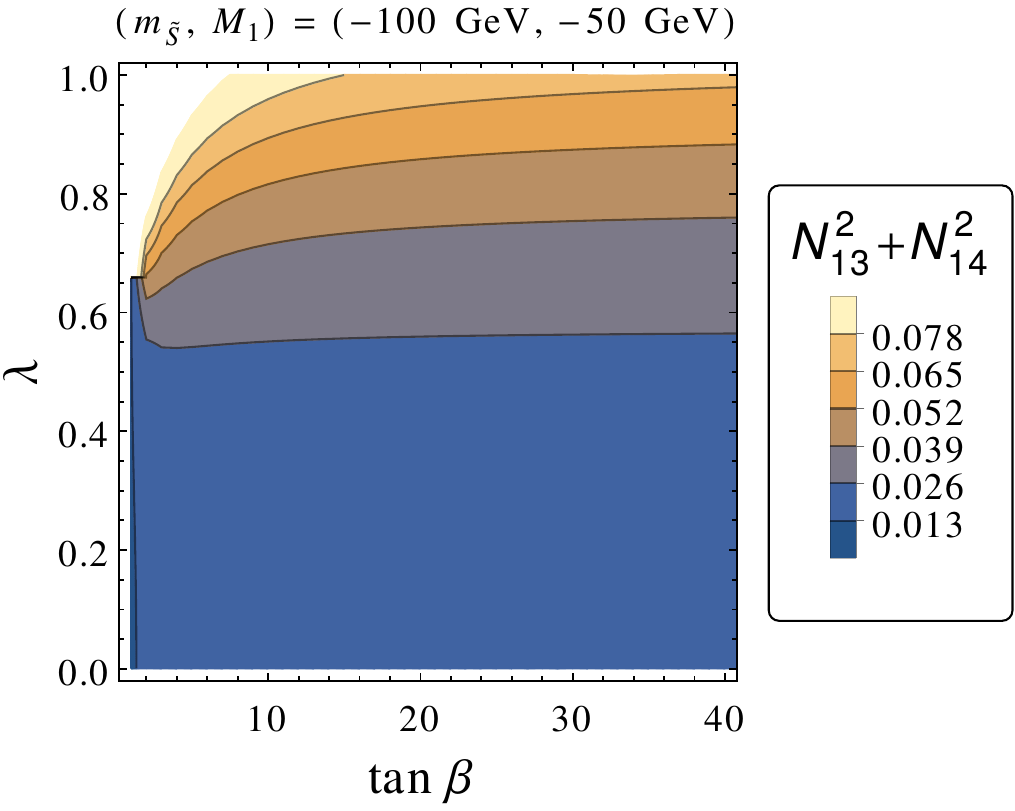}\\
\vskip 10pt
\hspace{-0.2cm}\includegraphics[width=0.24\linewidth]{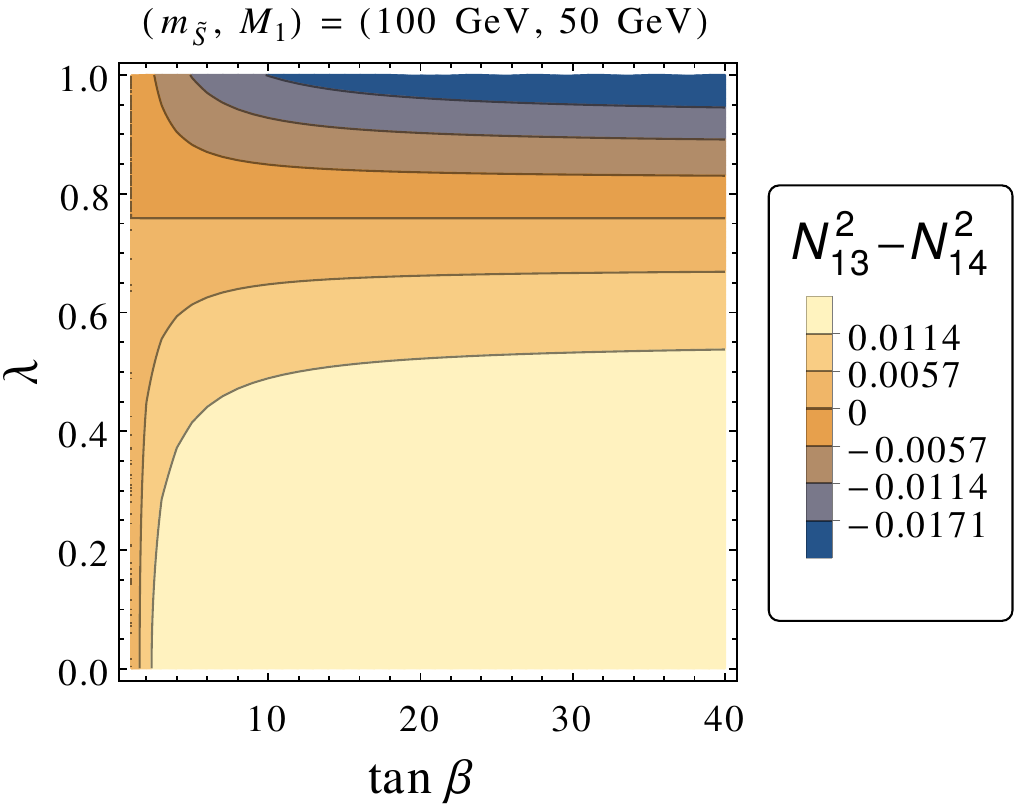}~~
\includegraphics[width=0.24\linewidth]{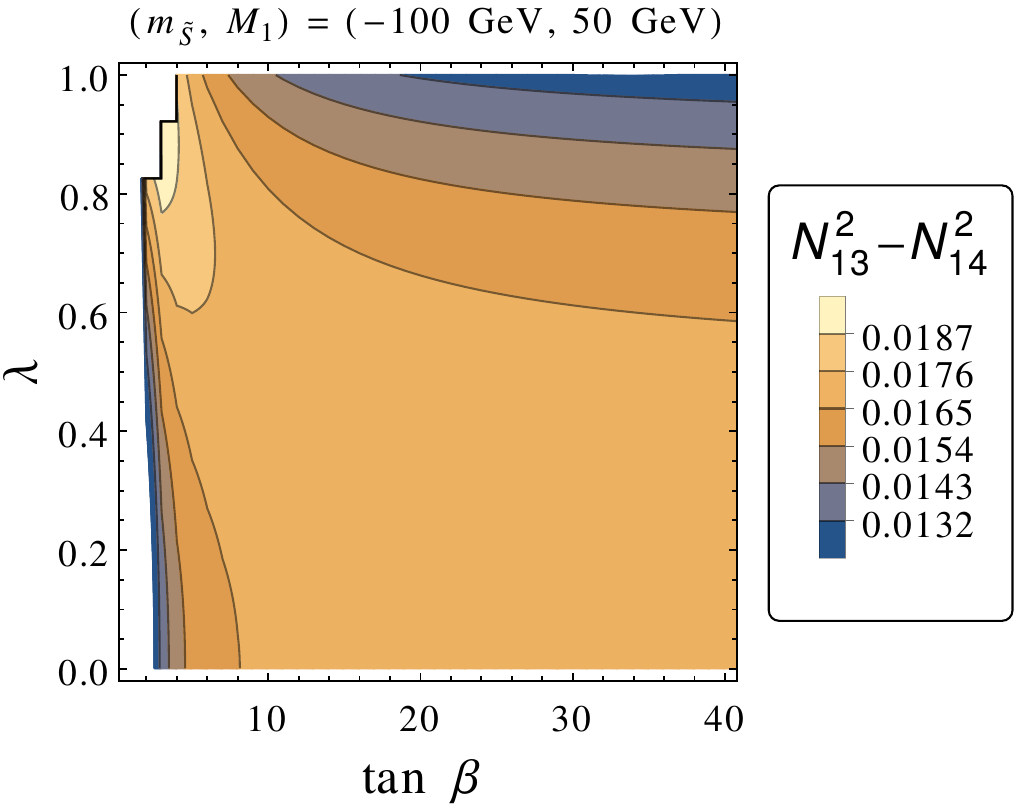}~~
\includegraphics[width=0.24\linewidth]{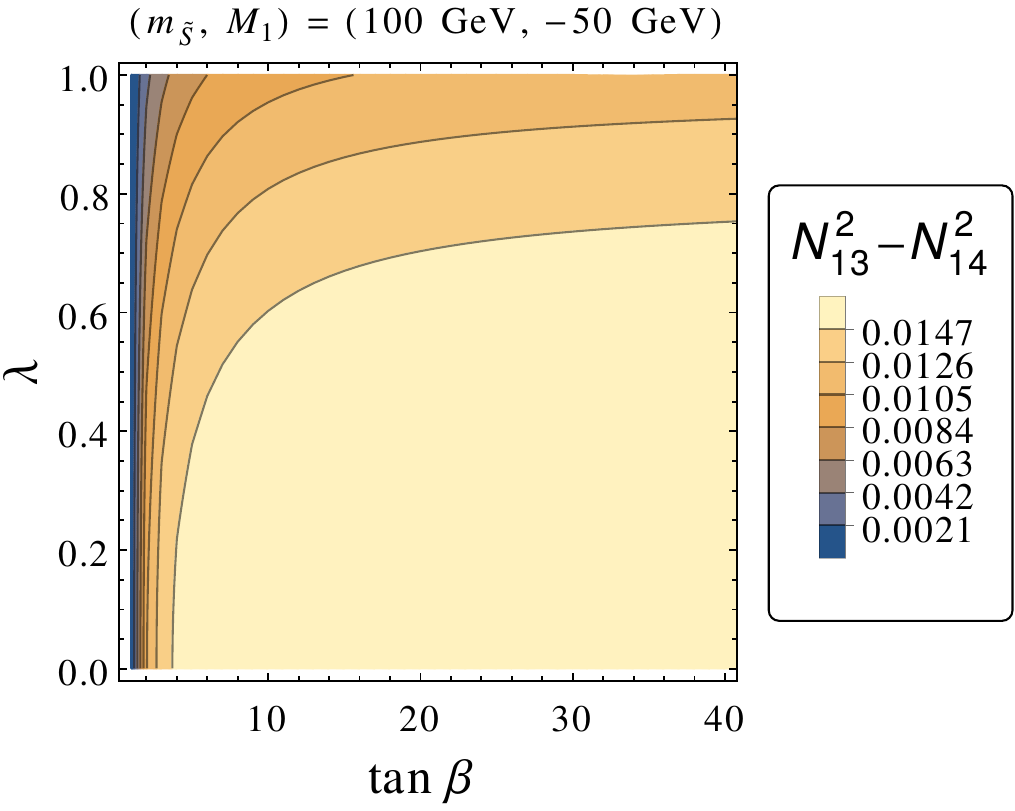}~~
\includegraphics[width=0.24\linewidth]{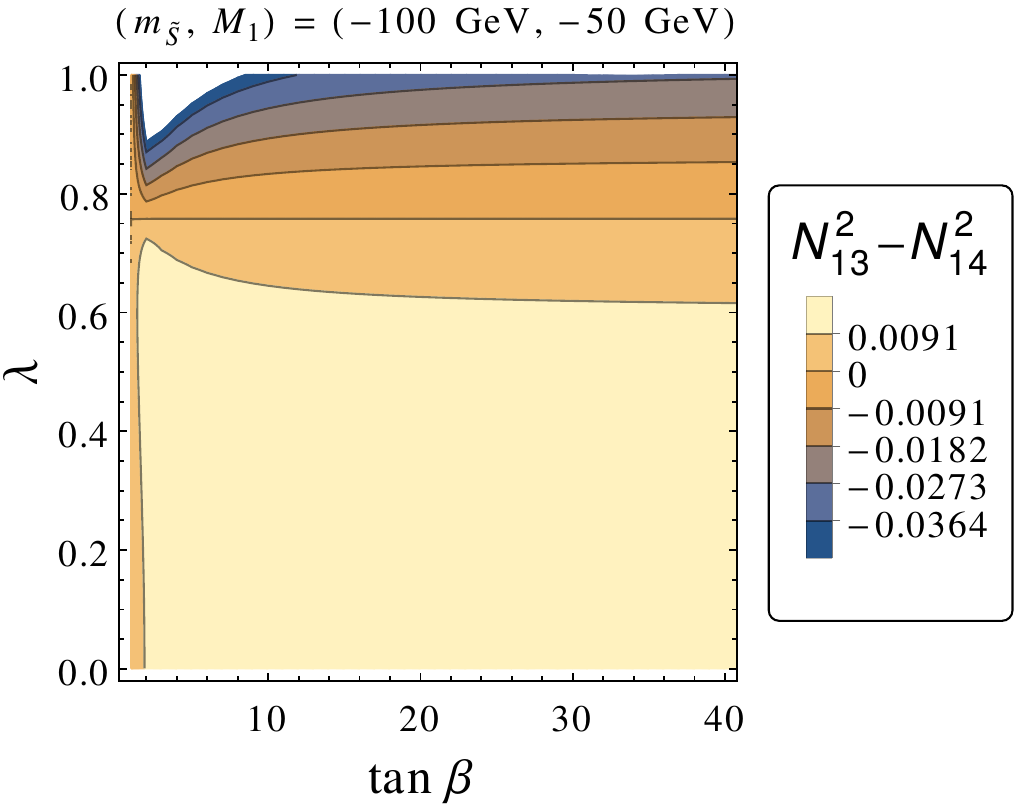}
\caption{Variations of the singlino ($N_{15}^2$, top panel), the total higgsino 
($N_{13}^2+N_{14}^2$, middle panel) admixtures and the quantity
$N_{13}^2-N_{14}^2$ (bottom panel) in an LSP (with at least 50\% bino-content) 
in the $\tanb$--$\lambda$ plane for a fixed value of
$\mueff$ (=\,350~GeV). In each row the first (last) two plots are 
with positive (negative) values of $\mone$ with $|\mone|=50$~GeV while
$|\msinglino|=100$~GeV with flipped signs on it, as indicated on the top of
each plot. Adjacent palettes indicate the magnitudes of the 
respective quantities whose variations are plotted.
$\mtwo$ is taken to be 2.5~TeV.}
\label{fig:higgsino-singlino}
\end{center}
\vspace{-0.7cm}
\end{figure}

In figure~\ref{fig:higgsino-singlino} we present such variations in the $\tanb
$--$\lambda$ plane by keeping $|\mone|$ ($=50$~GeV), $|\msinglino|$ ($=100$~GeV) and
$\mueff$ ($=350$~GeV) fixed. These fixed magnitudes are broadly representative of the phenomenologically interesting part of the parameter space we study in this work. Although these parameters could take arbitrary signs, it is found that due to the presence of a freedom in redefining the corresponding fields, only two such signs are physical.\footnote{A similar observation is made albeit in the MSSM context in ref.~\cite{Cheung:2012qy}.} This can now be clearly seen from eqs.~(\ref{eqn:n^2j5-value}), (\ref{eqn:sigma-SD-1}) and (\ref{eqn:total-higgsino}). Thus, we could 
fix the sign on one of these mass parameters (say, $\mueff$) while 
considering possible sign-combinations (four of them) for the rest two. 
The top row shows the variations of the singlino admixture ($N_{15}^2$) 
while the middle row does so for the higgsino admixture ($N_{13}^2+N_{14}^2$) in the LSP that has at least 50\% bino admixture, for various combinations of relative sign between $\mone$ and~$\msinglino$. In the bottom row we present similar
variations of the quantity {$N_{13}^2-N_{14}^2$}, the coupling factor that controls the 
annihilation of the LSP via the $Z$-funnel and the SD 
cross section. Following are a few important observations to make.
%
\begin{itemize}
\item In the top row, the singlino component in the LSP increases with increasing
$\tanb$ and~`$\lambda$' when $\msinglino$ is positive, irrespective of the sign
on $\mone$ (see the first and the third plots). In contrast, for negative
$\msinglino$ (see the second and the fourth plots), increasing $\tanb$ might decrease the singlino component in the LSP to some extent, in particular, for larger `$\lambda$' values. In the middle row, the higgsino admixture is found to have very
different patterns of variation for the four sign-combinations. In particular, note that  for relative signs between the mass parameters (the middle two plots) the higgsino content in the LSP could drop as `$\lambda$' grows unlike in the cases with no relative signs between the said mass parameters.\\[-0.7cm]
\item When variations of $N_{15}^2$ (top row) and $N_{13}^2+N_{14}^2$
(middle row) are compared, one can find the corresponding ones to be rather
similar when there is no relative sign between $\msinglino$ and $\mone$
(see the first and the fourth plots in the corresponding rows).\\[-0.7cm]
\item For a given hierarchy of input mass
parameters (as is the case here), there can be drastic differences in
the characteristic ranges over which the admixtures could vary for different
relative signs among these parameters. These absolute magnitudes of the admixtures are no less crucial for both DM and collider phenomenology
when compared to their patterns of variations over the parameter space.    
\item Plots in the bottom row have a generic feature in yielding
small to vanishing values of $N_{13}^2-N_{14}^2$ for $\tanb \to 1$ irrespective
of the value of `$\lambda$'. This is prominent in the third and the fourth plots from this row. Another generic feature is the vanishing values
of this quantity at a relatively large value of `$\lambda$', this time, being virtually independent of $\tanb$ and is favored for $\msinglino$ and $\mone$ having the same sign. Both these situations can be understood by studying eq.~(\ref{eqn:sigma-SD-1}). The value of `$\lambda$' for which such a vanishing of the quantity $N_{13}^2-N_{14}^2$ occurs depends on the extent of bino-singlino mixing. The larger this mixing is, i.e., the smaller the split between $\msinglino$ and $\mone$ is, the smaller the value of `$\lambda$' that suffices. Furthermore, on both sides of this `$\lambda$' value, $|N_{13}^2-N_{14}^2|$ increases with varying `$\lambda$'.
\end{itemize} 

The hierarchy among the three mass parameters $\mone$, $\msinglino$ and
$\mueff$ also have an interesting bearing on the admixtures of various 
components (gaugino, singlino and higgsino) in a given neutralino including 
those in the LSP. The differences in magnitudes among these mass parameters
$\mone$, $\msinglino$ and $\mueff$ control the said admixtures. For the present
study, it is useful to have an understanding of how various components in an otherwise bino-dominated 
LSP vary when, for a fixed smaller value of  $\mone$ and a fixed $\mueff$,
$|\msinglino|$ approaches $\mone$. As illustrated in figure~\ref{fig:higgsino-singlino}, the nature of these variations could well depend
on the chosen values of `$\lambda$', $\tanb$ and the relative signs among
the three mass parameters.
%
\begin{figure}[t!]
\begin{center}
\includegraphics[height=5.5cm,width=0.43\linewidth]{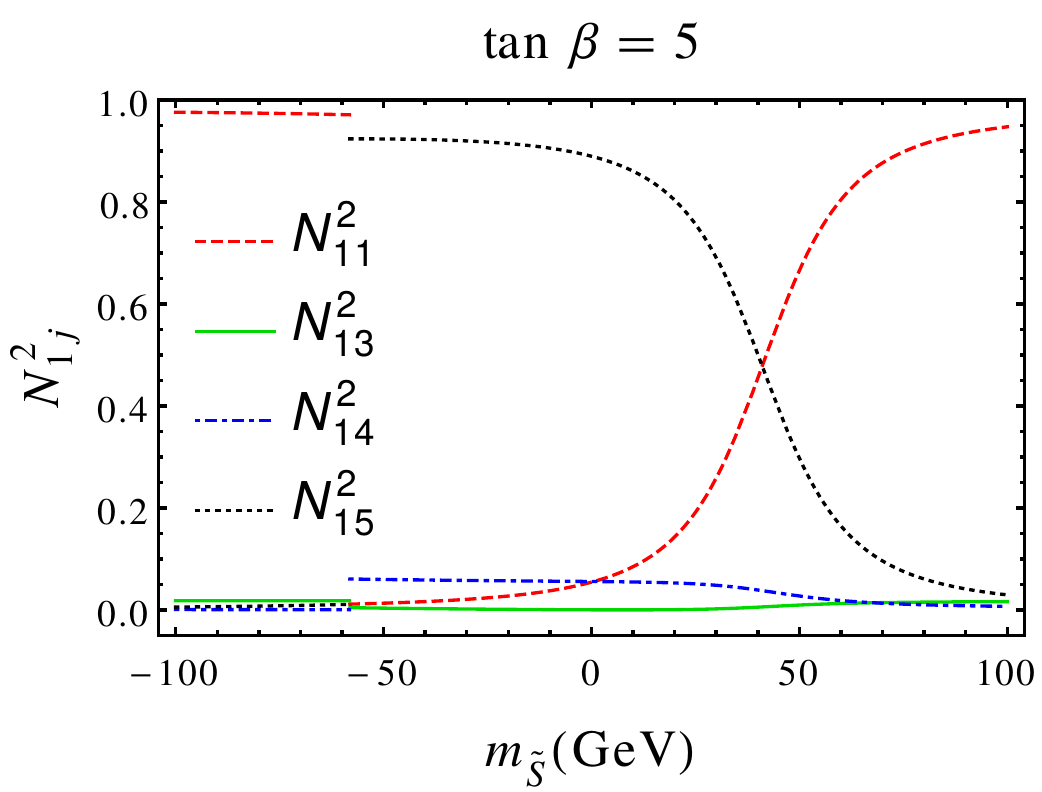}
\hskip 40pt
\includegraphics[height=5.5cm,width=0.43\linewidth]{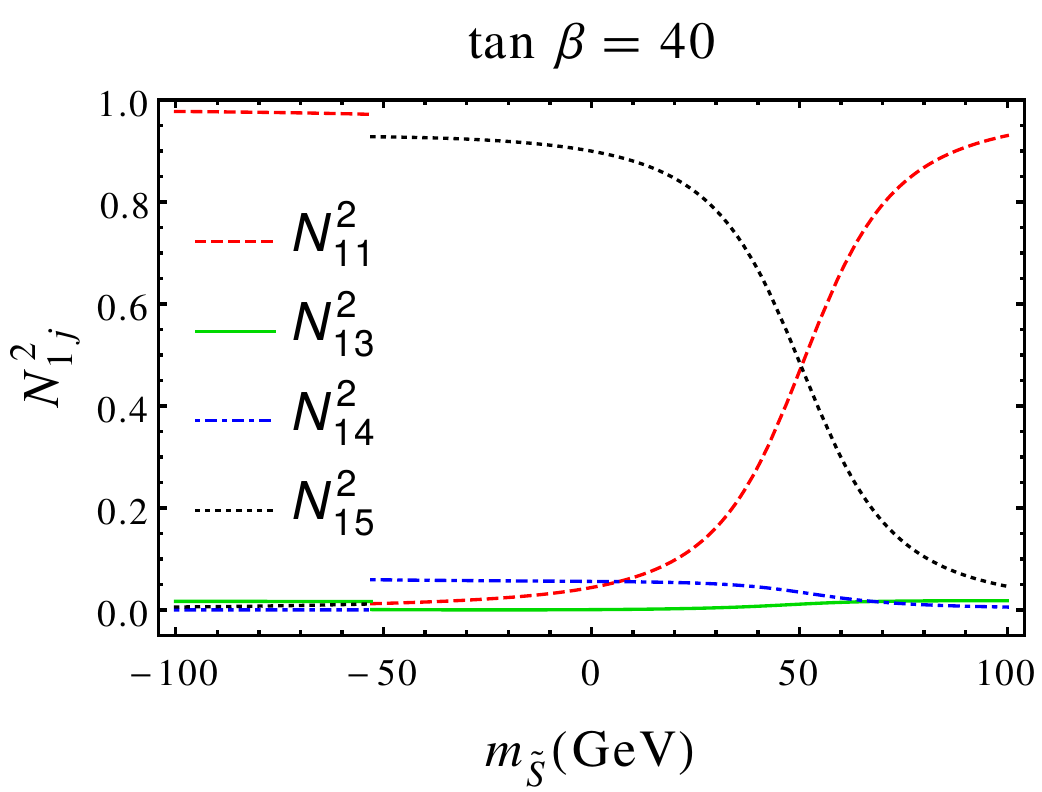}~~
\caption{Variations of various elements $N_{1j}$ of the matrix `$N$' in eq.~(\ref{eqn:mneut}) representing the admixtures of bino ($j$=1), higgsinos ($j$=3,4) and singlino ($j$=5) in the LSP as functions of $\msinglino$ with flipping signs over the shown range and for fixed values of $\mueff$ (=\,350~GeV) and $\lambda$ (=\,0.5) and two fixed values of $\tan\beta$ (= 5 (left) and 40 (right)). With $\mtwo$ set to a much larger value (2.5~TeV), the wino admixture in the LSP ($N_{12}$) is negligible and hence ignored. 
}
\label{fig:all-components}
\end{center}
\vspace{-0.5cm}
\end{figure}
%

In figure~\ref{fig:all-components} we present the variations of the various admixtures like bino, higgsino and singlino in the LSP as functions of
$\msinglino$ which can have a flipped sign and for fixed values of $\mone$ (=\,50~GeV),
$\mueff$ (=\,350~GeV), `$\lambda$' (=\,0.5) and for two fixed values of $\tan\beta$ (=\,5, 40). Given the large value of $\mtwo$ that we choose (2.5~TeV), the wino admixture in the LSP, for the chosen inputs, has to be negligible and hence ignored.
Discontinuities in the curves showing variations of bino and singlino admixtures mark the point when the singlino-like LSP with a negative mass-eigenvalue abruptly turns into a bino-dominated LSP with a positive 
eigenvalue. The important point to note here is that just about such transition points, the LSP (bino-like) and the NLSP (singlino-like) are rather close in mass while, unlike in normal circumstances, still largely remaining to be pure states. We exploit its phenomenological implication in section~\ref{subsec:relic}. We have also checked variations
of these $N_{1j}$ elements for a flipped sign on $\mone$ and for smaller values of `$\lambda$'. We find that, in these cases, their salient features remain unaltered while the
cross-over points could get shifted and/or individual variations could become
sharper/flatter as $\msinglino$ varies.

\begin{figure}[t!]
\begin{center}
\hskip 5pt
\includegraphics[width=0.235\linewidth]{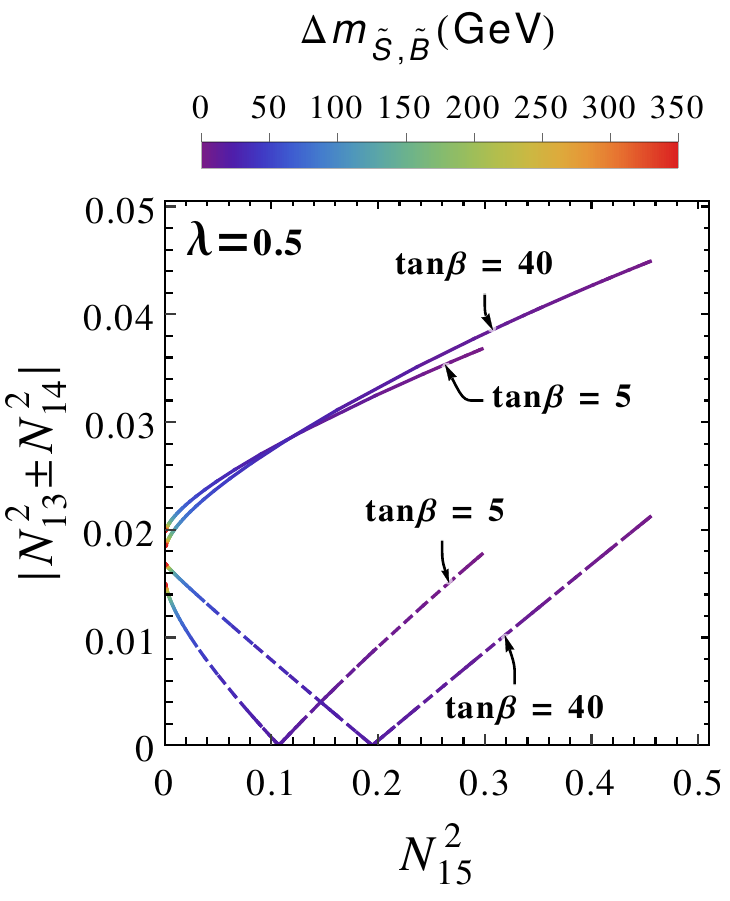}
\includegraphics[width=0.24\linewidth]{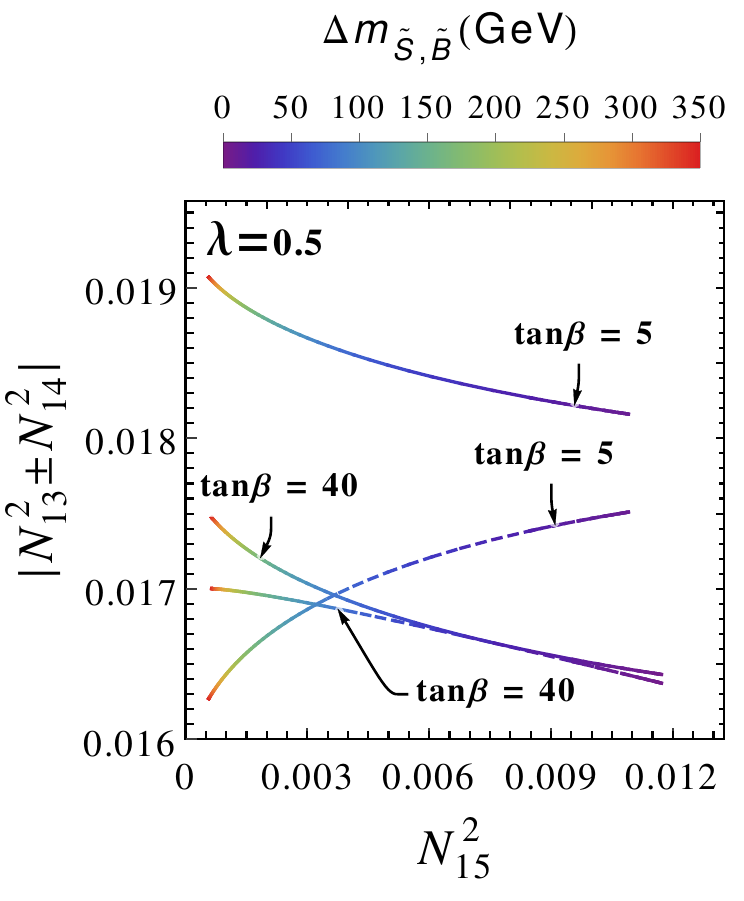}
\includegraphics[width=0.24\linewidth]{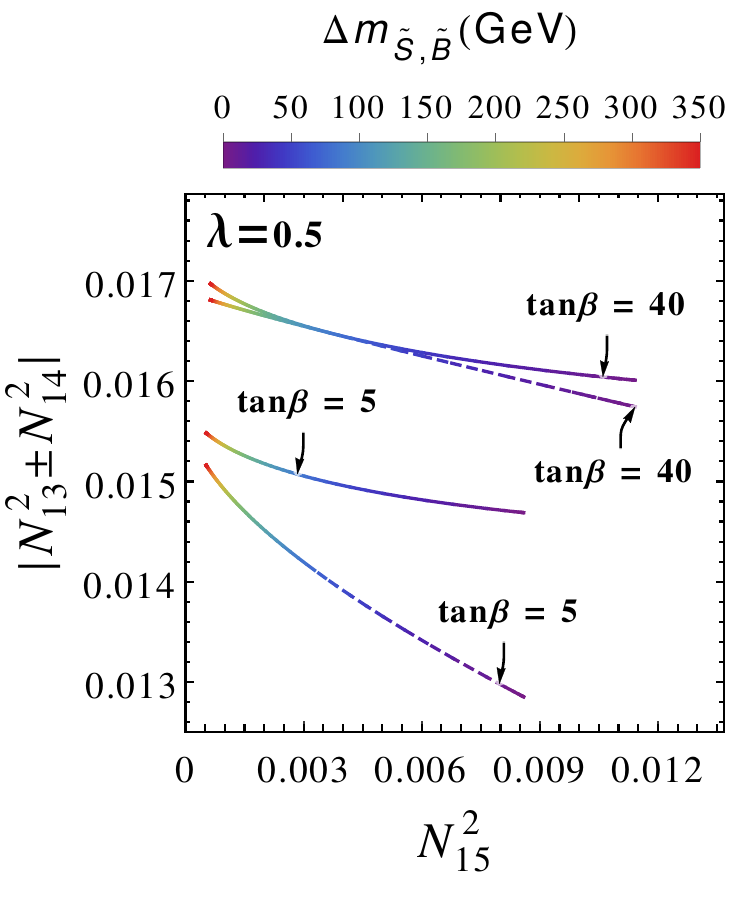}
\includegraphics[width=0.236\linewidth]{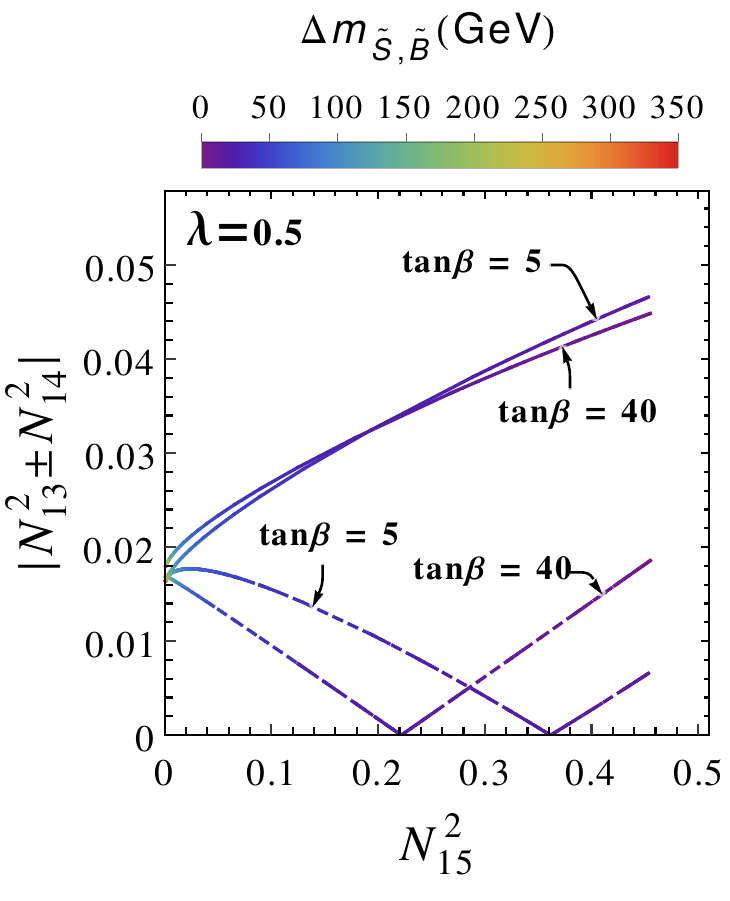} \\[0.25cm]
\includegraphics[width=0.24\linewidth]{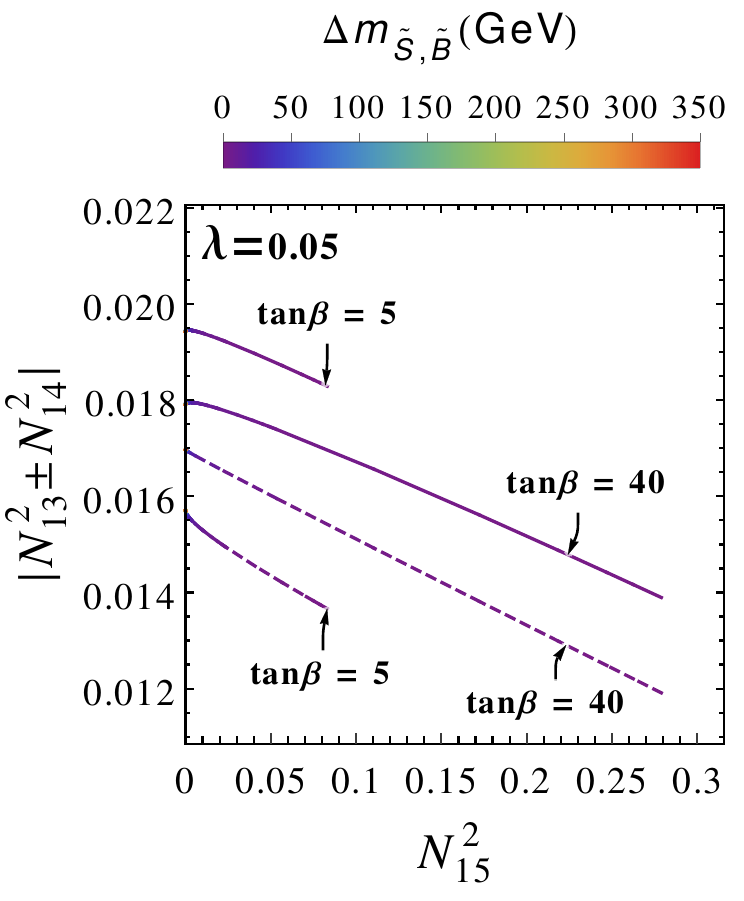}
\includegraphics[width=0.24\linewidth]{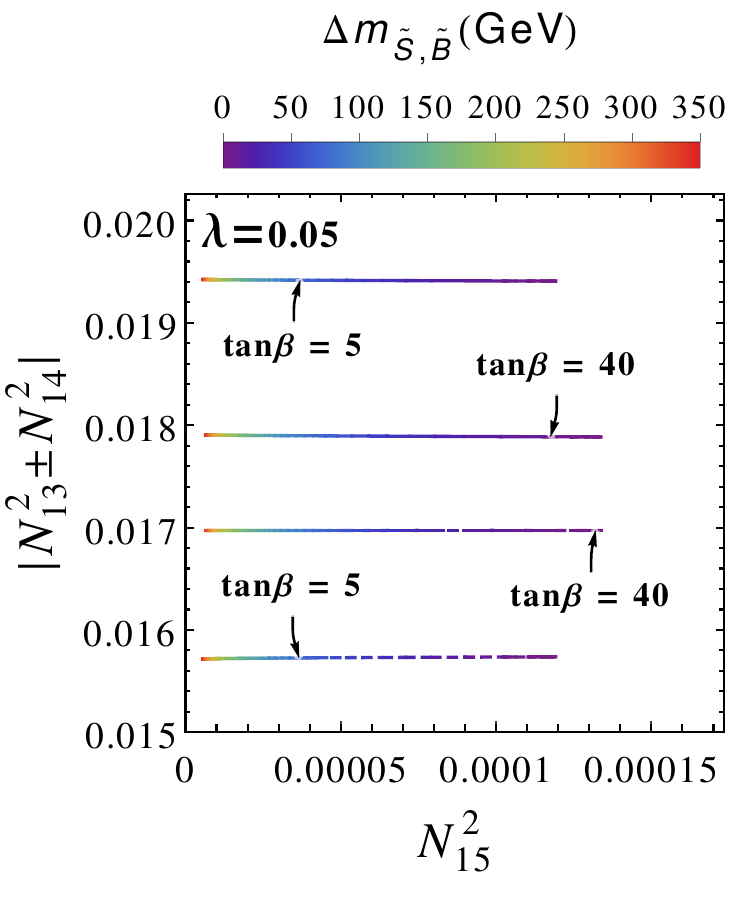}
\includegraphics[width=0.24\linewidth]{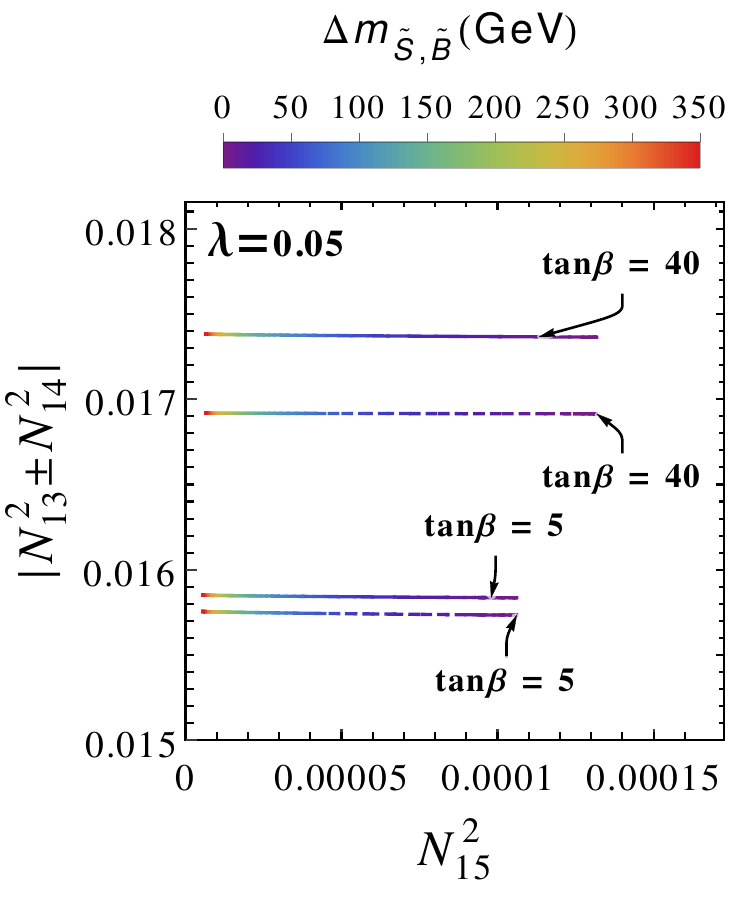}
\includegraphics[width=0.24\linewidth]{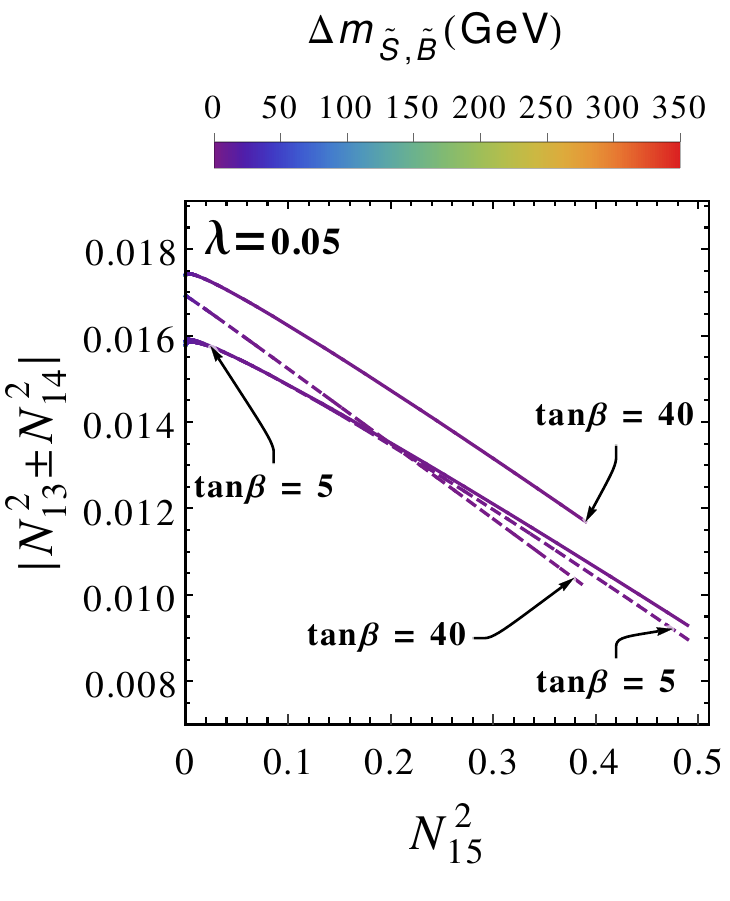}
\vspace{-0.25cm}
\caption{Associated variations of the singlino ($N_{15}^2$) and the overall 
higgsino admixtures ($N_{13}^2+N_{14}^2$; solid lines) 
and the difference of individual higgsino admixtures ($|N_{13}^2 - N_{14}^2|$; 
broken lines) in an LSP with at least 50\% bino admixture for varying $\Delta m_{\singlino,\bino}$ (in the palettes) arising from the variation of $\msinglino$ over the 
range $|\msinglino|<400$~GeV and for a fixed value of $\mueff$ (=350~GeV)
and for two values of $\tan\beta$ (=5, 40) with $|\mone|=50$~GeV.  
Plots in the upper (lower) panel are with $\lambda=0.5 \, (0.05)$. Signs on $\mone$ and $\msinglino$, from left to right in each row, are
sign$(\mone, \msinglino) \equiv (++, +-,-+, --)$. $\mtwo$ is kept fixed at 2.5~TeV.
}
\label{fig:deltam-admixtures}
\end{center}
\vspace{-0.5cm}
\end{figure}

In figure~\ref{fig:deltam-admixtures} we illustrate the simultaneous variations 
of the singlino admixture in an LSP with at least 50\% bino admixture ($N_{15}^2$, along the abscissa), the higgsino admixture in the same ($N_{13}^2 + N_{14}^2$; solid lines) and the 
difference of individual higgsino admixtures
($|N_{13}^2 - N_{14}^2|$; broken lines), both along the ordinate as $\Delta m_{\singlino,\bino} = |\msinglino| - |\mone|$ is 
varied (range shown in the adjacent palettes). This is done for various combinations of signs 
on $\mone$ and $\msinglino$ and for two different values of $\tanb$ (= 5, 40) 
while $|\mone|$ and $\mueff$ are kept fixed at 50~GeV and +350~GeV, 
respectively. In order to vary $\Delta m_{\singlino,\bino}$, $\msinglino$ is 
varied from 400~GeV down to a value close to the fixed value of $|\mone|$, i.e., 50~GeV such that the LSP always remains bino-dominated ($> 50$\%). 
Plots in the upper (lower) panel correspond to $\lambda=0.5 \, (0.05)$. In a 
given panel, plots from left to right represent cases where signs on $\mone$ and
$\msinglino$ appear in the following combinations: $(++)$, $(+-)$, $(-+)$ and $(--)$. Note that $N_{15}^2$ always increases with decreasing $\Delta m_{\singlino,\bino}$
for all the cases. This can be gleaned from eq.~(\ref{eqn:n^2j5-value}).

As can be found from the left- $(++)$ and the rightmost $(- -)$ plots in the upper 
panel of figure~\ref{fig:deltam-admixtures}, both $N_{13}^2 + N_{14}^2$ and 
$N_{15}^2$ increase as $\Delta m_{\singlino,\bino}$ drops. In all other cases
both $N_{13}^2 + N_{14}^2$ and $|N_{13}^2 - N_{14}^2|$ either decrease or remain
mostly unaffected (except for the second plot in the upper panel in which  $|N_{13}^2 - N_{14}^2|$ grows for $\tan\beta=5$) as
$\Delta m_{\singlino,\bino}$ decreases, while in all these cases $N_{15}^2$ still
keeps growing. Also, in general, $\tanb$ does not seem to play any dominant role.
Furthermore, scales of variations in these quantities are larger for
$\lambda=0.5$ as compared to $\lambda=0.05$. It may further be noted that,
irrespective of the magnitude of `$\lambda$', the variations involve smaller 
values of the plotted quantities when there is a relative sign between $\mone$ and~$\msinglino$ (the two plots in the middle of each panel).

Also noteworthy is the
fact that the quantity $N_{13}^2 - N_{14}^2$ could become vanishingly small and can undergo a flip in its sign as
$\Delta m_{\singlino,\bino}$ drops, when $\mone$ and $\msinglino$ have the same sign and when $\lambda=0.5$, i.e., larger `$\lambda$' (see appendix~\ref{appsec:blindspot}). This phenomenon is indicated by the inflections in the bottom pair of curves in the extreme left and right plots of the upper panel.
Given that the $Z$-boson coupling to a pair of
LSPs is proportional to $|N_{13}^2 - N_{14}^2|$, a blind spot in the same occurs in the vicinity of such inflection points thus, at the same time, critically affecting
the annihilation of the LSPs via $Z$-funnel and the rate of their SD scattering off the nuclei.
The rightmost plot in the upper panel
reveals that the higgsino admixture in the LSP can go up to about 5\% for the given set of input parameters when the singlino admixture reaches  $\sim 45$\%. On the contrary, when
$\mone$ and $\msinglino$ have a relative sign between them (the two plots in the
middle of the top panel), the combined admixture of singlino and higgsinos in the 
LSP does not exceed $\sim 3$\% and hence the LSP remains highly bino-like.
%
\section{Results}
\label{sec:results}
In this section, we discuss the prospects of a relatively light
($\lesssim 200$~GeV) highly bino-like LSP by referring to its implications for 
(i) the DM relic abundance, (ii) the DMDD rates
(both SI and SD cases) and that for (iii) the LHC. 

Results presented in this section draw heavily on what are discussed in section 
\ref{sec:scenario}. These comply with a multitude of crucial constraints 
implemented in {\tt NMSSMTools-v5.4.1}. The constraints include the 
theoretical ones like the requirements of non-tachyonic particle spectrum, 
absence of an unphysical global minimum and that of Landau poles in various 
Yukawa-type couplings like `$\lambda$', $`\kappa$', $y_t$ and $y_b$ and the 
experimental ones from the flavor sector and those coming from collider
experiments like the LEP, Tevatron and the LHC. In addition, up-to-date 
constraints pertaining to the observed Higgs sector are checked via dedicated 
packages like {\tt HiggsBounds-v5.4.0}~\cite{Bechtle:2008jh, Bechtle:2013wla, Bechtle:2020pkv} 
and {\tt HiggsSignals-v2.3.0}~\cite{Bechtle:2013xfa, Bechtle:2014ewa} and
the mass of the SM-like Higgs boson is considered within the range $122~\mathrm{GeV} <\mhsm <128~\mathrm{GeV}$
to account for the uncertainties in its theoretical estimates. However, 
with the mass of the smuons (and, in general, of all the sfermions) set to be in the multi-TeV domain, we do not attempt to reproduce 
the purported discrepancy between the theoretically predicted and the 
experimentally measured values of muon anomalous magnetic moment
(muon~$g-2$), pending a forthcoming, much improved experimental number (and 
the same for its theoretical counterpart) which might put the uncertainty 
finally at rest.

In the DM sector, our results are confronted by the observations at the Planck 
experiment~\cite{Ade:2015xua, Aghanim:2018eyx} on the DM relic abundance. We ensure strict compliance with the Planck-reported $2\sigma$ upper bound on the same, i.e.,
we consider $\Omega h^2 < 0.131$ (to avoid overclosure of the Universe). On the other side, we  do not confine ourselves to its reported $2\sigma$ lower
bound (0.107) and allow $\Omega h^2$ to go below the same. We, however, choose to remain agnostic on
what exactly accounts for the remaining part of the invisible relic in such an
under-abundant DM regime. We also seek compliance with the constraints 
on the DM-nucleus (elastic) scattering cross sections obtained from the SI and the SD DD experiments
\cite{Aprile:2018dbl, Aprile:2019dbj, Amole:2019fdf}. The latest bounds on the SI cross 
sections~\cite{Aprile:2018dbl} are already incorporated in
{\tt NMSSTools-v5.4.1} while the same on the SD DM-neutron~\cite{Aprile:2019dbj} and
the SD DM-proton~\cite{Amole:2019fdf} cross sections are not.  Hence we impose them by hand in our 
analysis. For all the above-mentioned DM observables, their values are obtained from 
a dedicated package like {\tt micrOMEGAs-v4.3}
\cite{Belanger:2006is, Belanger:2008sj, Barducci:2016pcb} as adapted in
{\tt NMSSMTools-v5.4.1}. As customary, we allow for scaling of the SI and the SD cross 
sections by the actual DM relic abundance; but only when the latter is in the Planck-allowed ($2\sigma$) band, i.e., when $0.107 < \Omega h^2 < 0.131$. However, in a conservative approach, we do not scale these cross sections when the relic abundance turns out to be below the said band, i.e.,
no point appears in our scan that would have an SI and/or SD rate(s) below their respective bounds just because of such a downward scaling of these rates for points yielding DM relic abundance below the Planck-allowed band.\footnote{Since such a downward scaling of the DD rates would pick up new points from our scan that now comply with the DD constraints, viable new possibilities, viz., new mixing patterns among the ewinos might show up.}

The present study is based on a random scan of the $Z_3$-symmetric pNMSSM 
parameter space using \nmssmtools ~where the scan is parallelized via {\tt T3PS}~\cite{Maurer:2015gva} with $10^9$ points generated. 
Unless otherwise mentioned, we subject this set of points to the constraints discussed 
above, all at a time. In particular, with all other constraints in force, a 
relaxed constraint requiring only an upper bound on $\Omega h^2$ ($< 0.131$) 
results in $\sim 9 \times {10^4}$ allowed points while demanding
$\Omega h^2$ to be strictly within the Planck-reported $2\sigma$ band brings
the number of allowed points down to $\sim 2.4 \times 10^4$.
%
\subsection{Relic abundance}
\label{subsec:relic}
To meet the Planck constraints on the DM relic abundance, i.e., $0.107 \leq \Omega h^2 \leq 0.131$ (at $2 \sigma$ level)~\cite{Ade:2015xua, Aghanim:2018eyx}
a pair of bino-like LSPs 
has to primarily annihilate resonantly in our scenario. Towards this, in
addition to the usual annihilation funnels involving the $Z$-boson and the Higgs bosons that are already present in the MSSM, the bino-like DM would now have access to 
funnels via the singlet-like scalars, $\as$ and $\hs$, of the NMSSM. 
Furthermore, the bino-like DM neutralino might coannihilate with the NLSP 
neutralino which is singlino-dominated~\cite{Baum:2017enm} with some higgsino admixtures. In practice, given that the LHC searches now have placed stringent lower bounds on the 
masses of the heavier MSSM-like Higgs bosons ($m_{H,A} \gtrsim 400$~GeV~\cite{Aaboud:2017sjh, Sirunyan:2018zut, Bahl:2018zmf, Aad:2020zxo}), these could hardly provide efficient DM annihilation funnels for $m_{\mathrm{DM}} \leq 200$~GeV that we are interested 
in. Further, its coannihilations
with sfermions are also ruled out as we assume the latter to be  rather heavy ($\sim 5 
$~TeV).

It should be noted that even in the presence of these new agents (the 
singlet scalars and the singlino) that aid DM-annihilation, one key requirement remains to be there as it does in the MSSM case, i.e., a pure bino DM, as such, cannot exploit them
and it takes a certain amount of admixtures of the higgsinos and/or the singlino as can be seen from eq.~(\ref{eqn:generic-sinjnk}) (or equivalently, from  eqs.~(\ref{eqn:hinjnk-reduced-without-approximation}) and~(\ref{eqn:ainjnk-reduced-without-approximation})) and eq.~(\ref{eqn:znjnk}) for the purpose. While funneling through
$Z$-boson or $\hsm$ necessarily requires some higgsino admixture in 
the otherwise bino-dominated LSP, the superpotential in eq.~(\ref{eqn:superpot})  already points to the fact that exclusive admixtures of 
either higgsinos ($\lambda$-dependent interaction) or the singlino
($\kappa$-dependent interaction) would do if a pair of LSPs were to funnel via 
singlet-like scalars of the NMSSM. In the latter case, however, the LSP has to have a significant singlino admixture which would require $\msinglino$ to be on the smaller side thus requiring a smaller `$\kappa$'. Clearly, this would pull down the $S \singlino \singlino$ coupling itself. Hence, practically, it is not possible to enhance the strength of this interaction indefinitely.

Since most of the interactions of the LSP DM that facilitate their 
annihilation thereby offering an acceptable relic abundance also tend to
increase their DD (SI or SD) rates, the observed upper bounds on the 
latter bring forth the well-known tension between the two. In the 
NMSSM, there is one exception to this when the light singlet-like pseudoscalar, $\as \sim \aone$, 
works as an efficient annihilation
funnel. This is since $\as$ could contribute only to the SD rate for which the same is velocity ($p$-wave) suppressed while it could still provide an efficient DM annihilation funnel which proceeds in $s$-wave~\cite{Freytsis:2010ne,Boehm:2014hva,Kumar:2013iva,Escudero:2016gzx}.
An important set of issues concerning a bino-like LSP DM in the 
present scenario, when seen through the lens of relic 
abundance,~are
\begin{itemize}
\item how efficient various funnels are for self-annihilating highly bino-like LSP DM,
\item how plausible its coannihilation with a neutralino NLSP is,
\item what kind of (minimal) admixtures of higgsinos and singlino in such a bino-like
      DM are viable in relation to its mass and
\item how the masses and the compositions of the other relatively light neutralinos (higgsino- or
      singlino-like) turn out to be.      
\end{itemize}
\begin{figure}[h!!!]
\begin{center}
\includegraphics[width=0.56\linewidth]{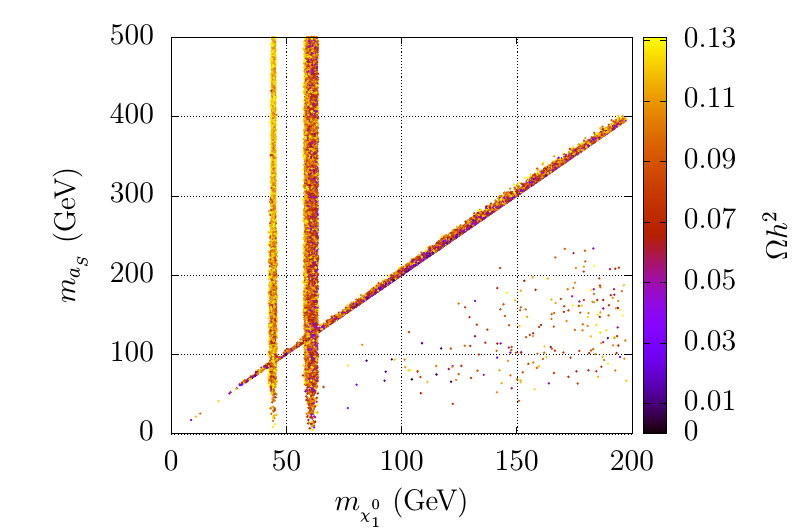}
\caption{Dominant modes of annihilation of a highly bino-like LSP DM: $Z$-, $\hsm$- and
$\as$-funnel regions along with the coannihilation region are shown
for $\Omega h^2 < 0.131$ in the $\mntrlone$--$\mas$ plane with the palette indicating the magnitude of $\Omega h^2$.}
\label{fig:funnels}
\end{center}
\vspace{-0.5cm}
\end{figure}

In figure~\ref{fig:funnels} we illustrate these issues by allowing for 
a relaxed range of relic abundance with only an upper bound on $\Omega h^2$ ($<0.131$). This shows the presence of three prominent funnel regimes of which the vertical bands stand for the ones involving the
SM $Z$-boson~(left) and $\hsm$~(right) while the 
inclined band represents funnel annihilation overwhelmingly attributed to $\as$ and which proceeds in $s$-wave. Over this band $\mas \approx 2 \mntrlone$. In contrast, the $\hs$-funnel is $p$-wave suppressed and hence is generally not as efficient.
Compliance in the scantly populated regions in the lower right part of the plot is aided mostly by
the bino-like DM coannihilating with the singlino-like NLSP.

Note that the diagonal and the 
vertical bands in figure~\ref{fig:funnels} intersect thus implying the possibility of two simultaneous 
funnels for a given DM mass: (i) $\as$ along with the $Z$-boson and (ii) $\as$ along with $\hsm$.
A priori, these give rise to the possibility of interference between processes with nearly resonant funnel-states as propagators. In the first case,
the dominant interference is between the purely $s$-wave amplitude of the $\aone$-mediated diagram with the longitudinal polarization (the Goldstone mode) of the $Z$-boson~\cite{Baum:2017enm}. 
It is noted that due to a 
comparatively larger total decay width of the $Z$-boson ($\Gamma_Z \sim 2.49$ 
GeV), the $Z$-funnel could kick in even when $2 \mntrlone$ is not exactly on the
$Z$-pole. Hence there is a somewhat enhanced scope of interference between $\as$ and
$Z$-mediated processes with $\mas \approx 2\mntrlone \sim m_Z \mp \Gamma_Z$. Such interference could 
occasionally be important. In contrast, no such interference is possible in
the second case since the processes involve different partial waves~\cite{Baum:2017enm}.

Furthermore,
for the range of bino-like DM mass under consideration ($\mntrlone < 200$~GeV), it 
is impossible to find it coannihilating with a higgsino-like state with its mass 
close-by since both DD and the LHC experiments have put stringent lower bound on the 
mass of the latter, as mentioned earlier. On top of that, since both sfermions and the wino are assumed to be much heavier, a 
coannihilating state (NLSP) can thus only be singlino-like.
However, with such proximity of their masses, the LSP and the NLSP neutralinos would tend to have significant admixtures of bino and singlino 
states. This could potentially jeopardize compliance with the experimental bounds on the DMDD rates. We find that avoiding a significant mixing but still 
allowing for the required proximity of these masses could be arranged when $\mone$ and $\msinglino$ have a relative sign between them. 
This can be seen in the plots of figure~\ref{fig:all-components} in the 
regions with discontinuities in the curves over which bino-singlino mixing 
remains negligible in spite of these states being nearly mass-degenerate. Our scan 
reveals that an efficient enough LSP-NLSP coannihilation
is possible when the NLSP mass is within~15\% of the mass of the LSP.

We further observe that in such a coannihilation regime the so-called `assisted 
coannihilation'~\cite{Ellwanger:2009dp} might become the key mechanism. This is a situation when the LSP and the NLSP have to be in a 
thermal equilibrium with comparable populations and the two equilibrium processes 
${\rm LSP} + X \rightleftharpoons {\rm NLSP} +Y$, where `$X$' and `$Y$' are the SM quarks or leptons, and 
${\rm NLSP}+{\rm NLSP} \rightarrow X$ happen to have competing overall rates thus 
simultaneously diluting the abundances of both the LSP and the NLSP (the 
essence of coannihilation) leading to a reconciliation of the observed relic
abundance. With the NLSP being singlino-like,
the latter process might be rather efficient in the presence of a singlet-like (pseudo)scalar with its mass close to $2\mntrltwo$ and for a large enough
`$\kappa$' (see eq.~(\ref{eqn:generic-sinjnk})). Thus, assisted coannihilation could overwhelm (by contributing up to a level of 90\%) the conventional one.

It may further be noted from the plots in figure~\ref{fig:funnels} that an LSP DM
mass below $\lesssim 30$~GeV (more precisely, $ \lesssim \mhsm/4$) is barely viable. This is since such an 
LSP invariably requires light scalar(s) with mass(es) 
below $\mhsm/2$ to funnel through which attracts stringent constraints from the 
studies of decays of the SM Higgs boson at the LHC~\cite{Aaboud:2018fvk, Aaboud:2018gmx, Aaboud:2018iil, Aaboud:2018esj, Aad:2020rtv, Sirunyan:2018mbx, Sirunyan:2018mot, Sirunyan:2020eum}, unless these
scalars get to be nearly pure singlet states. While the issue is a generic one
irrespective of the nature of the LSP, a bino-like LSP is affected exclusively because of the feebleness of its coupling to such singlet states.
In contrast, a singlino-dominated LSP of a similar mass could more readily funnel
through a singlet-like scalar in a region of parameter space with moderately~large~`$\kappa$'.  
%

\begin{figure}[t]
\begin{center}
\includegraphics[width=0.50\linewidth]{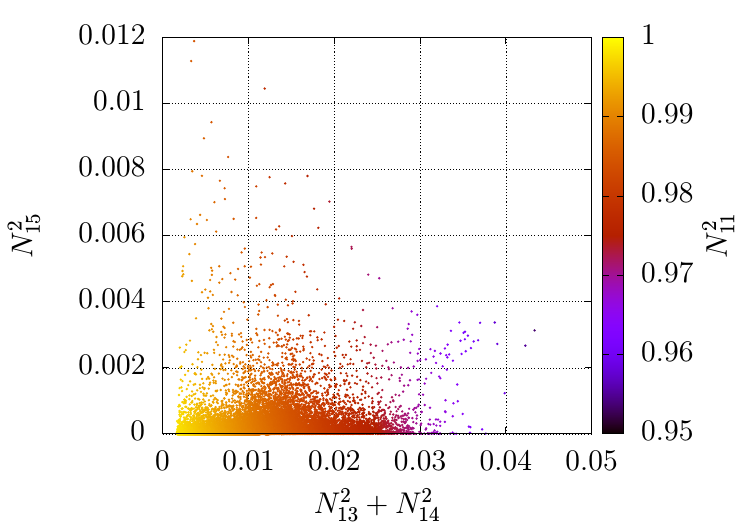}~~
\includegraphics[width=0.50\linewidth]{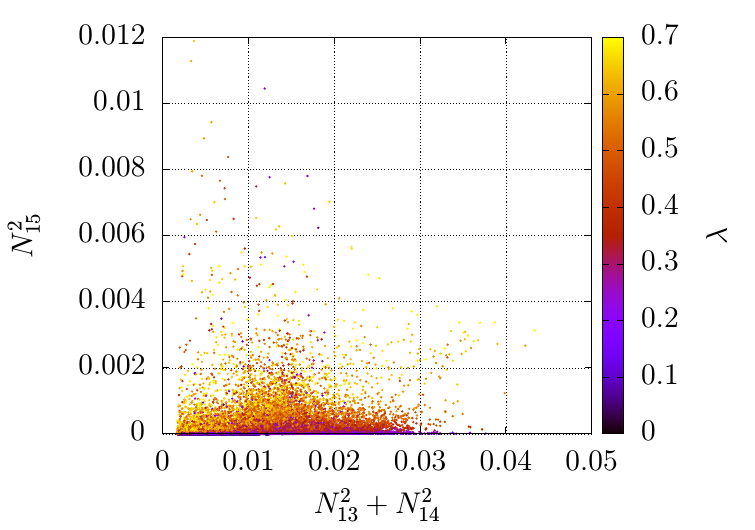}
\caption{Scattered points showing simultaneous variations of the higgsino
($N_{13}^2+N_{14}^2$) and the singlino ($N_{15}^2$) admixtures in the LSP with the 
palettes indicating the extent of bino admixture ($N_{11}^2$) in the LSP (left plot) and the magnitude of `$\lambda$'
(right plot).
}
\label{fig:lsp-admixtures}
\end{center}
\end{figure}
In figure~\ref{fig:lsp-admixtures} we present the corresponding admixtures an otherwise bino-like LSP DM could have while conforming to the experimental
constraints discussed earlier including those on the DD rates.
The allowed points are shown in the plane of higgsino ($N_{13}^2+N_{14}^2$) and singlino ($N_{15}^2$) admixtures in the LSP. For the plot on the left, the palette indicates the bino admixture ($N_{11}^2$) while for the one on the right it presents the magnitude of `$\lambda$'. From the left plot it is seen that for a viable bino-like LSP with a mass not exceeding 200~GeV, its bino-content cannot go below 95\% while its maximum higgsino and singlino contents could reach $\sim 5\%$ and $\sim 1\%$, respectively.
The right plot shows that the singlino admixture in the LSP is virtually absent for small values of `$\lambda$' and it reaches up to 1\% for larger
`$\lambda$' values while the higgsino admixture can be up to 5\% irrespective of the value of `$\lambda$'.
It may also be noted that the higgsino admixture in the LSP never vanishes. This is since 
there is always a $\lambda$-independent mixing in the bino-higgsino sector which is driven by the gauge coupling, $g_1$.
%
\subsection{Direct detection cross sections}
\label{subsec:ddrates}
Upper bounds on the cross sections of DM particles scattering elastically off the 
nucleus in DD experiments have emerged to be increasingly 
constraining over the last few years. However, as pointed out earlier, when 
compared to their drastic implications for the MSSM, a definite edge is enjoyed 
by the NMSSM. This is due to the presence of a (possibly light) singlet-like pseudoscalar
in the NMSSM spectrum which could provide an efficient funnel for DM 
annihilation to find the relic abundance in the right ballpark while
contributing at best feebly to the DD processes.
In this subsection, we discuss the processes/mechanisms that control
SI and SD scattering of a highly non-relativistic 
bino-like LSP DM and the constraints these attract for the NMSSM from the
respective experiments that currently provide us with the most stringent of 
the bounds~\cite{Aprile:2018dbl, Aprile:2019dbj, Amole:2019fdf}.
To be more specific, at 90\% confidence level,
\begin{itemize}
\item SI DM-nucleon elastic scattering cross section (for DM mass above 6~GeV) receives the most stringent upper bound of 
$4.1 \times 10^{-47} \mathrm{cm}^2$ at a DM mass of 30~GeV from the XENON1T experiment~\cite{Aprile:2018dbl},
\item SD DM-neutron elastic scattering cross section (for DM mass above 6~GeV) finds the most stringent upper bound of $6.3 \times 10^{-42} \,
\mathrm{cm}^2$ at a DM mass of 30~GeV, again from the XENON1T experiment~\cite{Aprile:2019dbj}. The same for the DM-proton
scattering (for DM mass above 3~GeV) comes from the PICO-60 experiment and is $2.5 \times 10^{-41} \, \mathrm{cm}^2$ for a DM mass of 25~GeV~\cite{Amole:2019fdf}.
\end{itemize}
In the subsequent subsections, we address the salient aspects of the SI and the SD scattering processes in reference to a highly bino-like LSP DM with mass below around 200~GeV.
%
\subsubsection{The SI scattering cross section}
\label{subsubsec:si}
With the squarks assumed to be rather heavy, the SI scattering of the DM off the detector nucleus predominantly 
proceeds in the $t$-channel mediated by the three $CP$-even scalars of the NMSSM 
scenario, $\hsm, H$ and $\hs$. These states interact with a LSP-pair at one vertex 
and with a pair of nucleons at the other. The former interaction is given by eq.~(\ref{eqn:generic-sinjnk}) while the latter, for the $i$-th scalar, is given by 
\cite{Badziak:2015exr}
\bea
\label{eqn:hiNN-badziak}
g_{_{h_i N N}}
 &= &
\frac{m_N}{\sqrt{2}\, v}\left({S_{i1} \over \cos \beta}  F^{(N)}_d+ {S_{i2}  \over \sin \beta} F^{(N)}_u \right),
\eea
where $m_N$ is the mass of the nucleon, $F^{(N)}_{d,u}$ are the combinations of various nucleon form factors appropriate for the nucleon `$N$' as presented in refs.~\cite{Belanger:2013oya, Badziak:2016qwg}.
It should be noted here that the coupling of the singlet-like $CP$-even Higgs boson
with the nucleon pair ($g_{_{\hs NN}}$) is suppressed due to small doublet admixture
in the former. In any case, the individual and the collective contributions of the above couplings to the SI rate are known to be rather intricate and vary across the parameter space. The SI DM-nucleon scattering in the $s$-channel involves rather heavy squarks as mediators and hence could be safely ignored. The SI cross section is then given by~\cite{Cao:2018rix}
\beq
\label{eqn:SI-equation-general}
\sigma^{\rm SI}_{\ntrlone-(N)} = \frac{4 \mu^2_r}{\pi} \left|f^{(N)}\right|^2, \hspace{0.5cm} f^{(N)} = \sum_{i=1}^3 {\frac{g_{_{h_{i} \ntrlone \ntrlone}}\, g_{_{h_{i} N N}}}{2 m^2_{_{h_i}}}}  \,, 
\eeq
where $\mu_r$ is the reduced mass of the DM-nucleon system. The reader is referred to appendix~\ref{appsec:blindspot} for a detailed discussion of the involved couplings and relevant expressions that are presented in a more convenient basis.

In the discussion of the DD rates, of crucial importance is the nature of the LSP DM. For a pair of highly bino-like
LSPs which is in the focus of the present work, their interactions to the Higgs-like scalar states depend much on the amount of higgsino and/or singlino admixtures the LSP contains. These can be read out from table~\ref{tab:interactions}.
The relevant interactions determining the SI rate are as follows.
\begin{itemize}
\item The doublet(-like) Higgs bosons ($\hsm$ and `$H$') couple to the bino and the higgsino (hence suppressed by $1/ \mueff$) admixtures from the respective LSP states with the gauge strength $g_1$. 
\item The NMSSM-specific interactions of $\hsm$ and `$H$' to the higgsino and the singlino components of the bino-like LSP are driven by `$\lambda$'. Their couplings to the higgsino components are suppressed by $1/ \mueff$ as expected while the same to the singlino component also have a similar dependence, other parameters remaining unaltered.
\item The interaction of the $CP$-even singlet(-like) scalar ($\hs$) with a pair of LSPs is driven by `$\lambda$' and involves their higgsino content only (thus, suppressed by $1/\mueff^2$).  
\end{itemize}

As discussed in section~\ref{subsec:relic}, while a pair of pure bino LSPs 
is unable to annihilate mutually through available funnels and hence would leave 
a large DM relic abundance, any admixture of active components in the form of
higgsino and/or singlino admixtures in such an LSP that would allow them to 
annihilate optimally must also ensure that their scattering cross sections off 
the nucleons in DD experiments to be sufficiently low to comply
with the constraints on the SI~\cite{Aprile:2018dbl} and SD~\cite{Aprile:2019dbj, Amole:2019fdf} rates. Such compliance is aided by the presence of a blind spot 
\cite{Cheung:2012qy, Huang:2014xua, Cheung:2014lqa, Badziak:2015exr,
Badziak:2016qwg, Badziak:2017uto, Cao:2018rix} in the DD processes. For the SI process, such a blind spot occurs when (i) the 
$\hsm$-$\ntrlone$-$\ntrlone$ coupling gets to 
be vanishingly small (the so-called `coupling blind spot') and 
(ii) there are destructive interferences among processes mediated by the $CP$-even scalars. For the first time, we derive
the conditions for these SI blind spots for the complete bino-higgsino-singlino system. These are presented in appendix~\ref{appsec:blindspot}
(see eqs.~(\ref{eqn:coupling-blindspot}) and~(\ref{eqn:NMSSM-BLINDSPOT-CONDITION-1})).

\begin{figure}[t]
\begin{center}
\hspace*{-0.8cm}
\includegraphics[height=6.0cm, width=0.53\linewidth]{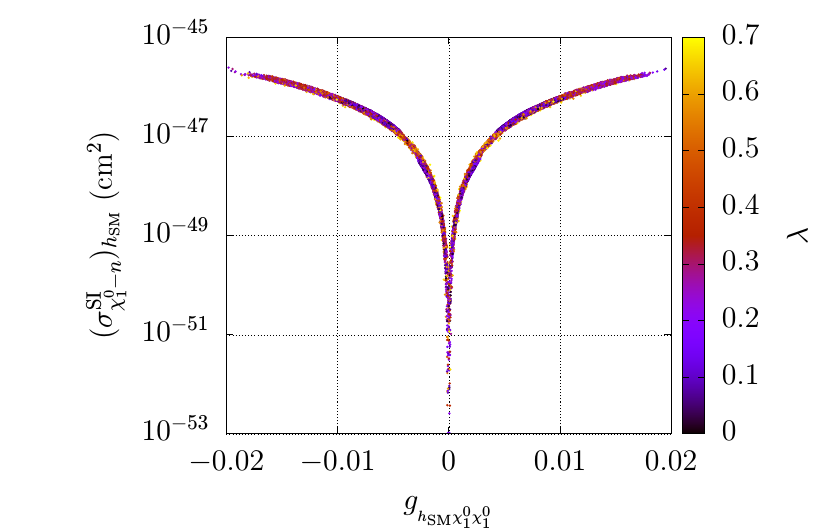}~~
\hskip -13pt
\includegraphics[height=6.0cm, width=0.55\linewidth]{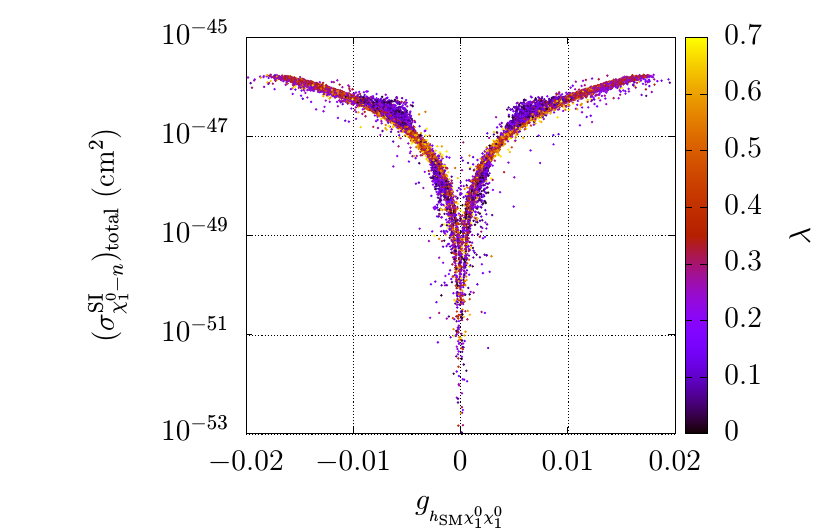}\\[-0.5cm]
\vskip 10pt
\hspace*{-0.8cm}
\includegraphics[height=6.0cm, width=0.53\linewidth]{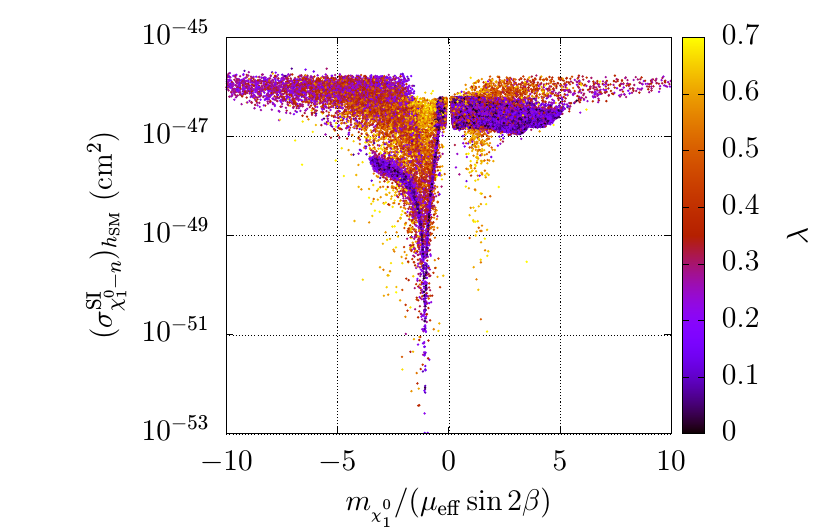}~~
\hskip -11pt
\includegraphics[height=6.0cm, width=0.55\linewidth]{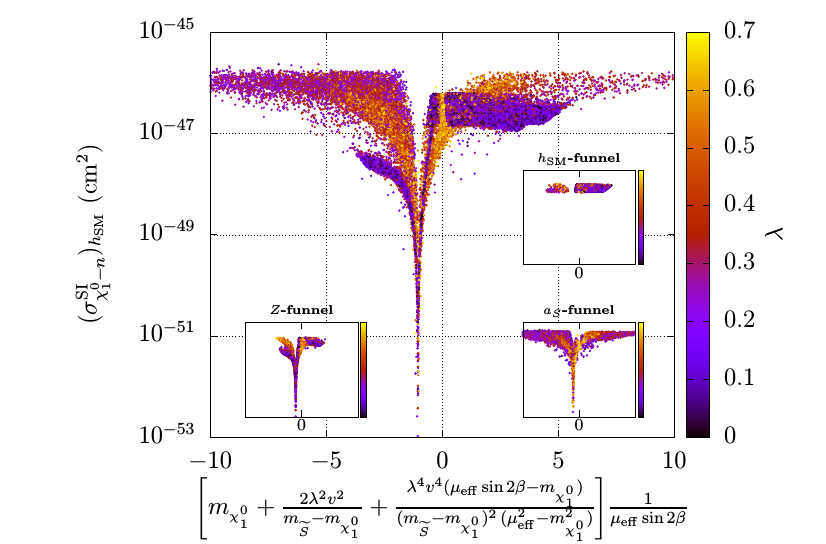}
\caption{Variations of $(\sigma^{\mathrm{SI}}_{\ntrlone-n})_{\hsm}$ (top, left) and 
 $(\sigma^{\mathrm{SI}}_{\ntrlone-n})_{\mathrm{total}}$ (top, right) as functions of
$g_{_{\hsm \ntrlone \ntrlone}}$ and variations of $(\sigma^{\mathrm{SI}}_{\ntrlone-n})_{\hsm}$ as a function of $\mntrlone/{(\mueff \sin 2\beta)}$
(bottom, left) and as a function of the full blind spot construct of eq.~(\ref{eqn:coupling-blindspot-text}) (bottom, right) in a regime with a highly bino-like $\ntrlone$ with $\mntrlone < 200$~GeV. Colors in the adjacent palettes indicate the magnitude of `$\lambda$'.}
\label{fig:coupling-blind-spot}
\end{center}
\vspace{-0.5cm}
\end{figure}

In the top panel of figure~\ref{fig:coupling-blind-spot} we demonstrate, using allowed points
from our scan, how the SI DM-neutron scattering cross section $(\sigma^{\mathrm{SI}}_{\ntrlone-n})$\footnote{For easier comparison of results obtained from various SI DD experiments that employ different target elements carrying nuclei with
characteristic numbers of protons and neutrons (nucleons), the experiments report
their bounds on the coherent DM-nucleus scattering rates in terms of the same on the DM-nucleon
scattering rates~\cite{Feng:2014uja} for which protons and neutrons are taken on equal footing,
as far as their couplings to the DM particle are concerned. In particular, the latter is a very good assumption for the neutralino DM~\cite{Tovey:2000mm}. Nonetheless, \nmssmtools ~computes (via \micromegas) the SI scattering rates for the DM-proton and the DM-neutron scatterings employing the most updated sets of appropriate nuclear form factors.
However, the resulting difference between the two numbers is not significant. We
choose the DM-neutron scattering rate $(\sigma^{\mathrm{SI}}_{\ntrlone-n})$ for
our presentations.} varies as a function of the coupling $g_{_{\hsm \ntrlone \ntrlone}}$ as presented in eq.~(\ref{eqn:hin1n1-2}). The said equation can be written down in terms of various input parameters by appropriate substitutions of the elements $N_{ij}$'s as follows:
\bea
\label{eqn:hin1n1-text}
\hskip -30pt
g_{_{\hsm \ntrlone \ntrlone}}
 &\simeq&
{g^2_1 v \over \sqrt{2} I_1} \Bigg[ \mntrlone + \mueff \sin 2\beta + {{2 \lambda^2 v^2} \over {\msinglino - \mntrlone}} + {\lambda^4 v^4 (\mueff \sin 2\beta - \mntrlone) \over (\msinglino - \mntrlone)^2 \, (\mueff^2 - \mntrlone^2)} \Bigg] \, .
\eea
The derivation and the details of eq.~(\ref{eqn:hin1n1-text}) are presented for the bino-higgsino-singlino system in appendix~\ref{appsec:blindspot} (see eq.~(\ref{eqn:hin1n1-3})). The condition for the coupling blind spot resulting from a vanishing $g_{_{\hsm \ntrlone \ntrlone}}$ follows and is given by (see appendix~\ref{appsec:blindspot})
\bea
\label{eqn:coupling-blindspot-text}
\Bigg[\mntrlone  + {{2 \lambda^2 v^2} \over {\msinglino - \mntrlone}} + {\lambda^4 v^4 (\mueff \sin 2\beta - \mntrlone) \over (\msinglino - \mntrlone)^2 \, (\mueff^2 - \mntrlone^2)} \Bigg]{1 \over \mueff \sin 2\beta} = -1 \, .
\eea
For the same set of points, the plots in the bottom panel of figure~\ref{fig:coupling-blind-spot} illustrate the variations of the SI cross section due to $\hsm$ with an approximate (left) and with the exact (right) blind spot constructs extracted from the left-hand side of eq.~(\ref{eqn:coupling-blindspot-text}). For both panels, palette-colors represent the variation of `$\lambda$'. It is clear from the plots in the upper panel that a falling
$|g_{_{\hsm \ntrlone \ntrlone}}|$ could cause the SI cross section nosedive below the most stringent upper bound on the same. Plots in the lower panel
just illustrate the same physics effect in terms of the blind spot~constructs. 

%
The plot on top, left (top, right) of figure~\ref{fig:coupling-blind-spot} 
shows the variation of SI scattering cross section due to $\hsm$-mediated 
process only (due to all three $CP$-even scalar-mediated processes, i.e., the total SI 
cross section) as a function of $g_{_{\hsm \ntrlone \ntrlone}}$.
The former shows how big $|g_{_{\hsm \ntrlone \ntrlone}}|$ could get
($\lesssim 0.02$) and how a decreasing $|g_{_{\hsm \ntrlone \ntrlone}}|$ could bring 
down the SI cross section below the most conservative upper bound on the same, eventually 
breaching the so-called `neutrino floor' ($\sim 10^{-49} \, \mathrm{cm}^2$)~\cite{Billard:2013qya}, even before it vanishes. In the latter plot,
points appearing above the wings of the underlying structure (as depicted 
in the left plot) are driven by relatively large $H$-induced contribution which 
could involve either a constructive or a destructive interference with the $\hsm$-induced process. On the 
other hand, points below the wings definitely result 
from destructive interference among the three scalar-mediated processes of which, as we will see soon, the dominant ones being 
between the processes mediated by $\hsm$ and `$H$' followed by the one between
$\hsm$ and $\hs$. Processes mediated by $\hs$ and `$H$' hardly reveal any appreciable interference effect for the range of SI cross section above the neutrino floor given their individual strengths are distinctly apart. It may further be noted that 
blind spots with origins in destructive interference(s) that aid compliance with 
stringent upper bounds on the SI cross section tend to be associated with smaller 
(larger) values of $\Omega h^2$ for larger (smaller) values of $|g_{_{\hsm 
\ntrlone \ntrlone}}|$ when the $\hsm$-funnel is in action. This may not be surprising since the size of $|g_{_{\hsm 
\ntrlone \ntrlone}}|$ dictates how efficient the $\hsm$-funnel is for DM 
annihilation once it is kinematically viable. In that sense, the presence of interference blind spots may offer more leg-room by allowing or higher annihilation cross sections via $\hsm$-funnel without getting into conflict with the upper bounds on the SI rate.

It is seen from eq.~(\ref{eqn:coupling-blindspot-text}) that the MSSM coupling blind spot condition (for $\hsm$), i.e., $\mntrlone/(\mueff \sin 2\beta)=-1$, can be retrieved for $\lambda \to 0$. The left plot in the bottom panel corroborates this fact where the dip around $-1$ (i.e., when there is a relative sign between $\mntrlone$ and $\mueff$) indeed happens for small values of `$\lambda$'
(in purple).
In addition, there is a dip for positive values of the blind spot
construct (i.e., for the same sign on $\mntrlone$ and $\mueff$) as well which occurs for larger values of `$\lambda$' (in yellow). This is clearly an NMSSM effect caused by
tempering of the bino-like LSP by a nearby singlino state when `$\lambda$' is large.
Also, note that when $\mntrlone/(\mueff \sin 2\beta)=0$ there is a narrow slit in the upper part of the plot which is not populated at all. This is expected~\cite{Baum:2017enm} as $|\mueff|$  has to be very large over this region such that effectively no higgsino admixture in the bino-like LSP is possible which prohibits efficient annihilation/coannihilation of the LSP DM  to comply with the Planck-observed upper bound on the DM relic abundance. 

The right plot in the bottom panel has got the full blind spot construct of the NMSSM of eq.~(\ref{eqn:coupling-blindspot-text}) along the horizontal axis.
The third term on the left side of eq.~(\ref{eqn:coupling-blindspot-text}) is usually subdominant (see appendix~\ref{appsec:blindspot}).
As expected, the SI rate is now dipping only around $-1$ as the blind spot condition of eq.~(\ref{eqn:coupling-blindspot-text}) dictates so for all values of `$\lambda$', including the points appearing in the yellow-colored dip for same-sign $\mntrlone$ and $\mueff$, as found in the plot on its left. It is further noted that for the points appearing in this dip `$\kappa$' has to be negative such that the NMSSM blind spot condition of eq.~(\ref{eqn:coupling-blindspot-text}) holds (see discussions in the context of eq.~(\ref{eqn:coupling-blindspot-approxi})). In the inset, we highlight the regions which associate three different funnel
processes (induced by $Z$-boson, $\as$ and $\hsm$) that control the DM relic abundance. Note that when the $\hsm$-funnel is responsible for an efficient DM annihilation, the SI rate (attributable to $\hsm$) cannot get smaller than a certain value.

At this point, a brief discussion on the relative contributions of `$H$' and $\hs$ 
to the SI cross section is in order.
The former, being from a doublet, has a much stronger coupling to the nucleons when compared to the latter. Thus, `$H$' may happen to contribute to the SI rate somewhat significantly even when 
it has a multi-TeV mass, including interfering with the process mediated by
$\hsm$. On the other hand, $\hs$
needs to be much lighter ($\sim \mhsm$) to contribute or to compete. In 
that case the sum-rule of eq.~(\ref{eqn:sumrule}) implies $\msinglino$ to be on the smaller side thus 
enhancing the possibility of a more involved bino-higgsino-singlino system.

\begin{figure}[t]
\begin{center}
\hspace*{-1.3cm}
\includegraphics[height=6.2cm, width=0.60\linewidth]{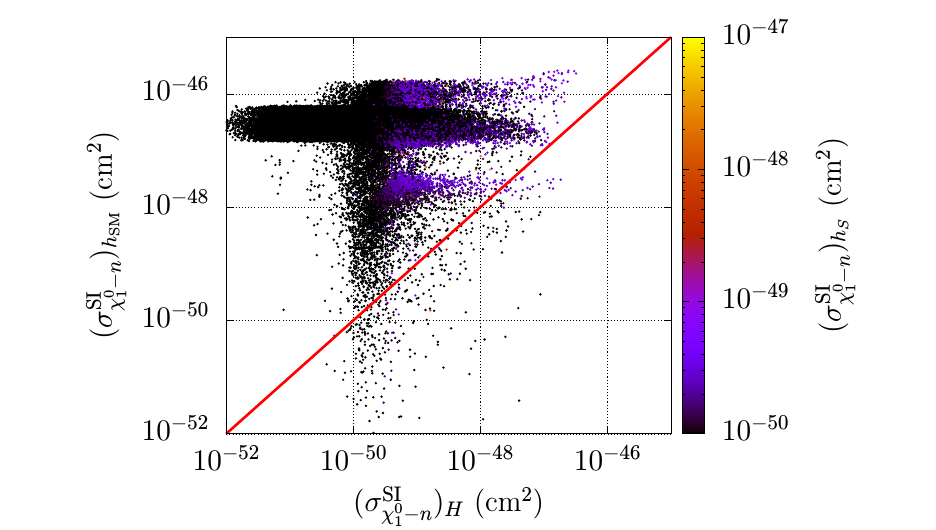}~~
\hskip -29pt
\includegraphics[height=6.2cm, width=0.60\linewidth]{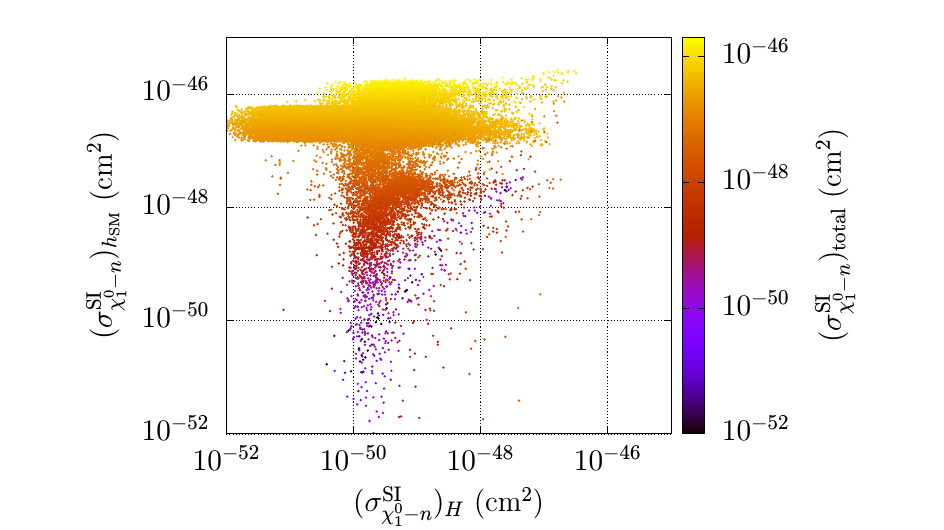} \\[-0.5cm]
\vskip 10pt
\hspace*{-1cm}
\includegraphics[height=6.2cm, width=0.48\linewidth]{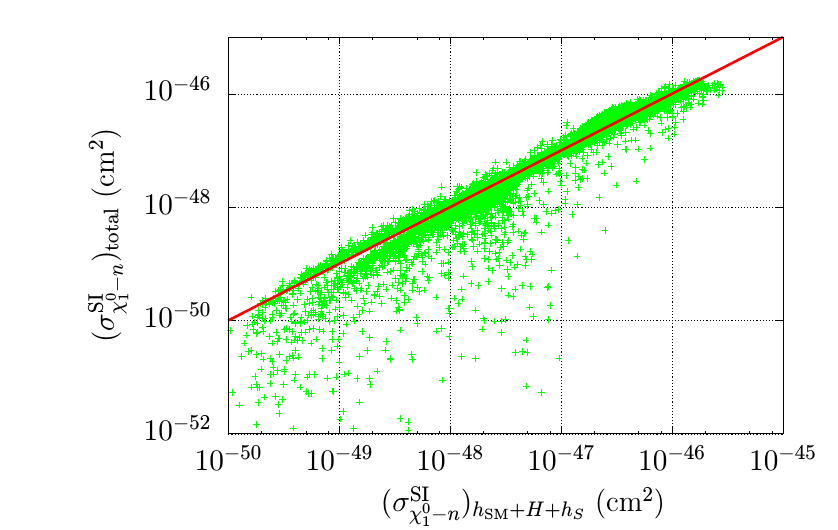}~~
\hskip 11pt
\includegraphics[height=6.2cm,width=0.55\linewidth]{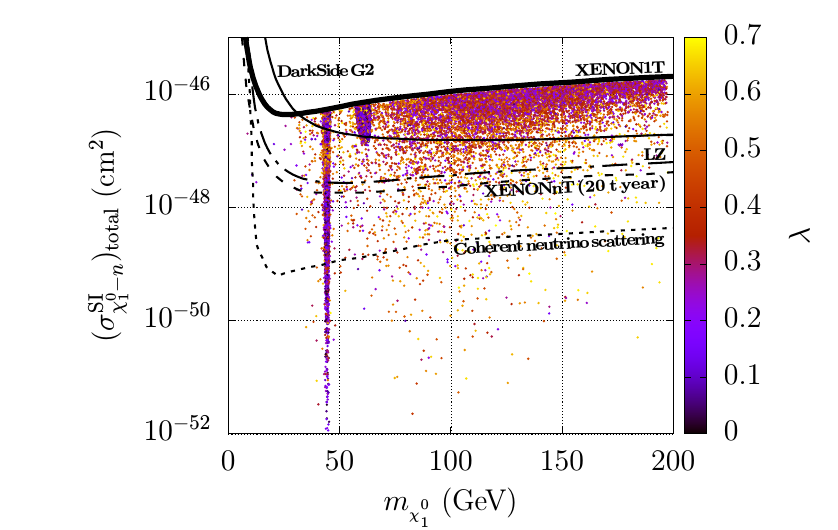}
\caption {Scattered points passing all relevant constraints and illustrating the anatomy and interplay of various contributions to SI DM-neutron scattering cross section (top panel and bottom, left) and the total SI cross section as a function of $\mntrlone$ vis-a-vis results/reach from/of various recent/future experiments with the corresponding magnitude of `$\lambda$' indicated by the palette (bottom,  right).}
\label{fig:si-individual}
\end{center}
\vspace{-0.5cm}
\end{figure}

Figure~\ref{fig:si-individual} presents scattered points that pass 
all experimental constraints 
 and illustrates the following: (i) individual contributions to the SI DM-neutron scattering cross section from various scalars (top, left)\footnote{Extracting the individual rates requires some tweaking in
{\tt NMSSMTools}.}, (ii) the total SI DM-neutron  cross section in relation to the two dominant individual ones (top, right), (iii) the same compared to the incoherent sum of the individual cross sections thus showcasing the points resulting from
constructive/destructive interference (bottom, left) and (iv) how such points are distributed in the $\mntrlone$--$(\sigma^{\mathrm{SI}}_{\ntrlone-n})_{\mathrm{total}}$ plane and juxtaposing them with the recent bounds from the XENON1T experiment and with the expected ones from various future experiments, accompanied by the associated magnitude of `$\lambda$' (bottom, right).
%
%

The top, left plot of figure~\ref{fig:si-individual} clearly shows that, as discussed earlier, the cross section for the process mediated by `$H$' can get to be of a similar order of magnitude  ($\lesssim 10^{-46} \, \mathrm{cm}^2$) as for the $\hsm$-induced one and could even exceed (the region below the diagonal) the latter. The palette shows that the rate for the process mediated by $\hs$ hardly exceeds
$\sim 10^{-49} \, \mathrm{cm}^2$ (actually goes down much below $\sim 10^{-50} \, \mathrm{cm}^2$ which is indicated by black color) and hence mostly gets overshadowed by the other two processes when the overall rate is within the range of sensitivity of the concerned experiment(s). The thick horizontal band that extends from left to right of this plot comprises of points for which the $\hsm$-funnel is active. As discussed earlier, this is the reason behind why the $\hsm$ contribution to the SI rate cannot get smaller than a certain magnitude
($\sim 10^{-47} \, \mathrm{cm}^2$). The other horizontal band below that mainly contains points having an active $Z$-funnel. The same is true along the lower part of the vertical band (in black).
Points that are scattered about the diagonal line in the lower part of the plot
have situations where the $Z$- or the $\as$-mediated funnel is at work.
On the other hand, the band with sparse points (in purple) above the one with
$\hsm$-funnel on the right side of the plot has predominantly $\as$ as the funnel state.

The plot on the top, right of figure~\ref{fig:si-individual} is the same as that on its left but for having the total SI rate indicated by the palette-color. Clearly,
as expected from the top, left plot, the total SI rate is dominated by contributions from the process mediated by $\hsm$ over most parts of the displayed region. However, close to the diagonal, where the
individual rates due to $\hsm$ and `$H$' 
are of comparable magnitudes, the appearance of dark (purple and black) points indicates a sharp drop of total SI cross section when
compared to the said two individual rates. These points result from the occurrence of blind spots due to destructive interference between processes mediated $\hsm$ and `$H$'. For such points, this plot also gives us a rough impression of the severity of the destructive interference by comparing the total cross section with the individual ones that are of similar magnitudes.

The plot on the bottom, left of figure~\ref{fig:si-individual} summarizes how frequently such
a destructive interference occurs. With the incoherent sum of
SI contributions from all three $CP$-even Higgs bosons and the total (coherent sum) presented along the abscissa and the ordinate, respectively, points above (below) the red bold line stand for those for which an overall constructive (destructive) interference
has taken place between the three individual processes among which
two processes driven by the doublet-like Higgs bosons are dominant.

The bottom, right plot of figure~\ref{fig:si-individual} presents our results in the conventional plane of
$\mntrlone$--$(\sigma^{\mathrm{SI}}_{\ntrlone-n})_{\mathrm{total}}$
with `$\lambda$' in the palette. The left (right) vertical funnel strip is attributed to the $Z$-boson ($\hsm$). The scattered points mostly have $\as$ as the funneling state while some such points might also comply with the constraints on relic abundance by means of DM coannihilation. It is clearly seen that an efficient
$Z$- or $\hsm$-funnel could be present for low values of `$\lambda$' while an
optimally efficient $\as$-funnel almost always requires larger `$\lambda$' values.
Note that the dwarfness of the $\hsm$-funnel strip indicates that for this funnel to work, the associated SI rate has to be bounded from below. 
The projected sensitivities of various future experiments like the DarkSide \!G2~\cite{Aalseth:2015mba}, the LZ~\cite{Aalseth:2015mba,Akerib:2018lyp} and the XENONnT (20 tonne-year)~\cite{Aprile:2015uzo, Aprile:2020vtw} along with the neutrino floor~\cite{Billard:2013qya} are indicated in the plot.
Only a handful of points is found to sink below the neutrino floor, some of which are due to severe destructive interference between processes with large individual contributions. 
%
\subsubsection{The SD scattering cross section}
\label{subsubsec:sd}
The SD scattering process also proceeds in the $t$-channel but, for all practical purposes, is mediated only by the SM $Z$-boson, in the scenario under study. An adequate suppression of its cross section and hence its compliance with the reported upper bounds on its rates~\cite{Aprile:2019dbj, Amole:2019fdf} can solely be achieved when a blind spot is hit in the $Z$-$\ntrlone$-$\ntrlone$ coupling itself. Again, derivation of the relevant blind spot condition is presented for the first time in the context of a bino-higgsino-singlino system in appendix~\ref{appsec:blindspot} (see eq.~(\ref{eqn:sigma-SD-1})).
%
\begin{figure}[t]
\begin{center}
\includegraphics[width=0.55\linewidth]{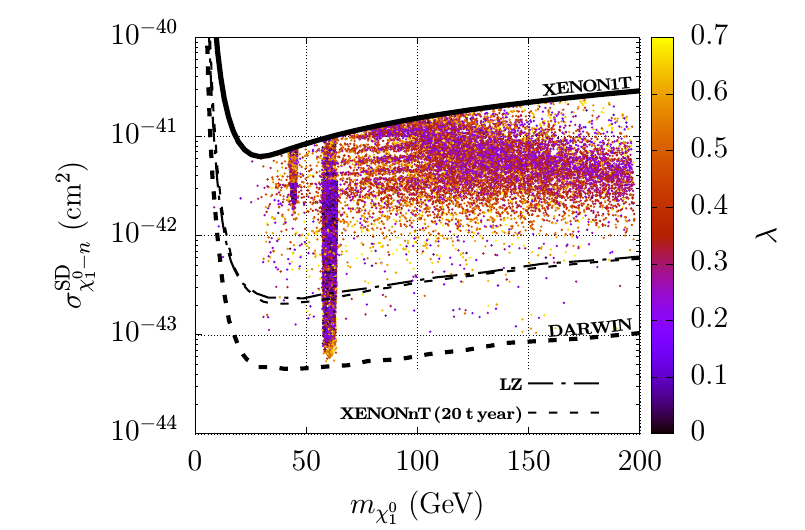}
\vspace{-0.5cm}
\caption {Same as in the bottom, right plot of figure~\ref{fig:si-individual} but for SD rate along the ordinate.}
\label{fig:sd-rate-mlsp}
\end{center}
\vspace{-0.5cm}
\end{figure}

Figure~\ref{fig:sd-rate-mlsp} is the same as the bottom, right plot of figure~\ref{fig:si-individual} but for the SD DM-neutron scattering cross section 
($\sigma^{\mathrm{SD}}_{\ntrlone-n}$) now 
being plotted along the ordinate. 
In sharp contrast to the former plot, here one finds 
the strip on the left hosting points for which $Z$-funnel aided DM annihilation 
is at work has a minimum
with a much larger value of SD cross section than that for the $\hsm$-driven one 
right next to it. The line of reasoning is the same as in the former case 
although the requirement is now just the opposite, i.e., here the strength of the 
$Z$-$\ntrlone$-$\ntrlone$ coupling has to be 
larger than a minimum for an efficient DM annihilation which, in turn, results in  
a $Z$-mediated SD rate bounded from below. This directly corresponds to
an upper bound on $|\mueff|$ which is $\lesssim 450$~GeV for this case. 
On the other hand, we find that the experimental upper bounds on the SD rates, under no circumstances, allow for $|\mueff| \lesssim 270$~GeV~\cite{Carena:2018nlf} while
on the $Z$-funnel the lower bound on $|\mueff|$ turns out to be $\sim 330$~GeV.  
Such a sensitivity originates in the $1/\mueff^2$ dependence of the involved
$Z$-$\ntrlone$-$\ntrlone$ coupling
(see  appendix~\ref{appsec:blindspot}).\footnote{Given that the $\as$-$\ntrlone$-$\ntrlone$ coupling also goes as $1/\mueff^2$, an upper bound
on $|\mueff|$ of roughly 600~GeV is obtained to have the $\as$-funnel to be efficient enough.} 
It may be noted that this bound has to be stronger than the one coming from the constraint on the SI rate since the latter
involves Higgs-$\ntrlone$-$\ntrlone$ interactions whose dominant parts go as
$1/\mueff$ and hence, on the SM-like Higgs funnel, the upper bound on DM relic abundance
could get satisfied even for $|\mueff| \lesssim 1$~TeV.
The horizontal streaks seem to have their origins in the pattern that is already present in the reported upper bounds on the yields in some specific final states as a function of the ewino mass, as obtained by an LHC experiment~\cite{Sirunyan:2018ubx} in its analysis of a search for the lighter ewinos. Such a pattern is
likely to have emerged from the combining of searches in those final states (as each of them has a different experimental sensitivity~\cite{private communication}) and has been extracted
from ref.~\cite{Sirunyan:2018ubx} and is implemented in {\tt NMSSMTools}~\cite{Ellwanger:2018zxt}.

Projected sensitivities of various future experiments~\cite{Aprile:2020vtw,Akerib:2018lyp,Aalseth:2015mba,Aalbers:2016jon} to the SD rates are also illustrated in
figure~\ref{fig:sd-rate-mlsp}. It can be seen that only the DARWIN experiment~\cite{Aalbers:2016jon} is expected to probe the entire range of
$\sigma^{\mathrm{SD}}_{\ntrlone-n}$ thus being sensitive to even the region of parameter space where the SI rate dips below the neutrino floor. 

It may be recalled (see section~\ref{subsec:relic}) that the rarity of allowed points below $\mntrlone \approx 30$~GeV (i.e., $\lesssim \mhsm/4$) in the last plot of figure~\ref{fig:si-individual} and in figure~\ref{fig:sd-rate-mlsp} is linked to stringent constraints from the LHC experiments on the light
(pseudo)scalar states with masses below $\sim \mhsm/2$.
%
\subsection{The role of $\tanb$}
\label{subsec:tanb}
It turns out that the value of $\tanb$ plays an important role in the 
phenomenology of a bino-dominated LSP DM. It is noteworthy that a small $\tanb$ 
could enhance the $\lambda$-dependent tree-level contribution to $\mhsm$ (see eq.~(\ref{eqn:hsmmass})) thus ensuring a smaller
$\Delta_{\mathrm{rad.\,corrs.}}$ (see eq.~(\ref{eqn:radcorr})) suffices to render 
$\mhsm$ in its observed range. Given that a smaller
$\Delta_{\mathrm{rad.\,corrs.}}$ implies lighter stop(s), NMSSM scenarios with 
small $\tanb$ might emerge as highly `natural' and hence have inspired some
serious studies in the recent years~\cite{King:2012is, King:2012tr, Beuria:2015mta, Beuria:2016mur, Baum:2017enm}. In particular, ref.~\cite{Baum:2017enm} discusses the phenomenology of a bino-dominated DM in such a 
setup, in some detail.

A salient observation made in ref.~\cite{Baum:2017enm} is that such a DM 
candidate could hardly have a mass below that of the top quark primarily since it leads to an unacceptably large SI rate. We thoroughly agree with that observation
within the scenario it addressed. This is something not quite unexpected since the
observation concerns a reasonably large value of $\mntrlone / \mueff$. If one refers to the dominant
component of the blind spot condition, i.e. $\mntrlone/\mueff + \sin 2\beta=0$,
this evidently requires an equally large $\sin 2 \beta$ and hence a smaller
value of $\tanb$ for the blind spot to work (with a mandatory relative sign between $\mntrlone$ and $\mueff$) leading to an acceptably small SI rate. Thus, from this
perspective, a possible way to reach out to lower values of $\mntrlone$
(thus, a smaller value of $\mntrlone / \mueff$) would be to allow for larger values of $\tanb$. In being able to go down sufficiently in $\mntrlone$ could also open up
hitherto inaccessible funnels in the form of $\hsm$ and the $Z$-boson boosting
the possibility of an efficient annihilation of the LSP DM.

However, allowing for larger values of $\tanb$ would compel one to consider heavier doublet-like Higgs bosons, i.e., `$A$' and `$H$'
to comply with the latest LHC bounds~\cite{Aad:2020zxo} as their decays into canonical (MSSM-like) final states, i.e., $b \bar{b}$ and $\tau \bar{\tau}$, are known to dominate~\cite{King:2014xwa} at large~$\tanb$. This would then drive us away from the scenario of {\it alignment without decoupling} adopted in ref.~\cite{Baum:2017enm}. A heavy `$H$' might also help comply with the stringent upper bound on the SI rate.
Thus, we open up the parameter space to larger values of $\tanb$ ($>5$) and systematically study its effect. Simultaneously, we also choose to expose our study to smaller values of `$\lambda$' and larger values of $\alambda$, all of which are indicated in table~\ref{tab:ranges}.

We now take a closer look into the role played by $\tanb$
in achieving a small SI cross section. Towards this, we use the general NMSSM blind spot construct involving $\hsm$ and~`$H$' in the bino-higgsino-singlino system as deduced in eq.~(\ref{eqn:NMSSM-BLINDSPOT-CONDITION-1}) and define its left-hand side as 
\bea
\label{eqn:NMSSM-BLINDSPOT-CONSTRUCT}
\hspace{-1cm}
F_\mathrm{BS} &=&
 {\mntrlone \over \mueff} + \sin 2\beta + {2 \lambda^2 v^2 \over \mueff \, (\msinglino - \mntrlone)} + {\lambda^4 v^4 \big(\sin 2\beta - \mntrlone / \mueff\big) \over (\msinglino - \mntrlone)^2 \, (\mueff^2 - \mntrlone^2)} \nonumber \\ 
&+& \frac{1}{2} \cos 2\beta \bigg(\tan\beta- \frac{1}{\tan\beta}\bigg) \Bigg[{\lambda^2 v^2 \left[\lambda^2 v^2 + 2 \mntrlone (\msinglino - \mntrlone) \right] \over (\msinglino - \mntrlone)^2 \, (\mueff^2 - \mntrlone^2)} - 1 \Bigg] {\mhsmsq \over m^2_{H}} \, ,
\eea
where the first four terms arise from the $\hsm$-$\ntrlone$-$\ntrlone$ interaction (see eq.~(\ref{eqn:hin1n1-3})) of which the first two terms are $g_1$-driven and the last two terms are $\lambda$-driven and singlino-tempered. Terms in the line that follows originate in $H$-$\ntrlone$-$\ntrlone$ interaction (see eq.~(\ref{eqn:Hn1n1-3})) and are suppressed by $m_H^2$. There also, the  $\lambda$-driven terms have the NMSSM origin while the ones proportional to unity have pure MSSM root. Obviously, the smaller the value of $|F_\mathrm{BS}|$ is, the smaller would be the SI rate.
\begin{figure}[t]
\begin{center}
\includegraphics[width=0.54\linewidth]{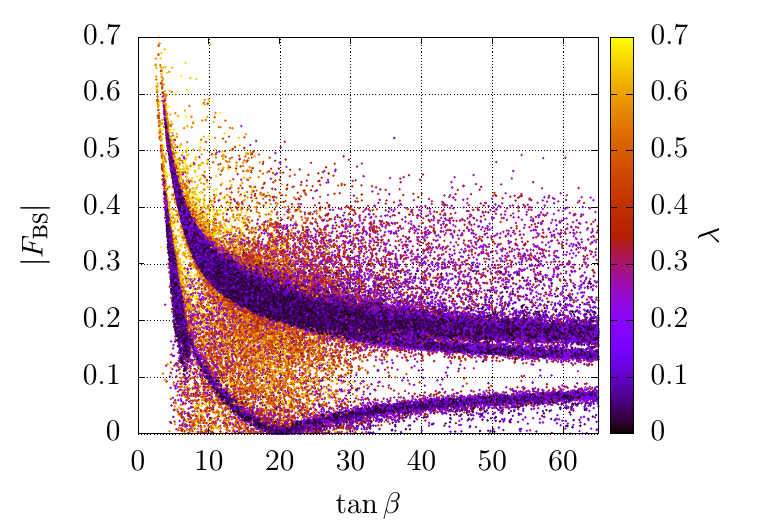}
\vspace{-0.4cm}
\caption{Variation of the blind spot construct of eq.~(\ref{eqn:NMSSM-BLINDSPOT-CONSTRUCT}) as a function of $\tanb$. Associated `$\lambda$' values are presented via
the palette. See text for details.}
\label{fig:mlsp-SI-tanb}
\end{center}
\vspace{-0.5cm}
\end{figure}

In figure~\ref{fig:mlsp-SI-tanb} we illustrate the variation of $|F_\mathrm{BS}|$ as a function of $\tanb$ for our scanned points which pass all relevant constraints. Associated `$\lambda$' values are indicated by the palette. It appears from this figure that smaller values of $|F_\mathrm{BS}|$ and hence smaller SI rates are found predominantly for $\tanb > 5$. 
The figure reveals three clear regions: two well-defined purple bands (i.e., having small `$\lambda$') in which mostly $Z$- and $\hsm$-funnels are active.
`$\lambda$' being on the smaller side, funneling via $\as$ is not so favored over these regions.  
Thus, such regions with small `$\lambda$' values, for large $m_H$ and/or small $\tanb$, to a very good extent, are describable via
\beq
\!\!\!F_\mathrm{BS} \xrightarrow[\mathrm{small}~\lambda]{} {\mntrlone \over \mueff} + \sin2\beta - {\mhsm^2 \over {2 m_H^2}} \cos2\beta \left(\tanb - {1 \over \tanb} \right) {\xrightarrow[\mathrm{small}\,\tanb]{\mathrm{large}~m_H~{\rm and/or}}}  {\mntrlone \over \mueff} + \sin2\beta \, .
\label{eqn:blind-spot-construct}
\eeq
On the other hand, the third is a broad region of scattered points mostly with $\lambda \gtrsim 0.2$. This reveals a predominance of points that associate larger (smaller) `$\lambda$' values with smaller (larger) values of $\tanb$.
Over this region, an efficient DM annihilation is achieved through either of $Z$-, $\hsm$- and $\as$-funnels or (to a much lesser extent) via coannihilation with a singlino-dominated NLSP.
Note that all these regions (irrespective of values of `$\lambda$', $m_H$ and $\tanb$) find their explanations in terms of the more general blind spot construct of eq.~(\ref{eqn:NMSSM-BLINDSPOT-CONSTRUCT}). We now take a brief, closer look into the workings of these regions.

For the lower (left) purple band, $\mntrlone \mueff < 0$.  Having already 
small values of `$\lambda$', for a sufficiently large $m_H$,
one finds $|F_\mathrm{BS}| \sim$ $|(\mntrlone/ \mueff) + \sin 2\beta|$
(see eq.~(\ref{eqn:blind-spot-construct})) which 
is just a measure of the $\hsm$-$\ntrlone$-$\ntrlone$
interaction strength. From our discussion under section \ref{subsec:ddrates}, this has to be bounded from below when the $\hsm$-funnel is in action. This is reflected in the abrupt (spiky) termination of a part of the vertical strand
in the lower (left)
purple band (for $\tan\beta \lesssim 8$ and $|F_\mathrm{BS}| \gtrsim 0.1$) where, given  $\mntrlone \mueff <0$, a possible cancellation between the terms
$\mntrlone/\mueff$ and $\sin2\beta$ weakens the $\hsm$-$\ntrlone$-$\ntrlone$ interaction. In fact, in such a case, it becomes increasingly difficult for the $\hsm$-funnel  
to be efficient enough once `$\lambda$' gets smaller~($\lesssim 0.2$) with
a growing $\tanb$ ($\gtrsim 8$).

On the other hand, this band hits the blind spot ($F_\mathrm{BS} \approx 0$) for $\tanb$ somewhere around~20 when the $Z$-funnel is in action.
As discussed in section \ref{subsec:ddrates} in the context of figure~\ref{fig:sd-rate-mlsp}, while on $Z$-funnel, the lower (330~GeV) and the upper (450~GeV) bounds on $|\mueff|$ arise from the constraints on SD rates and the $2\sigma$ upper bound on the observed relic abundance, respectively.  Simple algebra then reveals, for $\mntrlone \sim m_Z/2$ and  330~GeV $<|\mueff|<$ 450~GeV, cancellation between the said two terms requires $\tanb$ around~20. Also, for a given $\mntrlone$, the larger $|\mueff|$ is, the larger would be the value of $\tanb$ to comply with the upper bound on the SI rate.

The upper purple band appears for $\mntrlone \mueff >0$ which, though known to be not able to hit a blind spot for an effective bino-higgsino LSP system, does nonetheless exhibit a decreasing trend in $|F_\mathrm{BS}|$ as $\tanb$ increases. Closer inspections reveal two overlapping bands that arise when
$\mntrlone$ and $\mueff$ carry signs that are either both positive or both negative. Note that possible mechanisms that yield a small enough $|F_\mathrm{BS}|$ for larger values of `$\lambda$' are quite intricate and are encoded in eq.~(\ref{eqn:NMSSM-BLINDSPOT-CONSTRUCT}) (see also appendix~\ref{appsec:blindspot}).

In any case, it is clear from this plot that $\tanb \gtrsim 5$ is a pretty general requirement if a relatively light bino-like DM has to be viable. It may further be noted here that for~$\mntrlone < m_t$, DM annihilation via $\as$-funnel into a pair of `$b$'~quarks could find a welcome boost at large $\tanb$ for which the coupling strength $|g_{_{\as b \bar{b}}}|$ is enhanced~\cite{Ellwanger:2009dp}.
%
\subsection{Implications of direct searches for the ewinos at the LHC}
\label{subsec:lhc}
As pointed out in the Introduction, the most sensitive of the direct searches 
of relatively light ewinos at the LHC involve multi-lepton plus multi-jet final states, in 
particular, the rather clean final state with three leptons arising in the 
direct production of $\charonepm \ntrltwo$ cascading in the $WZ$-mode (see 
eq.~(\ref{eqn:wz-wh})) to $3\ell + \etmiss$ (where both $W^\pm$ and $Z$ 
bosons decay leptonically)~\cite{Sirunyan:2017lae, Sirunyan:2018ubx, Aaboud:2018jiw, Aaboud:2018sua, 
Aad:2019vvi, ATLAS:2020ckz} or to a state with $2\ell + 2$-${\rm jet} + \etmiss$ (where $W^\pm$ 
decays hadronically)~\cite{Sirunyan:2017lae, Sirunyan:2018ubx, Aaboud:2018jiw, Aaboud:2018sua}.
Assuming wino-like $\charonepm$ and $\ntrltwo$ and a bino-like LSP, the most stringent lower bound on $\mcharone (= \mntrltwo)$ obtained till date in the $WZ$-mode is 650~GeV~\cite{Sirunyan:2018ubx} for
a vanishing $\mntrlone$.

Of recent, there have been LHC analyses which further consider wino-like
$\charonepm$ and $\ntrltwo$ cascading not only in the $WH$-mode~\cite{Sirunyan:2017lae, Sirunyan:2018ubx, Aaboud:2018ngk, Sirunyan:2019iwo, 
Aad:2019vvf, Aad:2020qnn, ATLAS:2020ckz} as indicated in eq.~(\ref{eqn:wz-wh}) but also in a representative `mixed' mode where $\ntrltwo$ 
decays 50\% of the times to each of the $Z$-boson and
$\hsm$~\cite{Sirunyan:2018ubx}, i.e., in a mixed ($WZ$ and $WH$) mode. It is interesting to note that, in the $WH$-mode, by analyzing the final state $1\ell + 2b$-${\rm jet} +\etmiss$, where $b$-jets originate in the decay of~$\hsm$, a corresponding bound as stringent as $\mcharone \,(=\mntrltwo) > 740$~GeV is obtained~\cite{Aad:2019vvf} while
in the `mixed' mode a somewhat lower bound of 535~GeV has been reported~\cite{Sirunyan:2018ubx}.

We recollect from the Introduction that these severe bounds on the lighter
ewinos could get relaxed in our present pNMSSM scenario due to the following reasons.   
\begin{itemize}
\item Higgsino-like lighter ewinos that we consider in this work have production
      cross sections considerably smaller than those for the wino-like ones (of 
      the same mass) that the experiments consider. Note that such a 
      reduction does not get compensated in the presence of two higgsino-like
      near-degenerate neutralinos ($\chi^0_{\widetilde{H}_{1,2}}$), accompanied 
      by a higgsino-like chargino ($\chi^\pm_{\widetilde{H}}$), and the combined 
      cross section for
      $pp \to \chi^\pm_{\widetilde{H}} \, \chi^0_{\widetilde{H}_{1,2}}$ hardly 
      exceeds half the value of that for
      $pp \to \chi^\pm_{\widetilde{W}} \, \chi^0_{\widetilde{W}}$. 
      In fact, this is still true even when we include a singlino-like NLSP 
      ($\chi^0_{2 (\widetilde{S})}$) which is lighter than two nearly 
      degenerate higgsino-like neutralinos ($\chi^0_{3,4 (\widetilde{H})}$)
      thus referring to the production processes
      $pp \to \chi^\pm_{\widetilde{H}} \, \ntrltwothreefour$.
      This is since the production cross section for the individual 
      process
      $pp \to \chi^\pm_{\widetilde{H}} \, \chi^0_{2 (\widetilde{S})}$ is itself
      small as $\chi^0_{2 (\widetilde{S})}$ is singlino-dominated. A smaller 
      combined cross section leads to a smaller sensitivity to 
      experimental data. Hence one may expect relaxed lower bounds on the
      masses of these ewinos.\\[-0.6cm]
\item Bounds obtained by analyzing multi-lepton final states via the
      $WZ$-mode assume 
      both BR[$\charonepm \to W^\pm \ntrlone$] and 
      BR[$\ntrltwo \to Z \ntrlone$] to be 100\%.
      In our NMSSM scenario, in the presence of a singlino-like  
      NLSP neutralino, $\ntrltwo$, the higgsino-like $\charonepm$ 
      could, in addition, undergo the decay $\charonepm \to W^\pm \ntrltwo$. 
      Furthermore, since there could be light singlet-like scalars hanging around, the two 
      higgsino-like neutralinos, $\ntrlthreefour$ could further undergo  
      decays like $\ntrlthreefour \to \hs/\as + \ntrlonetwo$ while 
      $\ntrltwo$, arising in both $\charonepm$ and $\ntrlthreefour$ decays,
      could decay 100\% of the times to $\hs/\as + \ntrlone$. Given that
      such light $\hs$ and $\as$ decay dominantly in non-leptonic modes, 
      their appearance would surely deplete
      multi-lepton events thus eroding the lower bounds on the ewino
      masses reported by the LHC experiments.\\[-0.6cm]
\item Bounds obtained via the $WH$-mode, though have recently emerged to be 
      the most sensitive thus offering the strongest ever bound on the masses 
      of the wino-like $\charonepm$ and $\ntrltwo$, necessarily involve a pair
      of $b$-jets~\cite{Aad:2019vvf, Aaboud:2018ngk} or a pair of photons
     ~\cite{Aaboud:2018ngk} whose invariant masses are consistent with
      $\mhsm$. Thus, in our scenario, such bounds would apply only if $\hsm$ itself appears in the cascades of heavier neutralinos and/or 
      the masses of the participating singlet-like scalars fall into a window 
      about $\mhsm$.\footnote{The most sensitive decay mode of $\hsm$ is 
      found to be the $b\bar{b}$ mode and the invariant mass-window for a $b$-jet pair 
      employed by the experiments is $100~\mathrm{GeV} < m_{b\bar{b}} < 
      140~\mathrm{GeV}$~\cite{Aad:2019vvf}.} Hence regions of the 
      parameter space having suppressed branching ratios of the heavier 
      neutralinos to $\hsm$ could evade the experimental bounds.
\end{itemize}
\subsubsection{Decays of $\ntrltwothreefour$}
\label{subsubsec:individual-brs}
Observations made above prompt us to study the pattern of branching ratios of the heavier neutralinos and the lighter chargino over the otherwise allowed region of the pNMSSM parameter space.
Two phenomenologically relevant situations may arise depending on the relative magnitudes of $\mueff$ and $\msinglino$: one where a higgsino-like neutralino is the NLSP 
($|\mueff| \lesssim |\msinglino|$) and in the other case, the role of the NLSP is played by the singlino-like one ($|\mueff| \gtrsim |\msinglino|$).
\begin{figure}[t]
\vspace{-0.05cm}
\begin{center}
\includegraphics[height=0.30\textheight, width=0.40\columnwidth]{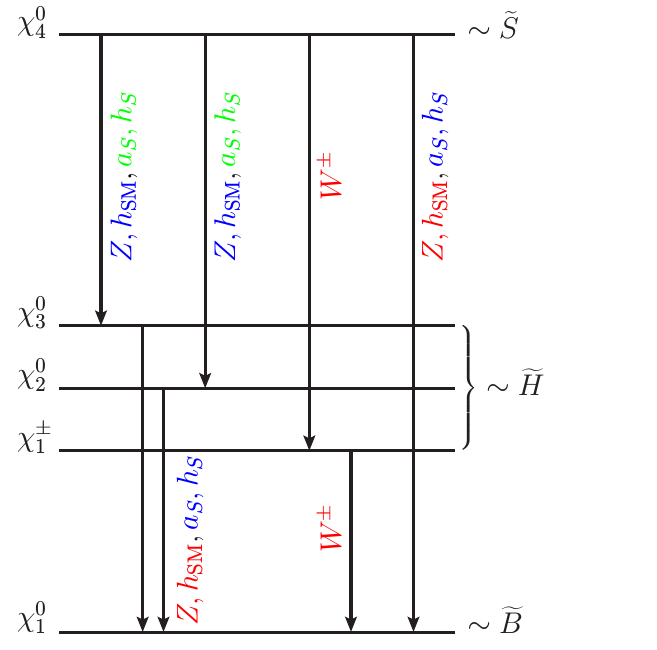}~~~~
\includegraphics[height=0.30\textheight, width=0.40\columnwidth]{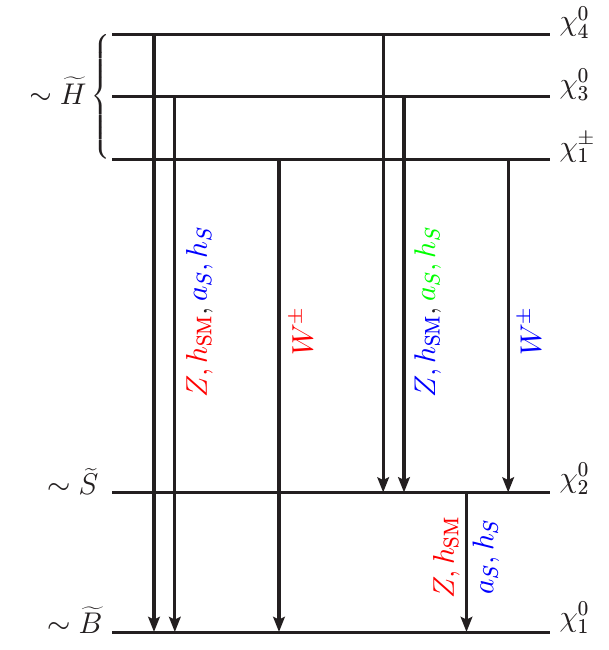} \\
\vspace*{0.5cm} 
\hspace*{-0.5cm}
\includegraphics[height=6.2cm, width=0.35\linewidth]{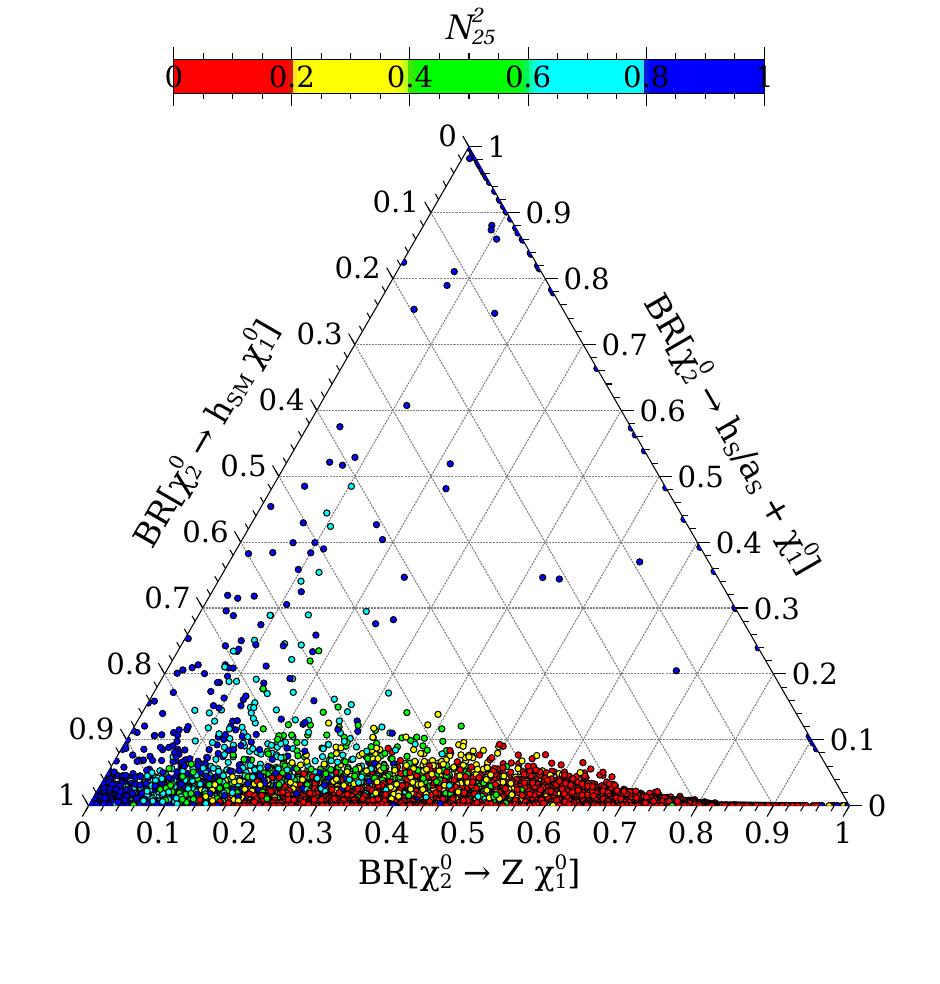}
\hskip -10pt
\includegraphics[height=6.2cm, width=0.35\linewidth]{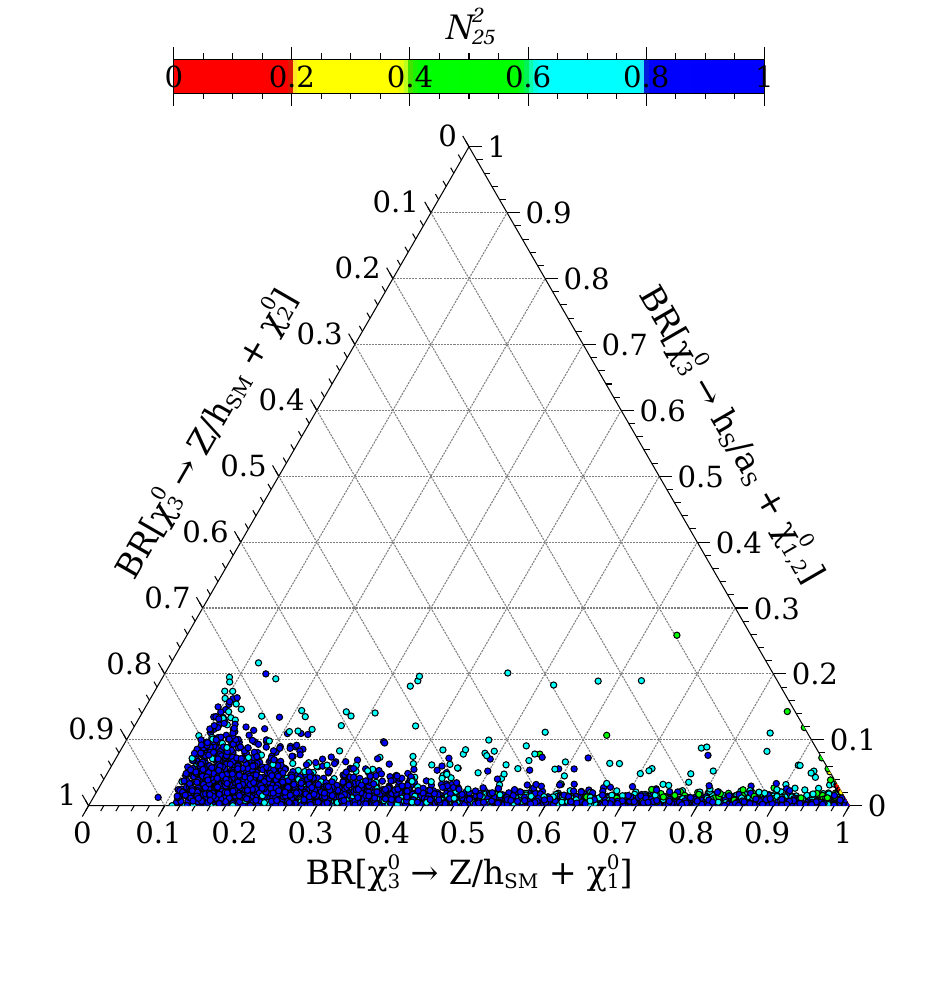}
\hskip -10pt
\includegraphics[height=6.2cm, width=0.35\linewidth]{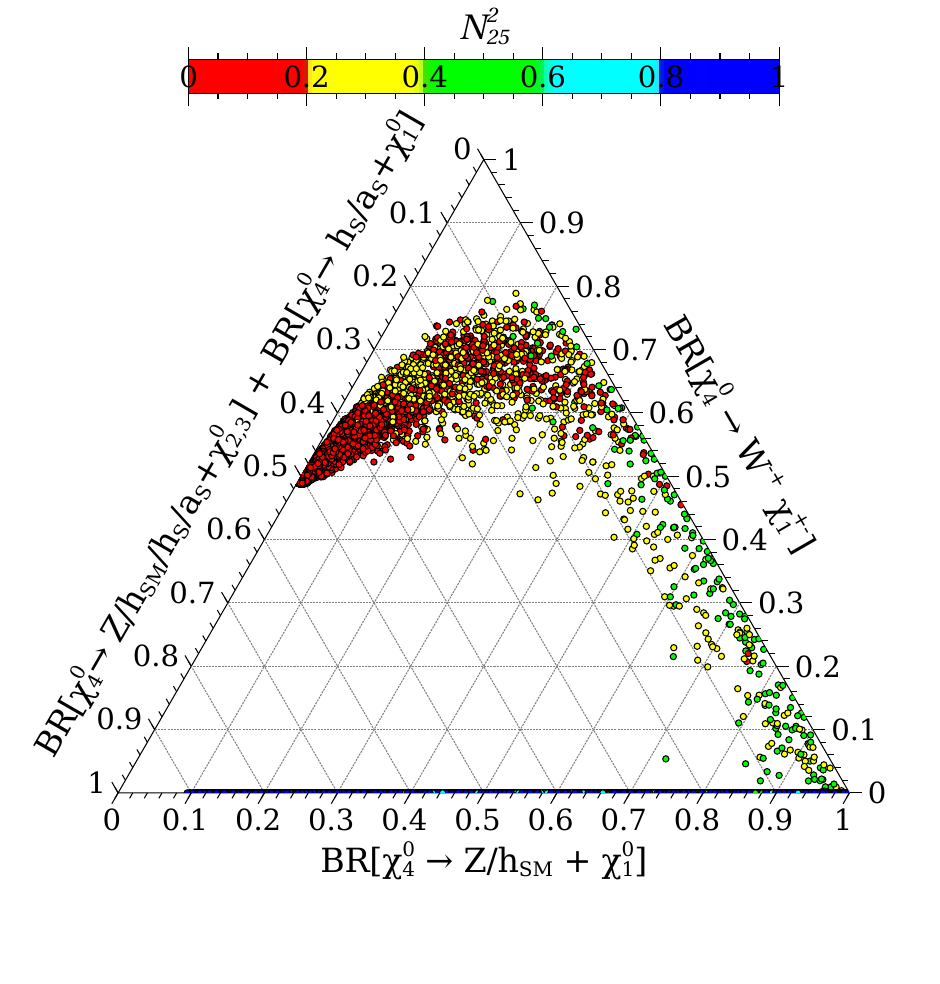}
\vspace{-1cm}
\caption{
Level diagrams showing two different neutralino hierarchies (top panel) with 
higgsino-like (singlino-like) NLSP on the left (right) and ternary scatter 
plots (bottom panel) presenting the branching ratios of the heavier 
neutralinos, $\ntrltwo$ (left), $\ntrlthree$ (middle), $\ntrlfour$ (right), for 
the allowed points, as functions of the singlino-content in the NLSP
($N_{25}^2$), as shown in the palettes. The wino-like $\ntrlfive$ is decoupled 
($\mntrlfive \sim \mtwo =2.5$~TeV). See text for details.}
\label{fig:ternary}
\end{center}
\vspace{-0.8cm}
\end{figure}

In the top panel of figure~\ref{fig:ternary} we illustrate these two possible hierarchies of the neutralinos and the lighter chargino via a pair of level diagrams
\cite{Potter:2015wsa, Abdallah:2019znp} that also indicate for the respective scenarios the viable two-body decays
of various ewinos. In the bottom panel of figure~\ref{fig:ternary} a triplet of ternary scatter plots presents various two-body branching ratios of the heavier neutralinos, $\ntrltwo$ (left), $\ntrlthree$ (middle) and $\ntrlfour$ (right),
respectively, along the three sides of the triangles and the associated magnitude of the singlino content ($N_{25}^2$) of the NLSP through the palette. These branching ratios are normalized over three suitably clubbed decay modes shown along the three axes of the equilateral triangles. These modes are classified as the ones containing the $Z$-boson, the scalars (Higgs bosons), the $W^\pm$-boson  and $\ntrltwo$. Parameter points yielding significant three-body branching ratios for these states are not so common in our scenario and hence are ignored. Thus, a transition from red to blue stands for a higgsino-dominated NLSP turning into a singlino-dominated one. Our focus would be on the regions where either the singlet-like scalars ($\hs$ and/or $\as$) and/or $\ntrltwo$ appear in the neutralino cascades. These would lead to enhanced lepton and/or jet activity in the final state arising from associated production of the lighter chargino and the neutralinos which, in turn, would help evade the existing stringent lower bounds on their masses.\footnote{In addition to various pertinent constraints from low energy experiments, the same from the LHC analyses of the observed SM-like Higgs boson and the ones from various DM experiments, {\tt NMSSMTools} also incorporates a simple-minded recast~\cite{Ellwanger:2018zxt} of an LHC  trilepton analysis~\cite{Sirunyan:2018ubx} to constrain masses of higgsino-like,
degenerate $\ntrltwothree$ and $\charonepm$. The scenarios we discuss in the present work have a third (or even a fourth) neutralino of a comparable mass and, in contrast to what is considered in ref.~\cite{Ellwanger:2018zxt}, the heavier neutralinos among them could decay to a lighter one in addition to their
decays to the LSP. Furthermore, in such scenarios, even the lighter
 chargino could decay to the NLSP neutralino, a mode that has not been considered in ref.~\cite{Ellwanger:2018zxt} but could easily erode the bounds derived there. Interestingly, for our present analysis, the strongest lower bound on the masses of such relatively light higgsino-like states seem to originate from the reported upper bounds on the SD cross section and is in the ballpark of 270~GeV (see section~\ref{subsec:ddrates}).}

The left plot in the bottom panel of figure~\ref{fig:ternary} reveals that the branching ratios of
$\ntrltwo$ are predominantly shared between final states with $Z$-boson and $\hsm$ while the same for the final states with $\hs$ and $\as$ are
mostly restricted to the 10\% level. As we graze along the horizontal axis from right to left, the gradual change in color (from red to blue) indicates increasing singlino-content in $\ntrltwo$ while branching to $\hsm$ increases  mostly  at the cost of the same to $Z$-boson. Such a change in the composition of $\ntrltwo$ is induced by an increasing $|\mueff|$ with respect to $|\msinglino|$. This strips  the higgsino components off both  $\ntrltwo$ and the LSP thus resulting in their weakened coupling to $Z$-boson and hence a dwindling BR[$\ntrltwo \to Z \ntrlone$]. Note that this is also accompanied by an increase in the coveted branching of $\ntrltwo$ to singlet scalars which could routinely reach the 50\% level and might occasionally exceed even 80\% as indicated by the blue scattered points in the upper half of the triangle.

The middle plot in the bottom panel of figure~\ref{fig:ternary} sheds light on various branching ratios of $\ntrlthree$ which is always higgsino-like in our scenario unlike $\ntrltwo$ or $\ntrlfour$, both of which can be singlino-like as well. The situation is necessarily different from the previous case since $\ntrlthree$ can undergo a richer set of cascades involving both $\ntrlone$ and $\ntrltwo$. This calls for a phenomenologically appropriate clubbing of the final states (which has to be different from the case of
$\ntrltwo$ decay) leading to an appropriate set of observables as presented along the three axes of this ternary plot. The combined quantity referring to decays to the $Z$-boson and $\hsm$ along with the LSP, plotted along the base axis, can be termed as the `canonical branching ratio' of $\ntrlthree$, given that these are the decay modes extensively studied by the LHC experiments, yielding stringent lower bounds on the masses of the ewinos. The left axis replaces the LSP by $\ntrltwo$ as a
`non-canonical' state whose further decay (see  the left plot) would supply additional leptons and/or jets (possibly, including $b$-jets) to the overall final state when compared to the ones to which the LHC studies are restricted. The right axis presents $\ntrlthree$ branching to states which, on top of $\ntrltwo$, exclusively contain the singlet scalars. With these, obviously, cascades that follow would result in events that would fall drastically away from the signal regions considered by the LHC experiments in their dedicated hunt for ewinos which have always looked for the $Z$-boson or $\hsm$ in the cascades.

Note that when $\ntrlthree$ is higgsino-like and nearly degenerate with $\ntrltwo$,
the only kinematically possible two-body decays that it can undergo are $\ntrlthree \to Z/ \hsm /\hs / \as \, + \ntrlone$ with a combined branching ratio equaling unity. 
Such a situation shows up as a small red region at the extreme right corner of the triangle. Any mass-split between $\ntrlthree$ and $\ntrltwo$ that allows the former to cascade to the latter requires the former to be higgsino-like (along with a  higgsino-like $\ntrlfour$) and the latter to be singlino-like with
$m_{\chi^0_{2(\widetilde{S})} } < m_{\chi^0_{3(\widetilde{H}_1)} } \approx m_{\chi^0_{4(\widetilde{H}_2)} }$. This is why the rest of the region shows up in a  bluish color indicating a singlino-rich $\ntrltwo$.

The decay branching ratios of a higgsino-like $\ntrlfour$ exhibits a behavior antipodal to $\ntrlthree$ which is also higgsino-like, i.e.,
the behavior of BR[$\ntrlthree \to Z \ntrltwo$] relative to BR[$\ntrlthree \to \hsm \ntrltwo$] is just the opposite to what it is for BR[$\ntrlfour \to Z \ntrltwo$] relative to BR[$\ntrlfour \to \hsm \ntrltwo$]. In other words,
the pairs of dominant branching ratios
(i) BR[$\ntrlthree \to Z \ntrltwo$] and BR[$\ntrlfour \to \hsm \ntrltwo$] and (ii) BR[$\ntrlthree \to \hsm \ntrltwo$] and BR[$\ntrlfour \to Z \ntrltwo$] behave in similar ways. A similar observation is made in ref.~\cite{Calibbi:2014lga} in a scenario with a bino-like LSP in the MSSM.

An altogether different situation may arise if $\ntrlfour$ is singlino-like and is heavier than the three nearly mass-degenerate higgsino-like states, $\ntrltwothree$ and $\charonepm$. This is when the decay $\ntrlfour \to W^\mp \charonepm$ might open up, and even could dominate, as shown in the rightmost plot of the bottom panel of figure~\ref{fig:ternary}. The blue line overlaid on the base axis clearly represents the spectra where $\ntrltwo$ is highly singlino-dominated (see palette) and hence $\ntrlthreefour$ are higgsino-like. An increase in $|\msinglino|$ (with respect to $|\mueff|$) is signalled by a gradual change in color (from blue to red) as one goes up along the right axis indicating a growing singlino-content in $\ntrlfour$ (or, equivalently, an increasing higgsino admixture in the NLSP) and an accompanying increase in BR[$\ntrlfour \to W^\mp \charonepm$]. Phenomenologically, however, such a possibility is not expected to have much impact given that it refers to a relatively heavy singlino-like state whose production at the LHC would be reasonably suppressed in the first place.
%
\subsubsection{Combined canonical branching ratios of $\ntrlthreefour$
and $\charonepm$}
\label{subsubsec:canonical}
%
In view of the above discussion, one way to anticipate the relative extent of relaxation that the existing bounds on ewino masses could undergo is to take note of the extent of depletion in the combined canonical branching ratio of the neutralinos defined~as
\be\label{CBR}
C_{\mathrm{BR}} \equiv \sum_{\substack{i=2,3,4}} C^{\ntrli}_\mathrm{BR}= \sum_{\substack{i=2,3,4}} \mathrm{BR}\!\left[\ntrli \to Z/\hsm + \ntrlone \right].
\ee
In addition, as pointed out in section~\ref{subsec:lhc}, the reduction in the
canonical branching ratio of $\charonepm$ into $W^\pm \ntrlone$, as the new
decay $\charonepm \to W^\pm \ntrltwo$ opens up, would also affect yields in various
important final states. What finally would matter is how small the product of
the canonical branching ratios of the neutralinos and the lighter chargino gets to be.

\begin{figure}[t]
\begin{center}
\hskip -43pt
\includegraphics[width=0.58\linewidth]{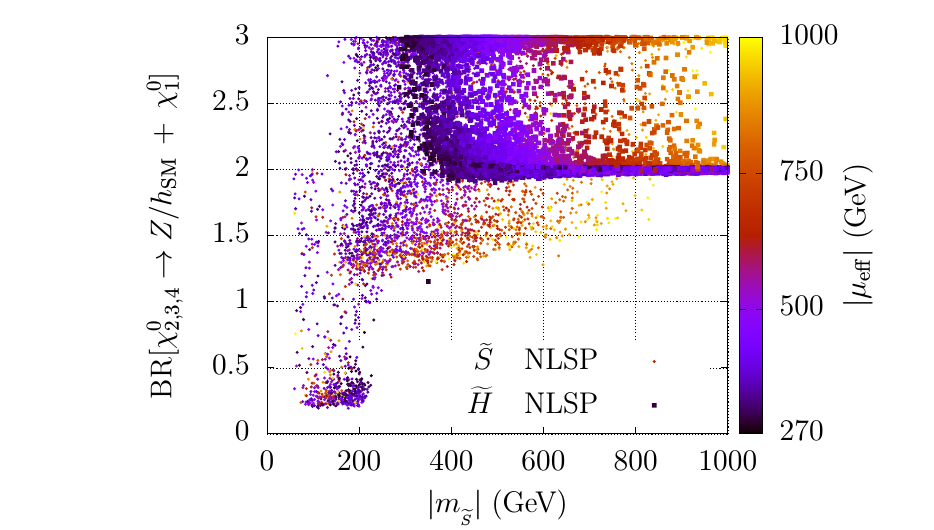}
\hskip -20pt
\includegraphics[width=0.55\linewidth]{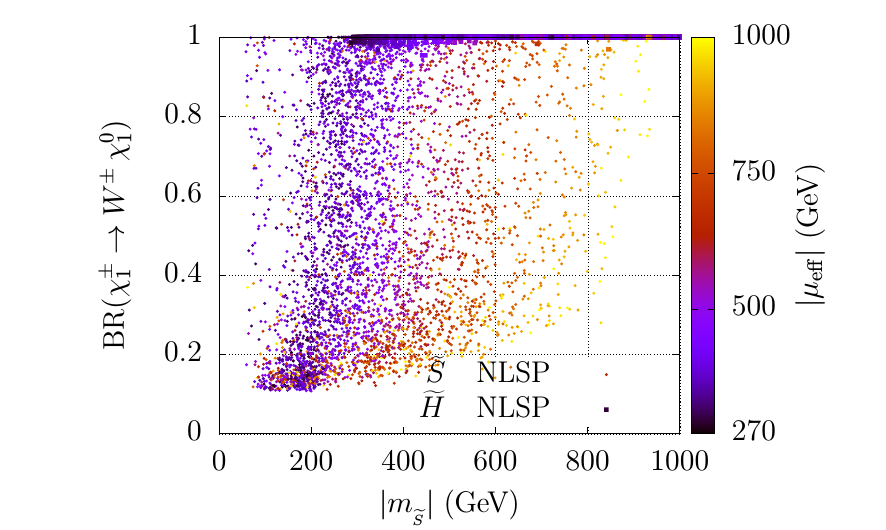}\\[-0.2cm]
\caption{Scatter plots (using the same set of points as in figure~\ref{fig:ternary}) presenting the combined branching ratio for the decays
$\ntrltwothreefour \to Z/\hsm +\ntrlone$ (left) and the branching ratio for the decay $\charonepm \to W^\pm \ntrlone$
(right) as functions of~$|\msinglino|$. Values of $|\mueff|$ are indicated via color palettes. The dominant nature of the NLSP (singlino-like or higgsino-like) is represented by two different point-types (`{\tiny$\bullet$}' or `$\sq$', respectively) as shown by the legends at the bottom of the plots.}
\label{fig:effbr}
\end{center}
\vspace{-0.5cm}
\end{figure}

The lower $C_{\mathrm{BR}}$ is, the higher are the possibility and the extent of relaxation. In the left plot of figure~\ref{fig:effbr} we illustrate how $C_{\mathrm{BR}}$ varies as a function of $|\msinglino|$ with the magnitude of associated $|\mueff|$ being indicated by the palette-color. It reveals that $C_{\mathrm{BR}}$ can range between~3 (i.e., the maximum, when all three of $\ntrltwothreefour$ decay only to $Z/\hsm+ \ntrlone$) and $\sim 0.25$ when other decay modes of these states (into lighter neutralinos and/or singlet-like scalars) become both accessible and favored. Salient features of this plot can be briefly described
as follows in terms of regions with a higgsino-like or a singlino-like NLSP.
%
%
%
\begin{itemize}
\item For points with a higgsino-like NLSP (i.e., with roughly degenerate
$\mntrltwothree \sim |\mueff|$), $C_{\mathrm{BR}}$ retains its defined maximum value of 3 for
$300~\mathrm{GeV} \lesssim |\msinglino| \lesssim 1~\mathrm{TeV}$. This 
happens if for the singlino-like $\ntrlfour$, BR$\left[\ntrlfour \to Z/
\hsm + \ntrlone \right] \approx 1$ which occurs when its decays to $\ntrltwothree
$ are kinematically prohibited as indicated by the varying color implying increasing
$|\mueff|$ (hence increasing $\mntrltwothree$) as $
|\msinglino|$ increases. This is 
since $\sum_{\substack{i=2,3}}\mathrm{BR}\!\left[\ntrli\! \to\! Z/\hsm\! +\! \ntrlone 
\right]\!\!\approx~\!\!2$ is mostly guaranteed in our scenario given the assured
mass-splitting(s) between the higgsino-like states and the LSP (i.e., $\mntrltwothree, \mcharone \, (\sim |\mueff|) \gtrsim 290~\mathrm{GeV}$ and $\mntrlone \leq 200$~GeV). This is broadly vindicated
by the purple horizontal band with $C_{\mathrm{BR}}$ value of $\approx 2$, extending over the 
range $400~\mathrm{GeV} \lesssim |\msinglino| \lesssim 1~\mathrm{TeV}$. 
This also implies that~$|\mueff|$ (and hence $\mntrltwothree$) remains small over this 
range thus opening up the decays $\ntrlfour \to Z/\hsm + \ntrltwothree$ 
(which are coupling-wise favored when compared to the decay to a bino-like LSP) and $\ntrlfour \to W^\mp \charonepm$ to take over completely. These ensure a nearly vanishing 
contribution to $C_{\mathrm{BR}}$. It is thus clear from this figure that in scenarios with a higgsino-like NLSP, $C_{\mathrm{BR}}$ can rarely get below the value of~2 (when there happens to be some appreciable branching of the higgsino-like neutralinos to singlet-like scalars).
\item In scenarios with a singlino-like NLSP, $\ntrltwo$, with $\ntrlthreefour$ being 
higgsino-like, it is seen that the maximum $C_{\mathrm{BR}}$ value of 3 is reached 
over the range $150~\mathrm{GeV} \lesssim |\msinglino| \lesssim 300~\mathrm{GeV}
$.  This happens when the splitting between $\ntrlthreefour$ and $\ntrltwo$ is not 
enough for the decay of the former to the latter to occur and hence each of $
\ntrlthree$ and $\ntrlfour$ may undergo the decays $\ntrlthreefour \to Z/ \hsm+ \ntrlone$ with $\sim 100$\% branching ratio, while the same for the decay $\ntrltwo \to Z/\hsm + \ntrlone$ tends to touch 100\%.
At even smaller $|\msinglino| \, (\lesssim 125)$~GeV, the decay $\ntrltwo \to Z/\hsm + \ntrlone$ gets closed even for the minimum allowed value of $\mntrlone \,(\gtrsim 30)$~GeV in our scenario. Here, a maximum $C_{\mathrm{BR}}$ value of $\approx 2$ is attained when both of $\ntrlthreefour$ decay~100\% of the times to $Z/\hsm+\ntrlone$. Conversely, $C_{\mathrm{BR}}$ touches its lowest value when the sum of these two contributions touches its minimum ($\approx 0.25$) as a result of an enhanced BR$\left[ \ntrltwo \to \as/\hs + \ntrlone \right]$ while
$\ntrlthreefour$ mostly decaying to $\ntrltwo$. It may be noted that an intermediate range of $C_{\mathrm{BR}}$ between $\sim 1.25$ and 2 does also exist
for $|\msinglino| \gtrsim 200$~GeV.
\end{itemize}

In the right plot of figure~\ref{fig:effbr} we present the variation of the canonical branching ratio of the lighter chargino, i.e., BR[$\charonepm \to W^\pm \ntrlone$], as a function of
$|\msinglino|$ with the magnitude of the associated $|\mueff|$ being indicated by the
palette-color. This branching ratio is 100\% when the mass-split between
$\charonepm$ and $\ntrltwo$ is less than $m_W$. Note that this always happens
for a higgsino-like NLSP which has to be nearly degenerate with a chargino of
similar nature. In contrast, a singlino-like NLSP with decreasing mass (increasing mass-split with the chargino) could bring down this canonical branching ratio even below
20\%, as is seen in this plot.

To summarize, we find that in a scenario with higgsino-like NLSP, there may only be very limited scope to evade recent lower bounds from the LHC on the masses of such near-degenerate ewinos. For, these higgsino-like heavier neutralinos ($\ntrltwothree$, with $\ntrltwo$ being the NLSP) would mostly decay to $Z/\hsm + \ntrlone$ assuming which the said bounds are derived in the very first place.
Furthermore, a singlino-like $\ntrlfour$ is not expected to play any role in the present context. For, in the first place, its production rate would be suppressed not only due to its large singlino-content but also since it has to be reasonably massive  to possess appreciable branchings in the non-canonical modes, i.e., $\ntrlfour \to Z/\hsm/\hs/\as + \ntrltwothree$.

In contrast, a scenario with a singlino-like $\ntrltwo$ NLSP with heavier higgsino-like neutralinos ($\ntrlthreefour$)  would have a much healthier prospect of evading such bounds. This is since a singlino-like NLSP can get much lighter when compared to the higgsino-like states without getting into conflict with the experimental findings. This paves the way for relatively light higgsino-like $\ntrlthreefour$ with large enough production cross sections and big enough splits with the singlino-like NLSP to which they can decay to and hence could act to erode the experimental lower bounds on the ewino masses.
%
\begin{figure}[t]
\begin{center}
\includegraphics[width=0.55\linewidth]{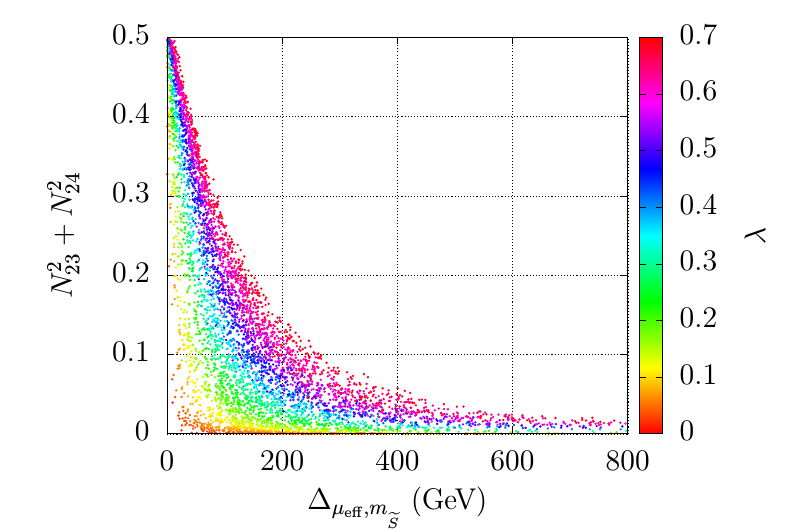}
\vspace{-0.4cm}
\caption{
Variation of the higgsino-content in an otherwise singlino-dominated NLSP
($\ntrltwo$) as a function of $\Delta_{\mueff , \msinglino} = |\mueff| - |\msinglino|$ and `$\lambda$'.
}
\label{fig:higgsino-content-NLSP}
\end{center}
\vspace{-0.8cm}
\end{figure}

The partial decay widths of the heavier higgsino-like neutralinos/lighter
chargino to the singlino-like NLSP also depend on the admixture of higgsinos in the latter state. This mixing depends primarily on two quantities: `$\lambda$' and the
mass-split, $\Delta_{\mueff , \msinglino}=|\mueff|-|\msinglino|$.
In figure~\ref{fig:higgsino-content-NLSP} we present how the 
higgsino admixture ($N^2_{23}+N^2_{24}$) in an otherwise
singlino-dominated NLSP ($\ntrltwo$) varies as a simultaneous function of 
$\Delta_{\mueff , \msinglino}$ (along the horizontal axis) and
`$\lambda$' (in the palette). It is clearly seen that for a given
$\Delta_{\mueff, \msinglino}$, the higgsino admixture in the NLSP increases with increasing `$\lambda$'.  This is also apparent from eqs.~(\ref{eqn:nj3-value}) and~(\ref{eqn:nj4-value}) where `$j$' refers to the singlino-dominated neutralino state.
It can also be gleaned from this figure that the said admixture drops with increasing $\Delta_{\mueff, \msinglino}$ when `$\lambda$' is held fixed.
\subsubsection{Effect of `$\lambda$' on combined non-canonical branching ratios of $\ntrlthreefour$ and $\charonepm$}
\label{subsubsec:lambda-effect}
The weakened canonical branching ratios for $\ntrlthreefour$
and $\charonepm$ find explanations in the magnitudes of various non-canonical branching ratios for these states that include final states containing the singlino-dominated (NLSP) $\ntrltwo$ and the singlet scalars, $\hs$ and $\as$. Thus, the (combined) non-canonical branching ratios for these neutralino states, i.e.,
$\sum_{\substack{i=3,4}}{\rm BR}[\ntrli \to Z/\hsm/\hs/\as + \ntrltwo$] and the non-canonical branching ratio for the lighter chargino, BR[$\charonepm \to W^\pm 
\ntrltwo$] become measures of to what extent the canonical 
decay modes of these heavier ewinos to $Z/\hsm/W^\pm + \ntrlone$ (steadfastly assumed by the LHC experiments to be 100\%) are starved that lead to 
the weakening of the experimental bounds on their masses.

The above-mentioned non-canonical branching ratios crucially depend on the value of~`$\lambda$'. `$\lambda$' not only 
drives the higgsino-singlino-Higgs boson (scalar) interaction involved in 
the decay $\ntrlthreefour \to \hsm \, \ntrltwo$, but also controls the 
mixing in the higgsino-singlino sector that affects the involved 
couplings in a crucial way. On the other hand, the $\lambda$-dependence of
the $Z$-boson coupling to the neutralino states only appears through the latter (mixing). 
Further, the said mixing is a function of $\Delta_{\mueff , \msinglino}$ which
is already discussed in section~\ref{subsubsec:canonical} (see figure~\ref{fig:higgsino-content-NLSP}). The 
actual forms of the dependencies of the squared effective couplings $g^2_{_{Z\ntrltwo \ntrlthreefour}}$ and $g^2_{_{\hsm \ntrltwo \ntrlthreefour}}$ (that enter the calculations of the partial decay widths of $\ntrlthreefour$) on `$\lambda$' and
$\Delta_{\mueff , \msinglino}$ are rather subtle, given that the latter also controls the available phase space for the decays. The expressions for these couplings are presented in eqs.~(\ref{eqn:chij-chi-z-coupling}) and~(\ref{eqn:coupling-hsmnjnk}), respectively, for a $(3\times 3)$ higgsino-singlino system.
\begin{figure}[t]
\begin{center}
\includegraphics[width=0.45\linewidth]{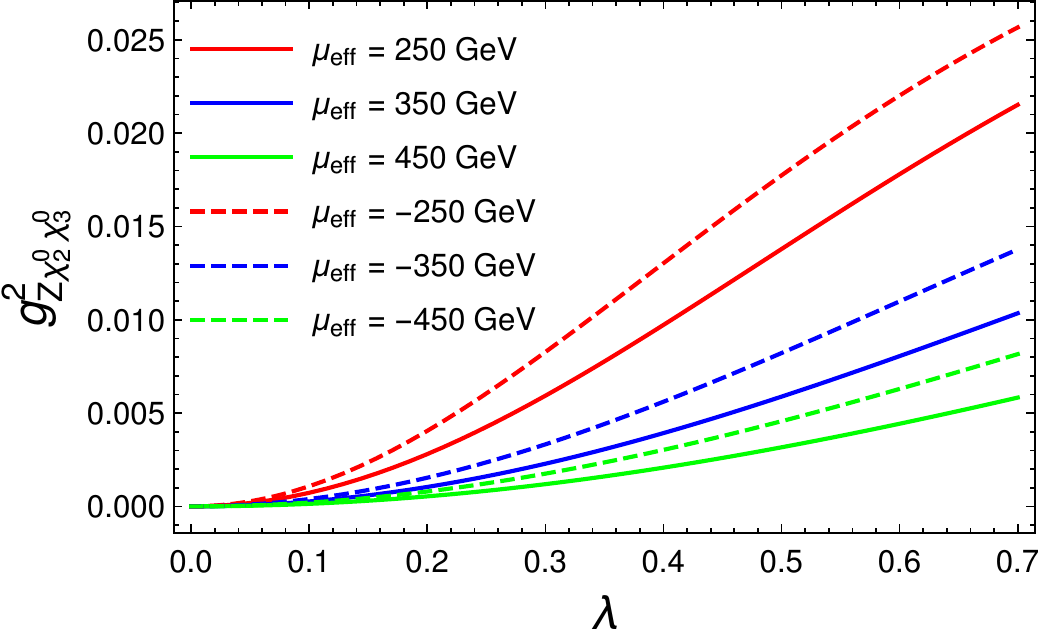}~~
\hskip 10pt
\includegraphics[width=0.45\linewidth]{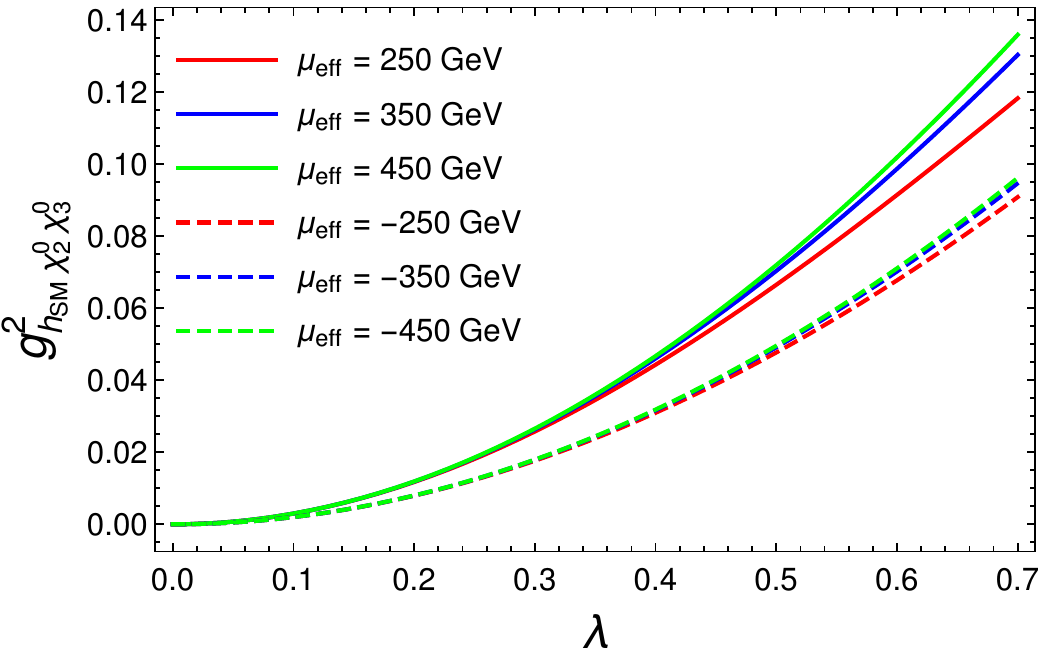}
\vspace{-0.3cm}
\vskip 15pt
\hspace*{-0.15cm}
\includegraphics[width=0.45\linewidth]{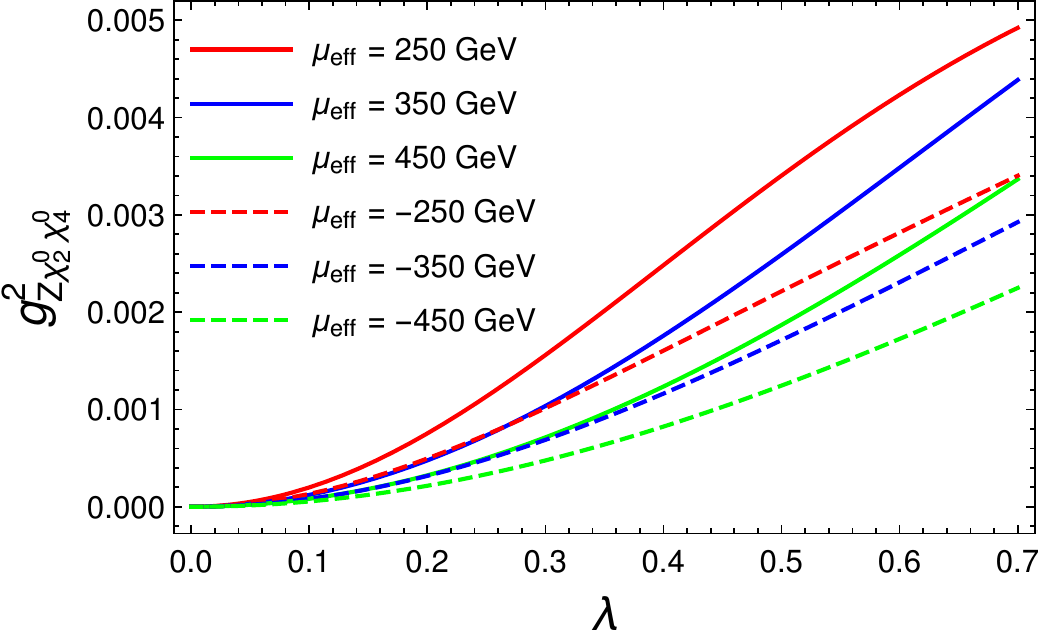}
\hskip 15pt
\includegraphics[width=0.45\linewidth]{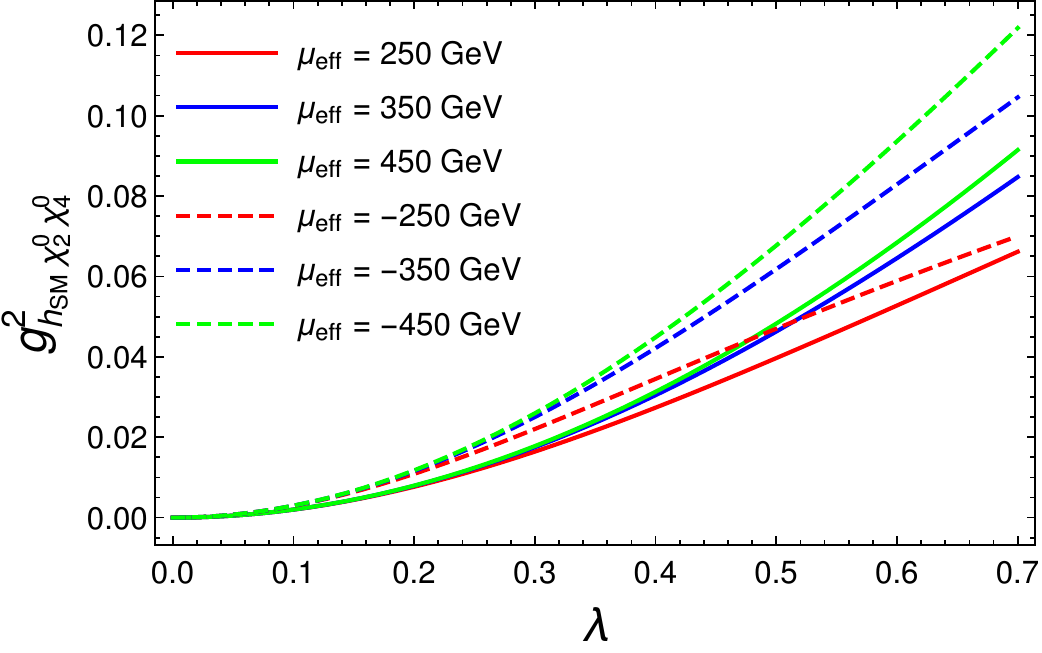}
\vspace{-0.2cm}
\caption{Variations of the couplings $g^2_{_{Z \ntrltwo \ntrlthree}}$ (top, left), $g^2_{_{\hsm \ntrltwo \ntrlthree}}$ (top, right), $g^2_{_{Z \ntrltwo \ntrlfour}}$ (bottom,~left) and $g^2_{_{\hsm \ntrltwo \ntrlfour}}$ (bottom, right) with respect to `$\lambda$' for various values of $\mueff$ using relations~(\ref{eqn:chij-chi-z-coupling}) and~(\ref{eqn:coupling-hsmnjnk}). $\ntrltwo$ is singlino-dominated while $\ntrlthreefour$ are higgsino-dominated. For all the plots $\msinglino = 100$~GeV while $\tanb=10$. See text for details.} 
\label{fig:coupling-lam-mueff}
\end{center}
\vspace{-0.5cm}
\end{figure}

The variations of these couplings, for $\ntrltwo$ ($\ntrlthreefour$) as the singlino-like (higgsino-like) state(s), with `$\lambda$' are presented in figure~\ref{fig:coupling-lam-mueff} for various values of $\mueff$ and for $\tanb=10$ and $\msinglino=100$~GeV. Note that these plots are obtained using relations~(\ref{eqn:chij-chi-z-coupling}) and~(\ref{eqn:coupling-hsmnjnk}) which are derived considering only the $(3 \times 3)$ higgsino-singlino sub-system but still referring to the actual neutralino eigenstates ($\ntrltwothreefour$) they roughly correspond to while $\ntrlone$ is all the way bino-like.
As noted earlier, a varying $\mueff$ alters $\Delta_{\mueff , \msinglino}$ which, in turn, affects the couplings. Note that these couplings also have implicit dependencies on `$\lambda$' via neutralino mass eigenvalues. It happens to be that for the higgsino-singlino system under consideration and for the concerned range of `$\lambda$', for all the cases shown in figure~\ref{fig:coupling-lam-mueff}, the squared effective couplings, for a fixed $\mueff$, i.e., a fixed $\Delta_{\mueff , \msinglino}$, increase with increasing `$\lambda$'. Furthermore, as can be gleaned from these plots, the sign on $\mueff$ has a nontrivial impact.

On the other hand, the interaction among the higgsino-dominated chargino ($\charonepm$), the singlino-dominated neutralino ($\ntrltwo$) and the $W^\pm$-boson is given by eq.~(\ref{eqn:gamma-cha-nj-w}) which tells us that the same grows
with increasing higgsino admixture in the otherwise singlino-dominated state.
As we have already seen (see section~\ref{subsubsec:canonical}; figure~\ref{fig:higgsino-content-NLSP}), this happens when `$\lambda$' increases for a fixed mass-split between the singlino- and the higgsino-like states. As we will see soon, this phenomenon would directly influence the decay $\charonepm \to W^\pm \ntrltwo$.

The partial widths for the decays $\ntrlthreefour \to \hs/\as + \ntrltwo$ receive contributions from both $\lambda$- (for the $\hs/\as$-$\higgsino$-$\higgsino$ case) and $\kappa$-driven (for the $\hs/\as$-$\singlino$-$\singlino$ case) interactions.
However, in the interesting situation where the singlino-like NLSP ($\ntrltwo$)
is not so heavy, `$\kappa$' has to be on the smaller side. On the other hand, the dominant decay modes of $\ntrlthreefour$ to $\ntrltwo$ turn out to be the ones involving the $Z$-boson and $\hsm$ while those with $\hs/\as$ remaining
subdominant.
\begin{figure}[t]
\begin{center}
\hspace*{-1.1cm}
\includegraphics[height=6cm, width=0.56\linewidth]{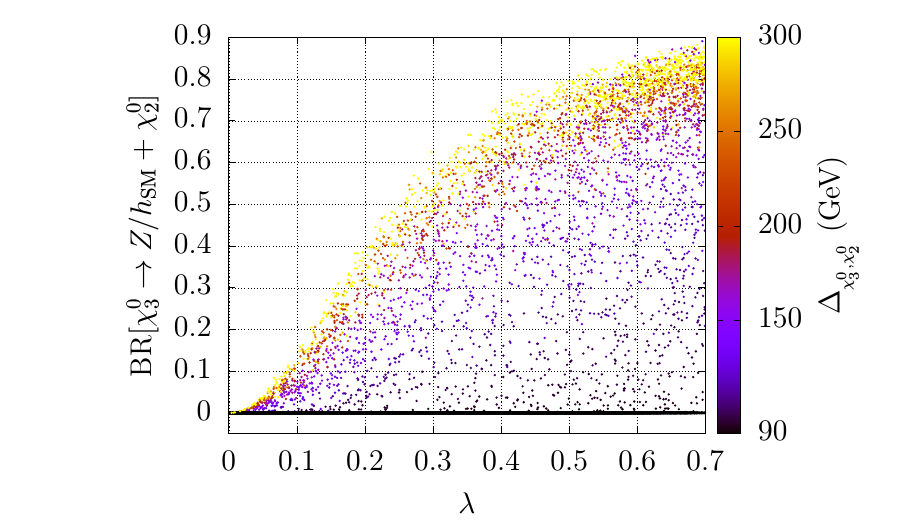}~~
\hskip -15pt
\includegraphics[height=6cm, width=0.56\linewidth]{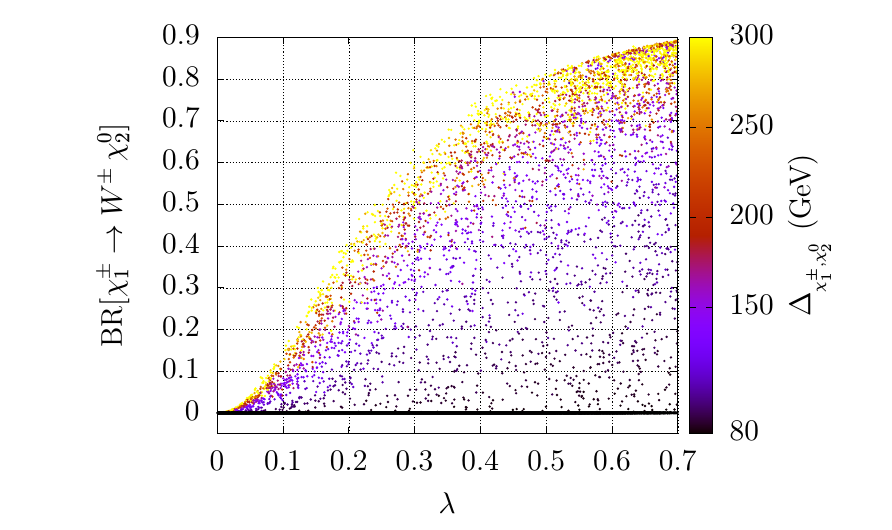}
\vspace{-0.8cm}
\caption{Variation of the branching ratios BR[$\ntrlthree \to Z/\hsm + \ntrltwo$] (left) and BR[$\charonepm \to W^\pm \ntrltwo$] (right)
as functions of `$\lambda$', $\Delta_{\ntrlthree, \ntrltwo}= \mntrlthree - \mntrltwo$ (left) and $\Delta_{\charonepm, \ntrltwo}= \mcharone - \mntrltwo$ (right). Palettes for the latter two plots are truncated at the top at a mass-split of 300~GeV (for which the available phase space for the decays under consideration already saturates) for added clarity in colors
albeit the plots themselves contain points with values larger than that.
}
\label{fig:lambda-var}
\end{center}
\vspace{-0.5cm}
\end{figure}

The left and the right plots of figure~\ref{fig:lambda-var} illustrate the 
variations of the key (combined) branching ratios BR[$\ntrlthree \to 
Z/\hsm + \ntrltwo$] and that of BR[$\charonepm \to W^\pm \ntrltwo$], as 
functions of `$\lambda$' (now, along the horizontal axis) and the mass-splits $
\Delta_{\ntrlthree, \ntrltwo}= \mntrlthree - \mntrltwo$ and $\Delta_{\charonepm, \ntrltwo}= \mcharone - \mntrltwo$, respectively, which are the actual measures 
of the available phase space for the decays in context and are roughly 
synonymous to $\Delta_{\mueff, \msinglino}$ when the higgsino-singlino mixing 
is not large. The (combined) branching ratio BR[$\ntrlfour \to 
Z/\hsm + \ntrltwo$] behaves in a way similar to BR[$\ntrlthree \to 
Z/\hsm + \ntrltwo$] thanks to the antipodal behavior of $\ntrlthree$ and $\ntrlfour$, as pointed out earlier.

The contours of uniform colors roughly illustrate the variations of the branching ratios with `$\lambda$' for various fixed ranges of the involved mass-splits.
On the other hand, variations of colors for fixed values of `$\lambda$',
demonstrate the effect of varying mass-split, i.e., of changing phase space
volume for the concerned decay process(es). The overall profiles are the results of
the combined effects of these two on the respective branching ratios. These can be traced back to the profiles of the
corresponding partial widths in each case which, in turn, are found to be the
outcomes of nontrivial dependence of the involved interaction strengths on `$
\lambda$', as have been pointed out in the contexts of figures~\ref{fig:higgsino-content-NLSP} and~\ref{fig:coupling-lam-mueff}. To  summarize, larger values of~`$\lambda$' lead to large non-canonical branching ratios, albeit in a subtle way.

In the next subsection, we will present a few benchmark points with relatively
light higgsino-dominated neutralinos and a chargino in such a pNMSSM scenario
thanks to the enhanced non-canonical decay branching ratios of these states that help evade the recent stringent lower bounds on their masses from the LHC experiments.
%
\subsection{Benchmark scenarios}
\label{subsec:benchmark}
In this section, we discuss a few benchmark scenarios that satisfy all 
pertinent constraints from various low energy experiments, the bounds 
obtained from the LHC experiments on the observed Higgs sector and the 
experimental constraints on various DM observables like the relic abundance 
and the SI and the SD rates. On top of these, 
though generically found to be less constraining for the scenario under 
study, we now also take into account the constraints obtained from 
experiments dedicated to indirect detection of the DM, viz., the Fermi-LAT~\cite{Fermi-LAT:2016uux} and the AMS-02~\cite{Aguilar:2013qda,
Accardo:2014lma} and subsequently discussed in refs.~\cite{Ibarra:2013zia, Giesen:2015ufa} thanks to late time
small values of thermally averaged $\langle \sigma v \rangle$ for the DM  particles $\left( \langle \sigma v \rangle \lesssim {\cal O} (10^{-29}) \, \mathrm{cm}^3/{\rm s} \right)$.
Furthermore, we take advantage of the relaxed lower bounds on the masses 
of higgsino-like ewinos resulting from a preliminary recast of a relevant LHC 
analysis~\cite{Sirunyan:2018ubx} as incorporated in \nmssmtools.
The benchmark scenarios presented in table~\ref{tab:BPs} broadly belong to the 
allowed set from our random scan  of the parameter space. However, 
these scenarios possess optimized values of branching ratios for various 
decay modes of the ewinos. These not only help evade the experimental bounds 
presented in ref.~\cite{Sirunyan:2018ubx} under a recast (as mentioned 
above) for our scenario but are also expected to remain instrumental against recent diversified analyses~\cite{Aaboud:2018jiw, Aaboud:2018sua, Aaboud:2018ngk, Aad:2019vnb, Sirunyan:2019iwo, Aad:2019vvf, Aad:2019vvi, Aad:2020qnn, ATLAS:2020ckz} (and improved bounds thereof)
when those are eventually subjected to recasts.

The benchmark scenarios presented in table~\ref{tab:BPs} broadly belong to the 
allowed set from our random scan  of the parameter space. However, 
these scenarios possess optimized values of branching ratios for various 
decay modes of the ewinos. These not only help evade the experimental bounds 
presented in ref.~\cite{Sirunyan:2018ubx} under a recast (as mentioned 
above) for our scenario but are also expected to remain instrumental against recent diversified analyses~\cite{Aaboud:2018jiw, Aaboud:2018sua, Aaboud:2018ngk, Aad:2019vnb, Sirunyan:2019iwo, Aad:2019vvf, Aad:2019vvi, Aad:2020qnn, ATLAS:2020ckz} (and improved bounds thereof)
when those are eventually subjected to recasts.
%
%
\begin{table}[H]
\begin{center}
{\small\fontsize{7.5}{7.5}\selectfont{
\begin{tabular}{|@{\hspace{0.06cm}}c@{\hspace{0.06cm}}|@{\hspace{0.0cm}}c@{\hspace{0.06cm}}|@{\hspace{0.06cm}}c@{\hspace{0.06cm}}|@{\hspace{0.06cm}}c@{\hspace{0.06cm}}|@{\hspace{0.06cm}}c@{\hspace{0.06cm}}|@{\hspace{0.06cm}}c@{\hspace{0.06cm}}|@{\hspace{0.06cm}}c@{\hspace{0.06cm}}|@{\hspace{0.02cm}}c@{\hspace{0.02cm}}|}
\hline
\Tstrut
\multirow{2}{*}{\makecell{Input \\ parameters}} & \multicolumn{2}{c|}{\makecell{Singlet (pseudo)scalar \\ funnel}} {\hspace{0.06cm}}& \multicolumn{2}{c|}{\makecell{ $Z$-boson \\ funnel}}{\hspace{0.06cm}}& \multicolumn{2}{c|}{\makecell{ SM-like Higgs \\ funnel}}{\hspace{0.02cm}}& \makecell{Coannihilation \\ regime}\\
\cline{2-8}
\Tstrut
        & BP1    & BP2        & BP3    & BP4    &  BP5   &  BP6    &BP7     \\  
\hline
\Tstrut
$\lambda$        & 0.608    & 0.265        & 0.563    & 0.267    &  0.644   &  0.230    &  0.641     \\
$\kappa$         &$-0.110~~$& $-0.042~~$   & 0.093    & 0.030    & $0.137$  & $-0.031~~$&  0.142     \\
$\tan\beta$      & 19.72    & 15.34        & 26.23    & 17.64    &   28.43  &  23.76    &  9.160     \\
$A_t$~(TeV)      & 2.739    & 4.731        & 9.476    & 3.916    &   4.703  &  8.926    &  6.407     \\
$A_\lambda$~(TeV)& 8.219    & 5.653        &$-9.666~~$& 7.083    &   9.961  &  9.705    & $-3.472~~$     \\
$A_\kappa$~(GeV) & 46.83    & 38.42        & 42.16    & 0.423    &$-64.97~~$& $-1.033~~$&  2.647  \\
$\mu$~~(GeV)     & 381.9    & 350.8        &$-374.3~~$& 381.2    &   352.3  &  396.2    & $-386.9~~$     \\
$M_1$~(GeV)      & 37.21    & $-31.26~~$   & 43.14    &$-43.29~~$&$-58.04~~$&$-61.86~~$ &  169.4  \\
\hline
 & & & & & & &  \\[-0.1cm]
Observables & & & & & & &  \\
 & & & & & & &  \\[-0.1cm]
\hline

\Tstrut
$m_{\chi_1^0}$~(GeV)    & 37.023&  30.756  &  43.268  & 43.063  & 57.953 & 60.688  &  167.46   \\
$m_{\chi_2^0}$~(GeV)    & 126.88&  112.84  &  122.97  & 87.193  & 143.82 & 108.81  &  170.74   \\
$m_{\chi_3^0}$~(GeV)    & 406.68&  366.96  &  399.43  & 399.21  & 382.17 & 412.95  &  415.43   \\
$m_{\chi_4^0}$~(GeV)    & 421.21&  372.04  &  408.03  & 401.15  & 392.24 & 417.23  &  424.45   \\
$m_{\chi_1^\pm}$~(GeV)  & 395.58&  363.37  &  388.34  & 394.76  & 365.39 & 410.43  &  400.98   \\
$m_{h_1}$~(GeV)         & 123.92&  117.14  &  125.96  & 101.11  & 123.18 & 118.38  &  126.32   \\
$m_{h_2}$~(GeV)         & 185.97&  123.40  &  179.36  & 123.74  & 204.37 & 127.85  &  192.30   \\
$m_{a_1}$~(GeV)         & 77.641&  64.429  &  31.354  & 30.024  & 36.575 & 25.825  &  160.72   \\[0.05cm]
\hline

\Tstrut
$N_{11}$,$N_{21}$&$-0.99,~~0.07$&$~~0.99,~~0.07$&$-0.99,-0.07$&$0.99,-0.04$&$0.99,-0.07 $&$~~0.99,~~0.09$&$~~0.99,-0.04$\\
$N_{12}$,$N_{22}$&$~~0.00,-0.01$&$~~0.00,~~0.00$&$~~0.00,~~0.01$&$0.00,~~0.00$&$0.00,-0.01$ &$~~0.00,~~0.00$&$~~0.00,~~0.01$\\
$N_{13}$,$N_{23}$&$-0.11,-0.09$&$~~0.12,-0.04$&$-0.11,-0.07$&$0.11,~~0.02$&$0.12,~~0.11 $&$~~0.11,-0.01$&$-0.12,-0.09$\\
$N_{14}$,$N_{24}$&$~~0.00,-0.29$&$~~0.01,-0.14$&$-0.01,~~0.26$&$0.00,-0.12$&$0.00,-0.33$ &$~~0.02,-0.10$&$-0.03,~~0.30$\\
$N_{15}$,$N_{25}$&$~~0.07,~~0.95$&$-0.07,~~0.99$&$-0.06,~~0.96$&$0.04,~~0.99$&$0.06,~~0.93 $&$-0.09,~~0.99$&$~~0.04,~~0.95$\\[0.05cm]
\hline

\Tstrut
$\Omega h^2$  &  0.115  &  0.117  & 0.113 &  0.126 & 0.116& 0.115 &  0.124\\[0.10cm]

$\sigma^{\rm SI}_{\chi^0_1-p(n)}\times 10^{47}$~(cm$^2$)&$1.2 (1.3)$&$0.8 (0.8)$&$4.7 (4.8)$&$0.8 (0.8)$&$5.2 (5.3)$&$3.6 (3.6)$&$0.01 (0.01)$ \\[0.30cm]

$\sigma^{\rm SD}_{\chi^0_1-p(n)}\times 10^{42}$~(cm$^2$)&$5.5 (4.3)$&$7.7 (5.9)$&$5.5 (4.2)$&$5.5 (4.2)$&$7.2 (5.6)$&$5.0 (3.8)$ &$6.8 (5.2)$ \\[0.2cm]
\hline
\Tstrut
BR[$\chi^\pm_1 \to W^\pm \chi_1^0$]   
&0.16&  0.49 &  0.18  & 0.47 &  0.16& 0.56 &0.13\\[0.05cm]
BR[$\chi^\pm_1 \to W^\pm \chi_2^0 $]  &0.84&  0.51 &  0.82  & 0.53 &  0.84& 0.44 &0.87\\[0.05cm]
\hline
\Tstrut
 BR[$\chi^0_2 \to a_1 \chi_1^0$]      & 1.00 &  1.00 &  1.00  & 1.00 &  1.00 & 1.00 &  0.00    \\[0.05cm]
 BR[$\chi^0_2 \to \gamma \chi_1^0$]   & 0.00 &  0.00 &  0.00  & 0.00 &  0.00 & 0.00 &  0.91    \\[0.05cm]
\hline
\Tstrut
BR[$\chi^0_3 \to Z \chi_1^0$]    & \underline{0.09} & \underline{0.31} &  \underline{0.11} & \underline{0.25} & \underline{0.09} & \underline{0.40}   & \underline{0.05} \\[0.05cm]
BR[$\chi^0_3 \to Z \chi_2^0$]    & 0.66 & 0.41 &  0.58 & 0.22 & 0.64 & 0.31   &0.65 \\[0.05cm]
BR[$\chi^0_3 \to h_1 \chi_1^0$]  & \underline{0.06} & 0.01 &  \underline{0.07} & 0.00 & \underline{0.06} & 0.01   &\underline{0.08} \\[0.05cm]
BR[$\chi^0_3 \to h_1 \chi_2^0$]  & 0.14 & 0.00 &  0.18 & 0.00 & 0.12 & 0.00   &0.14 \\[0.05cm]
BR[$\chi^0_3 \to h_2 \chi_1^0$]  & 0.00 & \underline{0.18} &  0.00 & \underline{0.21} & 0.00 & \underline{0.17}   &0.01 \\[0.05cm]
BR[$\chi^0_3 \to h_2 \chi_2^0$]  & 0.01 & 0.09 &  0.00 & 0.31 & 0.00 & 0.11   &0.00 \\[0.05cm]
BR[$\chi^0_3 \to a_1 \chi_1^0$]  & 0.00 & 0.00 &  0.00 & 0.01 & 0.00 & 0.00   &0.00 \\[0.05cm]
BR[$\chi^0_3 \to a_1 \chi_2^0$]  & 0.04 & 0.01 &  0.06 & 0.00 & 0.09 & 0.00   &0.08 \\[0.05cm]
\hline
\Tstrut
BR[$\chi^0_4 \to Z \chi_1^0$]    & \underline{0.09} &  \underline{0.23} &  \underline{0.09}  & \underline{0.26} &  \underline{0.08} & \underline{0.20}  &\underline{0.10}    \\[0.05cm]
BR[$\chi^0_4 \to Z \chi_2^0$]    & 0.22 &  0.15 &  0.27  & 0.34 &  0.21 & 0.16  &0.24    \\[0.05cm]
BR[$\chi^0_4 \to h_1 \chi_1^0$]  & \underline{0.06} &  0.01 &  \underline{0.08}  & 0.00 &  \underline{0.06} & 0.01  &\underline{0.02}    \\[0.05cm]
BR[$\chi^0_4 \to h_1 \chi_2^0$]  & 0.57 &  0.01 &  0.54  & 0.00 &  0.62 & 0.01  &0.61    \\[0.05cm]
BR[$\chi^0_4 \to h_2 \chi_1^0$]  & 0.00 &  \underline{0.25} &  0.00  & \underline{0.24} &  0.00 & \underline{0.35}  &0.00    \\[0.05cm]
BR[$\chi^0_4 \to h_2 \chi_2^0$]  & 0.04 &  0.33 &  0.00  & 0.16 &  0.01 & 0.26  &0.02   \\[0.05cm]
BR[$\chi^0_4 \to a_1 \chi_1^0$]  & 0.01 &  0.00 &  0.01  & 0.00 &  0.01 & 0.00  &0.00   \\[0.05cm]
BR[$\chi^0_4 \to a_1 \chi_2^0$]  & 0.01 &  0.00 &  0.01  & 0.01 &  0.01 & 0.00  &0.00   \\[0.05cm]

\hline
\rule{0pt}{4.2ex}
$C_{\rm BR}^{\chi^0_{3 (4)}}$ (see eq.~(\ref{CBR}))               & 0.15 (0.15) &  0.49 (0.48) &  0.18 (0.17)  & 0.46 (0.50) &  0.15 (0.14) & 0.57 (0.55) & 0.13 (0.12) \\[0.05cm]
$C_{\rm BR}^{\ntrlthreefour} \times {\rm BR}[\charonepm \to W^\pm \ntrlone]$
 & 0.048 &  0.475 & 0.063 & 0.451 & 0.046 & 0.627 & 0.033  \\[0.1cm]
\hline
\Tstrut
$\sigma \times {\rm BR}[\to 3\ell]$~(fb)~\cite{Ellwanger:2018zxt} & 0.88&12.44&1.27&8.40&1.10&10.19&0.55\\[0.1cm]
\hline
\Tstrut
$\sigma^{\rm CMS~upper~limit}$~(fb)~\cite{Sirunyan:2018ubx,Ellwanger:2018zxt} & 33.02&39.81&32.02&32.15&43.40&27.93&46.70\\[0.1cm]
\hline
\end{tabular}
}}
\vskip 5pt
\captionof{table}{\footnotesize{
Benchmark points (BPs) allowed by relevant theoretical and experimental constraints including those from the LEP, the low-energy sector and the studies on the SM-like Higgs boson at the LHC. Shown are the masses of the various relevant states, the compositions of the LSP and the NLSP ($N_{ij}$'s). Relevant individual and combined branching ratios involving the ewinos are also presented. The last two rows show the theoretical expectation of the yields in the trilepton final state and the LHC (CMS) upper limit on the same (figures~7 and 8a of \cite{Sirunyan:2018ubx}), respectively. Fixed parameters are as presented in table~\ref{tab:ranges}. All BPs assume a decoupled wino ($M_2=2.5$~TeV) while $m_A$ and $m_H$ range over a few~TeVs.
}}
\label{tab:BPs}
\end{center}
\end{table}
Note that even in a favorable situation with such a tailored set of branching 
ratios of the lighter chargino and the relevant neutralinos, it is not that 
straightforward to derive the relaxed lower bounds on their masses. The best
one can do in the present scope is to take a conservative approach, i.e., to 
look out for degraded branching ratios of the neutralinos to the $Z$-boson 
and $\hsm$ (in favor of decays to lighter singlet scalar(s) and/or to a 
lighter neutralino other than the LSP thus yielding a smaller 
$C_{\mathrm{BR}}$) and the same for the lighter chargino decaying to $W^\pm 
\ntrlone$ (in favor of $W^\pm \ntrltwo$).

As far as the LHC analyses that lead to the reported mass-bounds on the 
ewinos are concerned, what essentially matters are the magnitudes of the 
production cross sections of the pertinent ewino pairs times their effective 
branching ratios that lead to final states of interest.
Fortunately, from 
a previous study~\cite{Abdallah:2019znp} based on a rigorous {\tt CheckMATE} 
\cite{Dercks:2016npn} analysis, we are aware of their ballpark critical 
values in reference to an LHC analysis on $3\ell + \etmiss$ final state 
(arising from the $WZ$-mode) carried out in ref.~\cite{Sirunyan:2017lae}. Given that there have been steady improvements in 
the lower bounds on the masses of these ewinos since then (not only via the $WZ$-mode 
but also through the $WH$-mode leading to final states with $b$-jets, as 
mentioned earlier) while the concerned 
analyses are still to be implemented in a recast package like
{\tt CheckMATE}, we exercise an extra bit of caution by choosing these key 
branching ratios rather conservatively. It is worthwhile to note that, towards this, 
achieving a rather suppressed BR[$\charonepm \to W^\pm \ntrlone$] is likely 
to play a crucial role, besides having smaller $C_{\rm BR}$, in relaxing the reported lower bounds on the masses of the 
ewinos. This is since such an arrangement renders final states  with larger jet-multiplicity more frequent 
by depleting the ones (with multi-leptons) to which the current bounds owe the most. From 
the discussions in section~\ref{subsec:lhc} it has already become clear that 
suppression of the above key branching ratios is characteristically 
achieved only in scenarios with a singlino-like NLSP. Hence all the benchmark 
scenarios we present in table~\ref{tab:BPs} have the singlino-like state as 
the NLSP.

Four broad classes of benchmark scenarios are presented based on the mode of
annihilation of the highly bino-like DM. These include funnels involving (i) singlet
(pseudo)scalar, (ii) $Z$-boson, (iii) $\hsm$ and (iv) coannihilation with
a singlino-like NLSP neutralino. Scenarios presented in each of the first three classes are further 
subdivided into two categories, one having $\hone$ and the other having $\htwo$ 
as $\hsm$. For the coannihilation class, we present only one scenario in~which~$\hone \equiv \hsm$.

Various relevant branching ratios of $\charonepm, \ntrltwothreefour$ are 
listed for each of these benchmark scenarios. Note that
BR$\left[\chi^0_{2(\widetilde{S})} \to a_1 \ntrlone\right]=1$ for all the benchmark 
scenarios except for the last one in which DM coannihilation is in action and
BR$\left[\chi^0_{2(\widetilde{S})} \to \gamma \ntrlone \right]$ is dominant. 
On the other hand, the magnitudes of the individual branching ratios for the states $\ntrlthreefour$, as shown in table~\ref{tab:BPs}, are checked to be pretty representative of our full set of allowed scanned data.
As $\ntrltwo$ always has non-canonical decays for our BPs, to aid  tractability at the numerical level, we downgrade the general definition
of the combined (canonical) branching ratio presented earlier in section
\ref{subsubsec:canonical} (which included $\ntrltwo$ as well) to contain the canonical branching ratios of $\ntrlthreefour$ only, i.e., $C_{\mathrm{BR}} \rightarrow C_{\mathrm{BR}}^{\ntrlthreefour} =\sum_{\substack{i=3,4}}\mathrm{BR}\!\left[\ntrli \to Z / \hsm +  \ntrlone \right]$. In table~\ref{tab:BPs}, the branching ratios that contribute to $C_{\mathrm{BR}}
^{\ntrlthreefour}$ are  underlined. One can find that $C_{\mathrm{BR}}
^{\chi^0_{3(4)}}$ is on the smaller side being between $\sim 10\%-20\%$ for scenarios 
with larger $\lambda \sim 0.6$ (i.e., for BP1, BP3, BP5 and~BP7) 
and are between $\sim 45\%-60\%$ for scenarios in which `$\lambda$'  ($\sim 0.25$) is somewhat smaller  (i.e., for BP2, BP4 and BP6). This is since the interactions $Z/\hsm$-$\chi^0_{2(\widetilde{S})}$-$\chi^0_{3,4(\widetilde{H})}$ are
found to grow with `$\lambda$'.\footnote{The forms of these interactions are derived in appendix~\ref{appsec:ew-widths}. Expressions of the respective decay widths are presented in ref.~\cite{Choi:2004zx}.}
This is already discussed in section \ref{subsubsec:lambda-effect} in reference to the
left plot of figure~\ref{fig:lambda-var}.

Note that the phenomena described above already indicate a significant weakening of the canonical branching ratios of the neutralinos. On top of these, a reduced decay branching ratio BR[$\charonepm \to W^\pm \ntrlone$] (which again is seen to vary over a range similar to what mentioned above for $\ntrlthreefour$, the reasons being very similar as well; see the right plot of figure~\ref{fig:lambda-var}) ends up in suppressing the yields in the canonical final states down to $\sim 2\%-30\%$ of their values assumed by the LHC experiments. Add to this the smaller production cross section (smaller by 50\%--60\%)  of higgsino-like $\ntrlthreefour \charonepm$ when compared to that of wino-like $\ntrltwo \charonepm$
assumed by the LHC experiments. Thus, such a situation, a priori, has the potential to cause significant relaxation of the lower bounds on the masses of the ewinos as reported by the LHC analyses. Hence dedicated recast analyses are called for~\cite{waadsr}.

Note that BP7 representing the coannihilation regime has a (bino-like) LSP 
mass on a somewhat heavier ($\mntrlone \approx 167$~GeV) than what is considered for  all the other scenarios. The reasons are the following. Given that $\mntrltwo \gtrapprox 
\mntrlone$ is a necessary requirement for finding an efficient coannihilation, all of $\ntrlthreefour$ and
$\charonepm$ have competing decays to both $\ntrltwo$ and $\ntrlone$, 
accompanied by $Z$, $\hsm$, singlet scalars and $\wpm$, respectively. In addition, BR[$\ntrltwo \to \gamma \ntrlone$] could now get to be 
significant where the photon is expected to be too soft to be detected~\cite{Domingo:2018ykx}. Hence the cascades of $\ntrlthreefour$ and $\charonepm$ 
via $\ntrltwo$ might lead to similar final states as when these states decay 
directly to the LSP. Thus, virtually, a large effective branching ratio to the LSP for 
these states could be envisaged. Hence stringent experimental lower bounds on 
their masses better not be ignored in such a situation. Note that the chosen value of $\mntrlone$ ($\approx 
167$~GeV) still appears to be allowed (i.e., $\gtrsim 150$~GeV) by the latest experimental 
analyses~\cite{Sirunyan:2018ubx,Aad:2019vvf,Aad:2019vvi} for the accompanying values of $\mntrlthreefour \sim \mcharone$
($\gtrsim 400$~GeV).\footnote{Note, however, that this value of $\mntrlone$ is smaller than our targeted maximum of 200~GeV (in particular, smaller than the top quark mass, though  not far from it) and the restriction is predominantly of LHC origin. It may also be noted that  $\mntrlone \lessapprox \mntrltwo \sim 60$~GeV still passes some recent constraints incorporated in the package NMSSMTools. We choose to adopt a conservative line in this matter.}

In the last but one row of table~\ref{tab:BPs} we present, for all the benchmark points, the resultant values of cross section times the effective branching ratio, $\sigma \times {\rm BR}[\to 3\ell]$, from the processes
$pp \to \ntrlthreefour \charonepm \to (Z \ntrlone) (W^\pm \ntrlone)$ which eventually lead to the trilepton final state, as obtained from 
{\tt NMSSMTools}. Note that the production process $pp \to \ntrltwo \charonepm$ would not
be contributing to this trilepton final state and this can be understood from the possible decay
modes of $\ntrltwo$ shown in table~\ref{tab:BPs}. These numbers are compared with the
corresponding maximum values for this quantity that are allowed by the experiment~\cite{Sirunyan:2018ubx} and as extracted with the help of {\tt NMSSMTools} following~ref.~\cite{Ellwanger:2018zxt}. Largely, it is clear that the smaller the effective branching ratio $C_{\rm BR}^{\ntrlthreefour} \times {\rm BR}[\charonepm \to W^\pm \ntrlone ]$~is, the smaller is $\sigma \times {\rm BR}[\to 3\ell]$ when compared to its corresponding experimental upper limit. This, in turn, demonstrates how easily such higgsino-like $\ntrlthreefour$ and $\charonepm$ with masses in the range $\sim$ 350~GeV-- 400~GeV could have been missed by the LHC experiments when searched for in one of the most sensitive of the final states like the one with three leptons while the reported experimental lower bound
on these masses from the analysis of such a final state is in the ballpark of 500~GeV~\cite{Sirunyan:2018ubx}. Also, the differences between the corresponding numbers in the last two rows of table~\ref{tab:BPs} are indicative of a significant room for even lighter such states surviving the experimental constraints, under the circumstance.
\subsection{The issue of naturalness}
\label{subsec:naturalness}
In this section, we briefly comment on how `natural' the scenario under consideration could get to be. The well-known finetuning measure following Ellis~\cite{Ellis:1986yg} and Barbieri and Giudice~(BG)~\cite{Barbieri:1987fn} is adopted which exploits
the relationship between $m_Z$ (or the weak scale) and the fundamental SUSY
parameters given by
\beq
{1 \over 2} m_Z^2 = {{m_{H_d}^2 + \sum_d -(m_{H_u}^2 + \sum_u)} \over {\tan^2\beta -1}} -\mu^2 \,,
\label{eqn:mzmu}
\eeq
where $m_H{_{d(u)}}$ are the soft masses at the weak scale for the
two doublet Higgs fields, $H_{d(u)}$ and $\sum_{d(u)}$ stand for the radiative
corrections to the respective masses while `$\mu$' is the SUSY conserving higgsino
mass parameter of the MSSM. The equality tells one that unless for
large cancellations (i.e., large finetuning and hence less `natural'), the measured value of $m_Z$ can be obtained only if all the terms are not too different in magnitude from $m_Z$ itself. Note that, in the first place, it
has been a puzzle (known as the `$\mu$-problem') within the framework of the MSSM as to why a SUSY
conserving parameter would at all acquire such a weak-scale value, to which the NMSSM provides a dynamical solution where $\mu \to \mueff= \lambda \vs$. In any case, the finetuning measure itself is given by
\beq
\Delta_{\mathrm{BG}(Z)} \equiv \mathrm{max}_i \,  \left | \frac{\partial \log m_Z^2}{\partial \log p_i}  \right |  \,,
\eeq
where $p_i$ denotes the set of Lagrangian parameters that enter eq.~(\ref{eqn:mzmu}).
%
%
\begin{figure}[t]
\begin{center}
\includegraphics[width=0.57\linewidth]{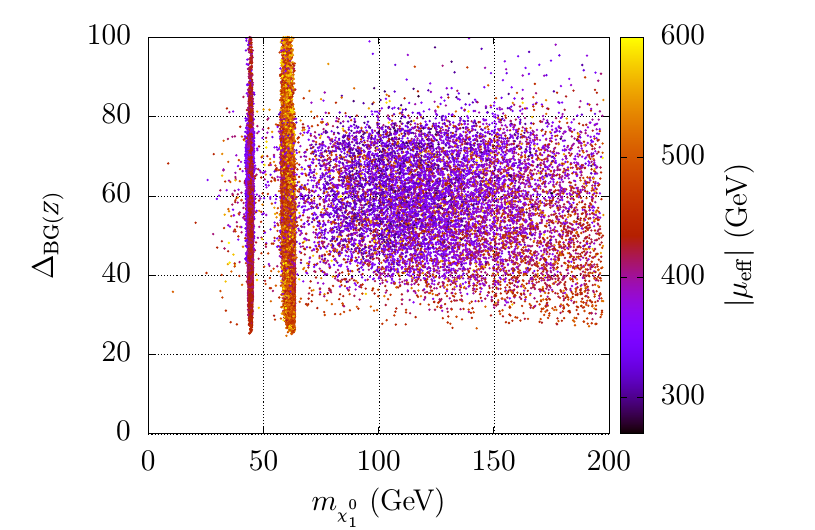}
\caption{Variation of the finetuning parameter $\Delta_{\mathrm{BG}(Z)}$ as
a function of the LSP mass $\mntrlone$. On the vertical axis
$\Delta_{\mathrm{BG}(Z)}$ is truncated at the value of $100$ and the palette shows $|\mueff|$ up to $600$~GeV.}
\label{fig:naturalness}
\end{center}
\end{figure}
%

In our analysis, the value of $\Delta_{\mathrm{BG}(Z)}$ is obtained from
\nmssmtools ~for each of the scanned points in the NMSSM parameter space.
In figure~\ref{fig:naturalness} we illustrate how $\Delta_{\mathrm{BG}(Z)}$
is distributed for the points allowed by all experimental data (as described in earlier sections) as a function of the LSP DM mass (along the horizontal axis) and the magnitude of $|\mueff|$ (indicated in the palette).
Note that $\Delta_{\mathrm{BG}(Z)}$ could reach a value as low as 30 (the
conventional benchmark maximum for a `natural' SUSY scenario) or even
lower. This is achieved notwithstanding  our consideration of rather heavy top squarks and the gluino ($\gtrsim 5$~TeV, which are the primary spoilers of `naturalness' through the terms $\sum_{d(u)}$ in eq.~(\ref{eqn:mzmu}) thanks to rather low values of $|\mueff|$ that are found to be still viable.  Nonetheless, we have checked explicitly that scenarios with top squarks and gluino as light as 2.5~TeV are seamlessly compatible with the broader physics situations we focus on in this work
while being able to pull $\Delta_{\mathrm{BG}(Z)}$ down to a value of 20 or
even less thus rendering the scenarios highly natural. 
%
\section{Conclusions}
\label{sec:conclusions}
A highly bino-like DM can be motivated from an ever diminishing upper bound on the DMDD cross sections. However, in the absence of comparably light sfermions and/or ewinos (as being hinted by experiments), a relatively light ($\lesssim 200$~GeV) LSP DM of such a kind has to look out for annihilation funnels induced by the SM $Z$-boson, the observed SM-like Higgs boson and various other light scalars of a SUSY scenario  that are yet to be ruled out experimentally. This inevitably requires the bino-like DM to have some minimal tempering with higgsinos, be it in the MSSM or in the pNMSSM, that does not spoil compliance with the DD constraints. Note that, with decoupled sfermions, the pure bino limit of a bino-dominated DM in a bino-higgsino(-singlino) system does not exist in the same sense as the pure higgsino limit for a higgsino-dominated DM in a higgsino-bino system does
with its mass around 1.1~TeV.

In the MSSM, such a tempering by higgsinos could allow for an efficient enough mutual annihilation of the highly bino-like DM states (lighter than $\sim$ 200~GeV) only via the well-known $Z$- and $\hsm$-funnels (with either $\mntrlone \approx m_Z/2$ or $\approx  \mhsm/2$) thus ensuring compliance with the experimental upper bound on DM relic abundance. 
However, stringent lower bounds on $|\mueff|$ that can be derived from recent direct searches for the light ewinos at the LHC might have started posing some reasonable threat to the viability of such a light
bino-dominated DM candidate, especially when it annihilates via $Z$-funnel.

In this work, we have demonstrated that the flexibility and the resulting richness of the ewino and the scalar spectrum of the pNMSSM, in tandem, could mitigate the situation and allow for a relatively light (below about 200~GeV, or more particularly, even below 100~GeV), highly bino-like
neutralino DM (with more than 95\% bino-content). Going beyond the MSSM, on the one hand, this scenario is known to offer additional DM annihilation funnels in the form of relatively light singlet-like scalars ($\as/\hs$) over a significant region of the pNMSSM parameter space thus allowing for a continuous range of the LSP mass which can have the right relic for suitable masses of $\as/\hs$, i.e., $\mntrlone \approx \mas/2$ or $\mntrlone \approx \mhs/2$.
A minimal tempering of a pure bino LSP is nonetheless necessary. For moderate to large values of `$\lambda$', not so heavy higgsino-like states and a singlino-like NLSP could induce such tempering  which in our case is shared between the higgsinos (up to 5\%) and the singlino (up to 1\%).  

The singlino-like NLSP plays a crucial role in the present study in at least three ways.
\begin{itemize}
\item First, in a favorable region of the pNMSSM parameter space, the singlino-like NLSP could serve as a viable coannihilating state for a highly bino-like LSP DM with mass below 200~GeV (and which is unable to funnel) thus ensuring DM abundance in the right ballpark over an extended range of the DM mass. Note that
recent LHC searches for ewinos,  even after their proper recast, are unlikely to allow higgsino-like such states with
mass $\lesssim 200$~GeV. On top of that, the latest upper bounds on the SD rates are found to put a stringent restriction on $|\mu_{\rm eff}|$ ($> 270$~GeV). Hence such higgsino-like states would be unable to play the role of coannihilating agents
for the DM in the said mass range.  The wino and the sfermions being considered to be 
rather heavy in this study, the singlino emerges as the only viable coannihilating state.
In principle, the DM wash out process in the early Universe could receive simultaneous contributions from both a funnel-annihilation and a coannihilation
process when $\mntrlone \approx m_{\mathrm{funnel}}/2 \approx m_{\mathrm{NLSP}}$.
\item Second, the tempering of the DM by a light singlino-like state provides new mechanisms in moderating the $\hsm$-$\ntrlone$-$\ntrlone$
and $Z$-$\ntrlone$-$\ntrlone$ couplings in such ways that the SI and the SD rates, respectively, could be tamed to comply with their experimentally allowed values away from the
well-known MSSM blind spot regimes.

\hskip 20pt In particular, in a light bino-higgsino system, for which the MSSM (gauge) coupling blind spot condition, relevant to the SI rate, could be satisfied only when there is a relative sign between $\mone$ and $\mueff$, now can have a suppressed overall
$\hsm$-$\ntrlone$-$\ntrlone$ interaction even
when these latter parameters have the same sign. In such a situation, the enhanced gauge coupling contribution ($\sim g_1$, for 
$\hsm$-$\bino$-$\higgsino$ interaction) to this effective interaction
strength could get canceled in the bino-higgsino-singlino system by the $\lambda$-dependent $\hsm$-$\singlino$-$\higgsino$ interaction, for moderate to large
`$\lambda$' and $\kappa < 0$.

\hskip 20pt On the other hand, the SD rate governed by the $Z$-$\ntrlone$-$\ntrlone$ interaction strength ($\sim |N_{13}^2-N_{14}^2|$) and which is suppressed in the MSSM when $\tanb$ is small (vanishing in the limit $\tanb \to 1$), could now
have new compensating terms in its expression in the presence of a light singlino-like state. These can then play their roles in suppressing the relevant coupling even for larger $\tanb$ values, again for a moderate to large value~of~`$\lambda$'.
\item Finally, the presence of such a light singlino-like NLSP, when accompanied by light singlet scalars of the scenario, could significantly alter the simplistic decay patterns (as assumed by the LHC experiments) of the immediately heavier ewinos which, in our case, are higgsino-like. This is particularly true for moderate to large values of `$\lambda$'. An already degraded lower bound on the masses of such higgsino-like states, derived via recast of the relevant LHC analyses which instead assume wino-like such states but still based on the same simple decay pattern as the latter, would then be further eroded. Hence a much smaller value of $|\mueff|$ than what is currently expected can still be allowed
thus keeping open the feasibility of a more `natural' scenario. 

\hskip 20pt
We demonstrated with examples how in various
different scenarios such a spectrum could erode yields
in the $3\ell+\etmiss$ and $1\ell + 2b$-${\rm jet} +\etmiss$ final states, the ones that traditionally appear as much cleaner and hence possess with high sensitivities to the LHC experiments.
\end{itemize}

It is further demonstrated that the viability of a
highly bino-like neutralino DM with a mass around
100~GeV (or less) depends much upon the magnitude of $\tanb$ and, as
can be expected, the smaller we go in the mass of this
DM neutralino, for a fixed value of $\mueff$, the larger the value of $\tanb$ we require
to keep a check on its SI scattering rates which are highly
constrained from experiments. Furthermore, for the DM mass below the top quark mass, larger values of $\tanb$ could also crucially aid DM annihilation to $b\bar{b}$ via the singlet-like pseudoscalar ($\as$) funnel as the involved coupling strength $|g_{_{\as b\bar{b}}}|$ increases with $\tanb$.

All these features are backed by analytical results that  we
obtained for the first time in the framework of a bino-higgsino-singlino system and are presented in a three-part appendix. We also observe that the phenomenology of such a scenario but with
a higgsino-like NLSP almost mimics an MSSM scenario with a similar spectrum.

In conclusion, it should be stressed that recent and upcoming analyses from the LHC experiments would shed important light on
the viability of such a relatively light, highly bino-like DM in the framework of pNMSSM by keeping track of the lower bound on $|\mueff|$ that these could extract. Under the current circumstances, this would involve a rather dedicated recast of the relevant LHC analyses involving not only the lepton-rich final
states which are traditionally known to be rather sensitive
to such new physics scenarios but also the ones with
the bottom quarks, the taus and the photons in the final state which arise in the cascades of the ewinos involving Higgs(-like) scalars and can occasionally be boosted as well. Nonetheless, the results obtained by us in a recent work using a limited recast and employed in this work already indicate that there is still much room left in the bottom
as far as a lower bound on $|\mueff|$ of the pNMSSM is concerned which can still be consistent with most other relevant constraints including those from the
DM experiments and could render the scenario rather `natural'.
However, it is the experimental constraint on the SD rate that puts the most stringent lower bound on $|\mueff| \, (\gtrsim 270~\mathrm{GeV})$
in the current scenario.

On the other hand, we find that the future DARWIN experiment 
would be able to probe the entire volume of the pNMSSM parameter space of such a scenario that gives rise to a highly bino-like LSP DM by studying the DM-nucleon SD scattering even when the DM-nucleon SI scattering cross section sinks below the neutrino floor.

Results presented in this work follow from a conservative approach of not
incorporating a downward scaling of the DD rates when the DM relic abundance falls 
below its 2$\sigma$ lower bound. However, when incorporated, such a scaling could
open up new allowed regions of the parameter space by pulling down the DD rates
below their respective upper bounds that assume the DM abundance to be equal to the central value obtained by the Planck collaboration. This, in turn, could allow for some increased level of mixings of the higgsino and the singlino states in an otherwise bino-dominated LSP DM. The analytical results
presented for the first time in this work by considering a coupled system of these three ewino states would aid a better understanding of phenomena that involve
these states in a nontrivial~way.
\acknowledgments{WA and SR are supported by the funding available from the Department of Atomic Energy~(DAE), Government of India for the Regional Centre for Accelerator-based Particle Physics~(RECAPP) at Harish-Chandra Research Institute (HRI). WA also acknowledges support from the XII Plan Neutrino Project funded by the DAE at HRI. The authors thank U.~Ellwanger, C.~Hugonie and T.~Stefaniak for useful discussions. They acknowledge the use of the High Performance Scientific Computing facility and RECAPP's Cluster Computing facility at HRI and would like to thank Rajeev Kumar and Ravindra Yadav for their technical help at these facilities.}
\vskip 30pt
\appendix
\section*{Appendices}
\addcontentsline{toc}{section}{\protect\numberline{}Appendices}%
\section{Scalar-neutralino-neutralino interactions}
\label{appsec:ew-sc-int}
Neutralino-neutralino-scalar interactions are given in their general forms in ref.~\cite{Ellwanger:2009dp}.
For the $CP$-even scalars ($h_i$) these are given by
\bea
g_{_{h_i \ntrlj \ntrlk}} &=& {\lambda \over \sqrt{2}}
(S_{i1} \Pi^{45}_{jk} + S_{i2} \Pi^{35}_{jk} + S_{i3} \Pi^{34}_{jk})
 - \sqrt{2} \, \kappa S_{i3} N_{j5} N_{k5} \nonumber \\
&+& {g_1 \over 2} (S_{i1} \Pi^{13}_{jk} - S_{i2} \Pi^{14}_{jk})
 -  {g_2 \over 2} (S_{i1} \Pi^{23}_{jk} - S_{i2} \Pi^{24}_{jk}),
\label{eqn:hinjnk1}
\eea
where $\Pi^{ab}_{jk} = N_{ja} N_{kb} + N_{jb} N_{ka}$ and the `$S$' matrix
diagonalizes the mass-squared matrix ${\cal M}_S^2$  of eq.~(\ref{eqn:cp-even-matrix}).
For the $CP$-odd scalars ($a_i$) the corresponding expression is
\bea
g_{_{a_i \ntrlj \ntrlk}} = &i& \bigg[ {\lambda \over \sqrt{2}}
(P_{i1} \Pi^{45}_{jk} + P_{i2} \Pi^{35}_{jk} + P_{i3} \Pi^{34}_{jk})
 - \sqrt{2} \, \kappa P_{i3} N_{j5} N_{k5} \nonumber \\
&-& {g_1 \over 2} (P_{i1} \Pi^{13}_{jk} - P_{i2} \Pi^{14}_{jk})
 +  {g_2 \over 2} (P_{i1} \Pi^{23}_{jk} - P_{i2} \Pi^{24}_{jk}) \bigg],
\label{eqn:ainjnk1}
\eea
where matrix `$P$'
diagonalizes the mass-squared matrix ${{\cal M}'}^2_P$  of eq.~(\ref{eqn:cp-odd-matrixp}). 

In a more convenient basis (of eq.~(\ref{eqn:hiEhhat})) for the $CP$-even 
scalars, and for the corresponding scalar states $h_i$, eq.~(\ref{eqn:hinjnk1}) reduces to
\bea
\label{eqn:hinjnk-reduced-without-approximation}
g_{_{h_i \ntrlj \ntrlk}}
 & = &
\Bigg[{\lambda \over \sqrt{2} } \big[ E_{h_i \hat{h}} N_{j5} (N_{k3} \sin\beta + N_{k4} \cos\beta)
 + E_{h_i \hat{H}} N_{j5} (N_{k4} \sin\beta - N_{k3} \cos\beta)  \nonumber \\
 & & \hskip 25pt + \; E_{h_i \hat{s}} (N_{j3}N_{k4} - {\kappa \over \lambda} N_{j5}N_{k5})\big] + {1 \over 2}\big[g_{1}N_{j1} - g_{2}N_{j2}\big] \big[E_{h_i \hat{h}} (N_{k3} \cos\beta - N_{k4} \sin\beta) \nonumber \\
 & & \hskip 25pt + E_{h_i \hat{H}} (N_{k3} \sin\beta + N_{k4} \cos\beta)\big]\Bigg] + \Bigg[j \longleftrightarrow k \Bigg]\, .
\eea
On the other hand, for the $CP$-odd scalars, $a_i$, interacting with a pair of neutralinos, $\ntrlj$ and $\ntrlk$, eq.~(\ref{eqn:ainjnk1}), in terms of diagonalization matrix ${\cal O}(2 \times 2)$ of eq.~(\ref{eqn:cp-odd-matrix}), is given by,
\bea
\label{eqn:ainjnk-reduced-without-approximation}
g_{_{a_i \ntrlj \ntrlk}}
 & = &
\Bigg[i\Big({\lambda \over \sqrt{2}} \big[{\cal O}_{iA} N_{j5} (N_{k4} \sin\beta + N_{k3} \cos\beta) + {\cal O}_{iS_I} (N_{j3}N_{k4} - {\kappa \over \lambda} N_{j5}N_{k5})\big]  \nonumber \\
 & & \hskip 25pt - \; {1 \over 2}\big[g_{1}N_{j1} - g_{2}N_{j2}\big] \big[{\cal O}_{iA} (N_{k3} \sin\beta - N_{k4} \cos\beta)\big]\Big)\Bigg] + \Bigg[j \longleftrightarrow k \Bigg] . \quad
\eea
In the present study, we consider a bino-like neutralino LSP and a rather heavy wino which is effectively decoupled. Consequently, wino admixtures $N_{j2}$ in the lighter neutralino states can be ignored.
%
\section{The bino-higgsino-singlino system and the SI and the SD blind spots}
\label{appsec:blindspot}
For a decoupled wino, the $(5\times 5)$, symmetric neutralino mass-matrix of eq.~(\ref{eqn:mneut}) effectively reduces to a $(4\times 4)$ matrix in the
basis $\psi^0=\{\bino,\higgsinod,\higgsinou,\singlino\}$, comprising of the bino, the higgsinos and the singlino, and is given by
\beq
\label{eqn:mneut-reduced}
{\cal M}_0 =
\left( \begin{array}{ccccc}
\mone & -\dfrac{g_1 \vd}{\sqrt{2}} & \dfrac{g_1 \vu}{\sqrt{2}} & 0 \\[0.4cm]
\ldots &  0 & -\mueff & -\lambda \vu \\
\ldots & \ldots & 0 & -\lambda \vd \\
\ldots & \ldots & \ldots & 2 \kappa \vs
\end{array} \right) \, .
\eeq
In the absence of $CP$-violation, symmetric matrix ${\cal M}_0$ can be diagonalized by an orthogonal $(4 \times 4)$ matrix `$N$',
i.e.,
\bea
\label{eqn:diagonalise-1}
N {\cal M}_0 N^T = {\cal M}_D = {\rm diag}(m_{{_{\chi}}_{_1}^0},m_{{_{\chi}}_{_2}^0},m_{{_{\chi}}_{_3}^0},m_{{_{\chi}}_{_4}^0})  \, . 
\eea
Sticking all through to the conventions/assignments of eq.~(\ref{eqn:mneut}) but for dropping the wino, the neutralino mass-eigenstates are again represented by
\bea
\ntrlj=N_{jk} \psi_k^0  \,,
\eea
with $j=1,2,3,4$ while $k=1,3,4,5$. Expressions for the elements $N_{jk}$ and the physical mass of the $j$-neutralino ($|\mntrlj|$) can be obtained by using 
\bea
\label{eqn:diagonalise-2}
N {\cal M}_0  - {\cal M}_D N = 0
\eea
and exploiting the orthogonality of the matrix `$N$', i.e.,
\bea
\label{eqn:diagonalise-3}
N^T N = 1 \,.
\eea
This leads to the following set of equations: 
\bea
\label{eqn:relation-equation-1}
(M_1 - \mntrlj)N_{j1} - {g_1 v_d \over \sqrt{2}}N_{j3} + {g_1 v_u \over \sqrt{2}}N_{j4} &=& 0 \, , \quad \\
\label{eqn:relation-equation-2}
- {g_1 v_d \over \sqrt{2}}N_{j1} -\mntrlj N_{j3} - \mueff N_{j4} - \lambda\, v_u N_{j5} &=& 0  \, , \quad \\
\label{eqn:relation-equation-3}
 {g_1 v_u \over \sqrt{2}}N_{j1} -\mueff N_{j3} - \mntrlj N_{j4} - \lambda\, v_d N_{j5} &=& 0 \, , \quad \\
\label{eqn:relation-equation-4}
- \lambda\, v_u N_{j3} - \lambda\, v_d N_{j4} + (\msinglino - \mntrlj) N_{j5} &=& 0 \,.
\eea
By adopting the approach of ref.~\cite{ElKheishen:1992yv} we divide the above four equations by $N_{j5}$ (assuming $N_{j5} \neq 0$) and pick up a suitable set of three eqs.~(\ref{eqn:relation-equation-2}),~(\ref{eqn:relation-equation-3}) and~(\ref{eqn:relation-equation-4}). We thus find the following relation between the bino ($N_{j1}$) and the singlino ($N_{j5}$) admixtures
 in the $j$-th neutralino and the similar ones for the two higgsino admixtures ($N_{j3}$ and $N_{j4}$) to the same singlino admixture:
\bea
\label{eqn:N11-N15relation-1}
{N_{j1} \over N_{j5}}
  &=& {\lambda^2 v^2(\mueff \sin 2\beta - \mntrlj) + (\msinglino - \mntrlj)(\mueff^2 - \mntrlj^2) \over {g_1 \over \sqrt{2}}\, \lambda\,   \mueff\, (v^2_u -v^2_d)} \,,\\
\label{eqn:N13-N15relation-1}
{N_{j3} \over N_{j5}}
  &=& {\lambda^2  v_d\, v^2 + (\msinglino - \mntrlj)(v_d\, \mntrlj + v_u\, \mueff) \over \lambda\,  \mueff \,(v^2_u -v^2_d)} \,, \\
\label{eqn:N14-N15relation-1}
{N_{j4} \over N_{j5}}
  &=& {-\lambda^2 v_u\, v^2 -(\msinglino - \mntrlj)(v_u\, \mntrlj + v_d\, \mueff) \over \lambda\,  \mueff \,(v^2_u -v^2_d)} \,.
\eea
By using the unitarity condition (for the $(4 \times 4)$ system)
\bea
\label{eqn:unitarity-relation-1}
N^2_{j1} + N^2_{j3} + N^2_{j4} + N^2_{j5} = 1 \,, 
\eea
expressions for the elements $N_{jk}$ can be obtained, e.g.,
\bea
\label{eqn:n^2j5-value}
N^2_{j5} = {Z^2 \over (\msinglino - \mntrlj)^2 \, (\mueff^2 - \mntrlj^2) \, I_j} \,,
\eea
where
\vspace{-0.5cm}
\bea
\label{eqn:Z^2-value}
Z =  {g_1 \over \sqrt{2}}\, \lambda\,   \mueff\, (v^2_u -v^2_d)
\eea
and
\bea
\label{eqn:I-value}
I_j &=& \mueff^2 - \mntrlj^2 +  {g^2_1 v^2 \over 2}\;{\mntrlj^2 + \mueff^2 + 2 \mntrlj \mueff \sin 2\beta \over {\mueff^2 - \mntrlj^2}} + {\lambda^2 g^2_1 v^4 (\mntrlj + \mueff \sin 2\beta) \over (\msinglino - \mntrlj) \, (\mueff^2 - \mntrlj^2)} \nonumber \\ 
 &+& {2 \lambda^2 v^2 (\mueff \sin 2\beta - \mntrlj) \over {\msinglino - \mntrlj}} + {Z^2 + \lambda^4 v^4\big[{g^2_1 v^2 \over 2} + (\mueff \sin 2\beta - \mntrlj)^2 \big] \over (\msinglino - \mntrlj)^2 \, (\mueff^2 - \mntrlj^2)} \,. \hskip 35pt \quad
\eea
Meanwhile, in the effective MSSM (decoupling) limit, i.e., $\lambda
\sim \kappa \rightarrow 0$ and $\vs \rightarrow \infty$~\cite{Ellwanger:2009dp} (and $\mntrlj \neq \msinglino, \,|\mueff|$), one recovers the $(3\times3)$ bino-higgsino system of the MSSM.
Thus, the expression in eq.~(\ref{eqn:I-value}) reduces to 
\[ I_j \rightarrow  \mueff^2 - \mntrlj^2 +  {g^2_1 v^2 \over 2}\;{\mntrlj^2 + \mueff^2 + 2 \mntrlj \, \mueff \sin 2\beta \over {\mueff^2 - \mntrlj^2}} \]
and this, when plugged in into eq.~(\ref{eqn:n^2j5-value}), results in
$N^2_{j5} \rightarrow 0$.
Furthermore, in this limit, from eqs.~(\ref{eqn:N11-N15relation-1}),~(\ref{eqn:N13-N15relation-1}) and~(\ref{eqn:N14-N15relation-1}), one can find
\bea
\label{eqn:N13-N11relation}
{N_{j3} \over N_{j1}} \, 
  \approx {g_1 v \over \sqrt{2}\mueff} {(\mntrlj/\mueff) \cos\beta + \sin\beta
                        \over {1 - (\mntrlj/\mueff)^2}}  \,,
\eea
\vskip -10pt
\bea
\label{eqn:N14-N11relation}
{N_{j4} \over N_{j1}}
 \, \approx {-g_1 v \over \sqrt{2}\mueff} {(\mntrlj/\mueff) \sin\beta + \cos\beta
                        \over {1 - (\mntrlj/\mueff)^2}} \,.
\eea
Eqs.~(\ref{eqn:N13-N11relation}) and (\ref{eqn:N14-N11relation}) agree with the corresponding expressions for the bino-higgsino system of the MSSM~\cite{Djouadi:2001kba,Pierce:2013rda}. 

To find the SI blind spot condition for the $(4\times4)$ bino-higgsino-singlino system we refer to the generic $CP$-even scalar-neutralino-neutralino coupling  
of eq.~(\ref{eqn:hinjnk-reduced-without-approximation}) which, for a pair of LSPs ($j$=$k$=1) and by dropping the wino-related term proportional to $g_2 N_{12}$, reduces to
\bea
\label{eqn:hin1n1-11}
g_{_{h_i \ntrlone \ntrlone}}
 & = &
{\sqrt{2} \lambda} \big[ E_{h_i \hat{h}} N_{15} (N_{13} \sin\beta + N_{14} \cos\beta)
 + E_{h_i \hat{H}} N_{15} (N_{14} \sin\beta - N_{13} \cos\beta)  \nonumber \hskip 35pt \\
 & & \hskip 12pt + \; E_{h_i \hat{s}} (N_{13}N_{14} - {\kappa \over \lambda} N^2_{15})\big] + g_{1}N_{11} \big[E_{h_i \hat{h}} (N_{13} \cos\beta - N_{14} \sin\beta) \nonumber \\
 & & \hskip 12pt + \hskip 3pt E_{h_i \hat{H}} (N_{13} \sin\beta + N_{14} \cos\beta)\big] \,.
\eea
From eq.~(\ref{eqn:hin1n1-11}) and
considering that there is no mixing between the singlet and the doublet scalars ($E_{\hsm \hat{s}}, E_{H \hat{s}} \sim 0$) and that the mixing between the doublet Higgs states could also be ignored (i.e., $E_{\hsm \hat{h}}, E_{H \hat{H}} \sim 1$ and $E_{\hsm \hat{H}} \sim 0$) for large enough $m_H$ and $\tanb$~\cite{Badziak:2015exr}, the coupling of the SM-like Higgs boson with an LSP pair is given by 
\bea
\label{eqn:hin1n1-2}
g_{_{\hsm \ntrlone \ntrlone}}
 &\simeq&
\sqrt{2} \lambda \, (N_{13} \sin\beta + N_{14} \cos\beta)N_{15} + g_{1} (N_{13} \cos\beta - N_{14} \sin\beta)N_{11} \nonumber \\
&=&
\Bigg[ {\sqrt{2} \lambda} \bigg({N_{13} \over N_{15}} \sin\beta + {N_{14} \over N_{15}} \cos\beta \bigg) + g_{1}\bigg({N_{13} \over N_{15}} \cos\beta - {N_{14} \over N_{15}} \sin\beta\bigg) {N_{11} \over N_{15}} \Bigg] N_{15}^2 \,.\hskip 30pt
\eea
Using eqs.~(\ref{eqn:N11-N15relation-1}),~(\ref{eqn:N13-N15relation-1}),~(\ref{eqn:N14-N15relation-1}) and~(\ref{eqn:n^2j5-value}), eq.~(\ref{eqn:hin1n1-2}) reduces to
\bea
\label{eqn:hin1n1-3}
\hskip -30pt
g_{_{\hsm \ntrlone \ntrlone}}
 &\simeq&
{g^2_1 v \over \sqrt{2} I_1} \Bigg[ \mntrlone + \mueff \sin 2\beta + {{2 \lambda^2 v^2} \over {\msinglino - \mntrlone}} + {\lambda^4 v^4 (\mueff \sin 2\beta - \mntrlone) \over (\msinglino - \mntrlone)^2 \, (\mueff^2 - \mntrlone^2)} \Bigg] \,.
\eea
A vanishing $g_{_{\hsm \ntrlone \ntrlone}}$ that
results in a (coupling) blind spot for the most dominant ($\hsm$) contribution to the SI scattering cross section of the DM (LSP) off the nucleus, as long as $\mntrlone \; \neq \; |\mueff|, \, \msinglino$ and $I_1 \neq 0$, thus implies
\bea
\label{eqn:coupling-blindspot}
\Bigg[\mntrlone  + {{2 \lambda^2 v^2} \over {\msinglino - \mntrlone}} + {\lambda^4 v^4 (\mueff \sin 2\beta - \mntrlone) \over (\msinglino - \mntrlone)^2 \, (\mueff^2 - \mntrlone^2)} \Bigg]{1 \over \mueff \sin 2\beta}= -1 \,.
\eea
For $|\msinglino|-|\mntrlone| \lesssim \lambda v$ and
$|\mueff| - |\mntrlone| \gg \lambda v$ the third term of the
left-hand side of eq.~(\ref{eqn:coupling-blindspot}) is suppressed
compared to the second term there. This is a situation leading to  at best a moderate splitting between the singlino-like NLSP and the bino-like LSP, a setup which is studied in detail in the present work. Under such a circumstance, the coupling blind spot condition
can be approximated as
\bea
\label{eqn:coupling-blindspot-approxi}
\bigg(\mntrlone + {2 \lambda^2 v^2 \over {\msinglino - \mntrlone}}\bigg) {1 \over \mueff \sin 2\beta} \simeq -1 \,.
\eea
Compliance with eq.~(\ref{eqn:coupling-blindspot-approxi}) under two specific situations may worth a mention. A closer inspection reveals that when $\mntrlone$ and $\mueff$ carry the same sign, such compliance requires $\msinglino$ and $\mueff$ to have a relative sign between them which, in turn, implies $\kappa < 0$. Thus, these together amount to a relative sign between $\mntrlone$ and $\msinglino$. Such a
requirement is indeed vindicated by our numerical scan (see the bottom, right plot of figure~\ref{fig:coupling-blind-spot}) when compliance with experimental bound on the SI scattering cross section is demanded. If, instead,
$\mntrlone$ and $\mueff$ carry a relative sign between them, the situation is a little more involved. `$\kappa$' could take both signs depending on the absolute magnitudes and signs on $\mntrlone$ and the term $\mueff \sin 2\beta$.

By its sheer appearance, eq.~(\ref{eqn:coupling-blindspot}) turns out to be
the suitable one for describing the blind spot condition for a bino-like
LSP in a bino-higgsino-singlino system of neutralinos.\footnote{Note that an equivalent condition could be arrived at by using the set of eqs.~(\ref{eqn:relation-equation-1}),~(\ref{eqn:relation-equation-2}) and~(\ref{eqn:relation-equation-3}) (instead of the set~(\ref{eqn:relation-equation-2}),~(\ref{eqn:relation-equation-3}) and~(\ref{eqn:relation-equation-4})) which will be suitable for describing the blind spot situation with a singlino-like LSP in such a system.} This is clearly demonstrated when one is able to recover the coupling blind spot condition for the bino-higgsino system of the MSSM 
\cite{Huang:2014xua} in the decoupling limit, i.e.,
\bea
\label{eqn:MSSMcoupling-blindspot}
{\mntrlone \over \mueff} \approx -\sin 2\beta \,.
\eea
Note that condition~(\ref{eqn:MSSMcoupling-blindspot}) has its sole origin in the gauge interaction of a pair of bino-dominated LSPs with the SM-like Higgs boson.

Note that while in the MSSM case of eq.~(\ref{eqn:MSSMcoupling-blindspot}) where the blind spot condition can only be satisfied for $\mntrlone$ and $\mueff$ carrying a relative sign between them, such a condition given by eq.~(\ref{eqn:coupling-blindspot-approxi}) for the NMSSM could get satisfied for $\mntrlone$ and $\mueff$ having the same sign as well. This happens for moderate to large `$\lambda$'
and negative `$\kappa$' (as discussed earlier). The former, in turn, tempers the bino-like LSP with some singlino admixture.

To study the blind spot in SI cross section with origin in the interference between the scattering diagrams mediated by $\hsm$ and the heavy MSSM-like $CP$-even Higgs boson, `$H$', one needs to know
the coupling of a pair of LSPs with the latter. The same is found from
the general expression for such couplings as given in eq.~(\ref{eqn:hin1n1-11}) and by considering $E_{H \hat{H}}$ $\sim$ 1 and $E_{H \hat{h}}, \, E_{H \hat{s}} \sim 0$ and is given by
\bea
\label{eqn:Hn1n1-2}
\hskip -25pt
g_{_{H \ntrlone \ntrlone}}
 & \simeq &
{\sqrt{2} \lambda} \, N_{15} \, (N_{14} \sin\beta - N_{13} \cos\beta) + g_{1}N_{11} \big(N_{13} \sin\beta + N_{14} \cos\beta\big)  \quad \nonumber \\
& = &
\Bigg[ {\sqrt{2} \lambda} \, \bigg({N_{14} \over N_{15}} \sin\beta - {N_{13} \over N_{15}} \cos\beta \bigg)  + g_{1}\bigg({N_{13} \over N_{15}} \sin\beta + {N_{14} \over N_{15}} \cos\beta\bigg) {N_{11} \over N_{15}} \Bigg] \, N_{15}^2 \,.
\eea
As before, substituting eqs.~(\ref{eqn:N11-N15relation-1}),~(\ref{eqn:N13-N15relation-1}),~(\ref{eqn:N14-N15relation-1}) and~(\ref{eqn:n^2j5-value}) into eq.~(\ref{eqn:Hn1n1-2}), one finds
\bea
\label{eqn:Hn1n1-3}
g_{_{H \ntrlone \ntrlone}}
 & \simeq&
{g^2_1 v \over \sqrt{2} I_1} \Bigg[{\lambda^2 v^2 \big[\lambda^2 v^2 + 2 \mntrlone (\msinglino - \mntrlone)\big] \over (\msinglino - \mntrlone)^2 \, (\mueff^2 - \mntrlone^2)} - 1 \Bigg] \mueff \cos 2\beta \, .
\eea
Now, the SI scattering cross section of the DM particle off a
nucleon `$N$' (proton ($p$) or neutron ($n$), in the nucleus of the detector material) is given by~\cite{Badziak:2016qwg, Cao:2018rix}
\beq
\label{eqn:SI-equation-1}
\sigma^{\rm SI}_{\ntrlone-(N)} = \frac{4 \mu^2_r}{\pi} |f^{(N)}|^2, \hspace{0.4cm} f^{(N)} = \sum_{i=1}^3 \frac{g_{_{h_{i} \ntrlone \ntrlone}} \, g_{_{h_{i} N N}}}{2 m^2_{_{h_i}}}  \,, 
\eeq
where $\mu_r$ is the reduced mass of the system comprising of the DM and the nucleon. The general expression for the coupling of a $CP$-even scalar with a pair of nucleons, $g_{_{h_{i} N N}}$ of eq.~(\ref{eqn:hiNN-badziak}), in the basis of eq.~(\ref{eqn:hiEhhat}) is given by~\cite{Badziak:2015exr}
\bea
\label{eqn:hiNN-newbasis}
g_{_{h_i N N}}
 &= &
\frac{m_N}{\sqrt{2} v}  \Big[ E_{h_i \hat{h}} (F^{(N)}_d + F^{(N)}_u) + E_{h_i \hat{H}} \big(\tan\beta F^{(N)}_d - \frac{1}{\tan\beta} F^{(N)}_u\big) \Big] \, ,
\eea
where $m_N$ is the mass of the nucleon, $F^{(N)}_{d,u}$ are the combinations of various nucleon form factors appropriate for the
nucleon `$N$' as presented in ref.~\cite{Belanger:2013oya, Badziak:2016qwg}.

Substituting eqs.~(\ref{eqn:hiNN-newbasis}),~(\ref{eqn:hin1n1-3}) and~(\ref{eqn:Hn1n1-3}) into eq.~(\ref{eqn:SI-equation-1}) and ignoring the contribution from the $CP$-even singlet-like Higgs (given the nucleons couple only to the minuscule doublet content within the singlet-like scalar), one finds
\bea
\label{eqn:SI-NMSSM-BLINDSPOT-1}
\hskip -230pt
f^{(N)} &\approx& 
 {g_{_{\hsm \ntrlone \ntrlone}} g_{_{\hsm N N}} \over \mhsmsq} + {g_{_{H \ntrlone \ntrlone}} g_{_{_{H N N}}} \over m^2_{H}} \nonumber \\
&=& {g^2_1 m_N F^{(N)} \over \mhsmsq \,I_1}\Bigg[\mntrlone + \mueff \sin 2\beta + {2 \lambda^2 v^2 \over \msinglino - \mntrlone} + {\lambda^4 v^4 (\mueff \sin 2\beta - \mntrlone) \over (\msinglino - \mntrlone)^2(\mueff^2 - \mntrlone^2)} \hskip 15pt \nonumber \\
&+&\frac{\mueff}{2}  \cos 2\beta \bigg(\tan\beta- \frac{1}{\tan\beta}\bigg) \bigg[{\lambda^2 v^2 (\lambda^2 v^2 + 2 \mntrlone (\msinglino - \mntrlone)) \over (\msinglino - \mntrlone)^2(\mueff^2 - \mntrlone^2)} - 1 \bigg] {\mhsmsq \over m^2_{H}}\Bigg] \, ,
\eea
where we have made the standard assumption $F^{(N)}_d \approx F^{(N)}_u \approx F^{(N)}$. Since for a bino-dominated LSP
$\mntrlone \; \neq \,|\mueff|, \, \msinglino$ and $I_1 \neq 0$, blind spot for the SI cross section arises when the
quantity within the big square bracket on the right-hand side of eq.~(\ref{eqn:SI-NMSSM-BLINDSPOT-1}) vanishes.
The last term represents the contribution from the heavy $CP$-even MSSM-like Higgs boson and is suppressed by its mass.
Dividing the entire content further by $\mueff$ to cast the resulting condition into a
familiar form, the latter appears as 
\bea
\label{eqn:NMSSM-BLINDSPOT-CONDITION-1}
& & \hspace{-1.4cm}
 {\mntrlone \over \mueff} + \sin 2\beta + {2 \lambda^2 v^2 \over \mueff \, (\msinglino - \mntrlone)} + {\lambda^4 v^4 \big(\sin 2\beta - \mntrlone / \mueff\big) \over (\msinglino - \mntrlone)^2 \, (\mueff^2 - \mntrlone^2)} \nonumber \\ 
& & \hspace{-1cm} + \frac{1}{2} \cos 2\beta \bigg(\tan\beta- \frac{1}{\tan\beta}\bigg) \Bigg[{\lambda^2 v^2 \left[\lambda^2 v^2 + 2 \mntrlone (\msinglino - \mntrlone) \right] \over (\msinglino - \mntrlone)^2 \, (\mueff^2 - \mntrlone^2)} - 1 \Bigg] {\mhsmsq \over m^2_{H}} \approx 0 \, .
\eea
As before, the corresponding blind spot condition pertaining to the bino-higgsino system of the MSSM with a bino-like LSP is
retrieved from eq.~(\ref{eqn:NMSSM-BLINDSPOT-CONDITION-1}) in the
decoupling limit and is given by
\bea
\label{eqn:MSSM-BLINDSPOT-CONDITION-3}
{\mntrlone \over \mueff} + \sin2\beta - \frac{1}{2}\cos2\beta \big(\tan\beta- \frac{1}{\tan\beta}\big){\mhsmsq \over m^2_{H}} \approx 0 \, .
\eea
For large $\tan\beta$ when $\frac{1}{\tan\beta}$ can be dropped and
$\cos2\beta \sim -1$, the blind spot condition~(\ref{eqn:MSSM-BLINDSPOT-CONDITION-3}) becomes identical to the one obtained for the
MSSM with a bino-like LSP~\cite{Huang:2014xua,Badziak:2017the}, i.e.,
\bea
\label{eqn:MSSM-BLINDSPOT-CONDITION-4}
{\mntrlone \over \mueff} + \sin2\beta + {\mhsmsq \over m^2_{H}} \, \frac{\tanb}{2} \approx 0 \, .
\eea
A corresponding blind spot condition for the SD scattering cross section can be obtained for the bino-higgsino-singlino system by considering
%
$\sigma^\mathrm{SD}
 \propto  (N^2_{13} - N^2_{14})^2$
%
where, by using eqs.~(\ref{eqn:N13-N15relation-1}),~(\ref{eqn:N14-N15relation-1}) and~(\ref{eqn:n^2j5-value}), one can find
\bea
\label{eqn:sigma-SD-1}
N^2_{13} - N^2_{14}
 =
{g^2_1 v^2 \over 2 I_1} \cos 2\beta \Bigg[-1 + {2 \lambda^2 v^2 \over \bigg({\msinglino \over \mntrlone} - 1 \bigg)(\mueff^2 - \mntrlone^2)} + {\lambda^4 v^4  \over (\msinglino - \mntrlone)^2(\mueff^2 - \mntrlone^2)} \Bigg] \, . \nonumber \\
\eea
An MSSM-like SD blind spot condition is trivially found when
$\tan\beta = 1$ whereby $\cos2\beta$ vanishes~\cite{Cheung:2012qy}. However, in the NMSSM,
it is also possible to find a small SD cross section if the terms within the square bracket of eq.~(\ref{eqn:sigma-SD-1}) could arrange to cancel each other. 
Note that the last term is always positive. The second term is positive (negative) if $\msinglino$ and $\mntrlone$ have the same (opposite) signs. In the first case, the cancellation has to occur between the first term and the sum of the rest two. In the second case, which may not be too natural, the cancellation should take place between the sum of the first two and the last term.
It must be noted, that neither $\msinglino$ nor $\mueff$ could become too large compared to the mass of the bino-like LSP if an overall  cancellation has to take place in the presence of the first term, i.e., $-1$. In figure~\ref{fig:deltam-admixtures} we encounter such a vanishing of the quantity $N^2_{13} - N^2_{14}$ for certain small splittings between $\msinglino$ and $\mntrlone$ when both carry the same sign. This may, however, induce a critically large bino-singlino mixing in the LSP which in turn could be disfavored by the constraints on the SI rates.
In any case, this is again purely an NMSSM phenomenon showcasing a tempering
of the bino-dominated LSP by the singlino which necessarily has to be assisted by the higgsinos.
Furthermore, in the limit $\mntrlone^2 \ll \mueff^2, \, \msinglino^2$ (i.e., for a 
highly bino-like LSP), while eq.~(\ref{eqn:SI-NMSSM-BLINDSPOT-1}) leads to
$\sigma^{\mathrm{SI}} \propto 1/\mueff^2$, eq.~(\ref{eqn:sigma-SD-1}) results in
$\sigma^\mathrm{SD} \propto 1/\mueff^4$~\cite{Carena:2018nlf}.

Another much relevant quantity in the current context is
$N^2_{13} + N^2_{14}$ which is a measure of the total higgsino admixture in the LSP and which controls scalar (Higgs boson) couplings to a pair of bino-like LSPs. Again, using eqs.~(\ref{eqn:N13-N15relation-1}),~(\ref{eqn:N14-N15relation-1}) and~(\ref{eqn:n^2j5-value}), one finds, for the bino-higgsino-singlino system
\bea
\label{eqn:total-higgsino}
N^2_{13} + N^2_{14}
 = 
{g^2_1 v^2 \over 2 I_1} \Bigg[{\mntrlone^2 + \mueff^2 + 2 \mntrlone \mueff \sin 2\beta \over \mueff^2 - \mntrlone^2} &+& {2 \lambda^2 v^2 \, (\mntrlone + \mueff \sin 2\beta) \over (\msinglino - \mntrlone)\, (\mueff^2 - \mntrlone^2)} \hskip 60pt \nonumber \\
&+& {\lambda^4 v^4  \over (\msinglino - \mntrlone)^2 \, (\mueff^2 - \mntrlone^2)} \Bigg] \, .\hskip 40pt \quad 
\eea
A little scrutiny reveals that eqs.~(\ref{eqn:n^2j5-value}),~(\ref{eqn:Z^2-value}),~(\ref{eqn:I-value}),~(\ref{eqn:sigma-SD-1}) and~(\ref{eqn:total-higgsino}) are all symmetric under simultaneous flips in signs on each of
$\mueff$, $\mntrlone \, (\mone)$ and $\msinglino$. Hence
the quantities $N^2_{15}$, $N^2_{13} - N^2_{14}$ and $N^2_{13} + N^2_{14}$ which the above-mentioned equations describe all exhibit this symmetry. We have pointed this out in our discussions on figures
\ref{fig:higgsino-singlino},~\ref{fig:all-components} and~\ref{fig:deltam-admixtures} in section~\ref{subsubsec:interactions}.

Before we end this section of the appendix it is worth taking a look at how the $(3 \times 3)$ higgsino-singlino system of the NMSSM follows from
the set of relations derived for the $(4 \times 4)$ bino-higgsino-singlino system. Towards this, dividing eqs.~(\ref{eqn:relation-equation-1}),~(\ref{eqn:relation-equation-2}),~(\ref{eqn:relation-equation-3}) and~(\ref{eqn:relation-equation-4}) by $N_{j1}$ (assuming  $N_{j1} \neq 0$) but this time picking up the three eqs.~(\ref{eqn:relation-equation-1}),~(\ref{eqn:relation-equation-2}) and~(\ref{eqn:relation-equation-3}) as the ones appropriate for the purpose in hand, one can find the following relations among $N_{jk}$'s:
\vspace{-0.25cm}
\bea
\label{eqn:N15-N11relation-1}
{N_{j5} \over N_{j1}}
  &=& {-{g^2_1 \over 2} v^2(\mueff \sin 2\beta + \mntrlj) + (M_1 - \mntrlj)(\mueff^2 - \mntrlj^2) \over {g_1 \over \sqrt{2}}\lambda\, \mueff\,(v^2_u -v^2_d)} \, , \\
\label{eqn:N13-N11relation-1}
{N_{j3} \over N_{j1}}
  &=& {{g^2_1 \over 2}v^2 v_u + (M_1 - \mntrlj)(v_u \,\mntrlj - v_d \,\mueff) \over {g_1 \over \sqrt{2}} \,\mueff \,(v^2_u -v^2_d)} \, ,\\
\label{eqn:N14-N11relation-1}
{N_{j4} \over N_{j1}}
  &=& {{g^2_1 \over 2}v^2 v_d  + (M_1 - \mntrlj)(v_d \,\mntrlj - v_u\, \mueff) \over {g_1 \over \sqrt{2}} \,\mueff \,(v^2_u -v^2_d)} \, .\quad  
\eea
Using eqs.~(\ref{eqn:N15-N11relation-1}),~(\ref{eqn:N13-N11relation-1}) and~(\ref{eqn:N14-N11relation-1}) one can now retrieve the following set of familiar~\cite{Badziak:2015exr} relations for the $(3 \times 3)$ higgsino-singlino system for decoupled $M_1$ (thus, $N_{j1} \rightarrow 0$) and $\mntrlj \neq M_1,|\mueff|$:
\vspace{-0.5cm}
\bea
\label{eqn:N13-N15relation}
{N_{j3} \over N_{j5}}
  \approx {\lambda v \over \mueff} {(\mntrlj/\mueff) \sin\beta - \cos\beta
                        \over {1 - (\mntrlj/\mueff)^2}} \, , \quad  \\
\label{eqn:N14-N15relation}
{N_{j4} \over N_{j5}}
  \approx {\lambda v \over \mueff} {(\mntrlj/\mueff) \cos\beta - \sin\beta
                        \over {1 - (\mntrlj/\mueff)^2}} \, . \quad  
\eea
This reduced system would suffice for our presentation of appendix~\ref{appsec:ew-widths}
where we discuss the $\lambda$-dependence of various partial widths of
heavier higgsino-like neutralinos decaying to a singlino-like
NLSP through the relevant ewino couplings. 
\vspace{-0.4cm}
\section{$\lambda$-dependence of the relevant ewino partial decay widths}
\label{appsec:ew-widths}
\vspace{-0.2cm}
Following section~\ref{subsec:lhc}, it is sufficient to study the partial widths of two types of decays of heavier higgsino-like 
neutralinos to a singlino-like NLSP: $\ntrlthreefour \to Z/\hsm + \ntrltwo$ 
and the same for the lighter higgsino-like chargino decaying to the NLSP:
$\charonepm \to W^\pm \ntrltwo$. The dependencies of these partial widths on `$\lambda$' are of crucial importance. Again, it suffices to consider such a dependence only through the involved couplings since the same through the phase space measures (i.e., the involved mass-splits) is generally subdominant unless at the boundaries. Thus, with only higgsinos and singlino involved, it would be sufficient to restrict ourselves to a $(3\times 3)$ higgsino-singlino system discussed at the end of appendix~\ref{appsec:blindspot}. 

The decay width of a neutralino ($\ntrli$) to another neutralino ($\ntrlj$) and a $Z$-boson is given~by
\bea
\label{eqn:gamma-chi-z}
\Gamma[\ntrli \rightarrow Z \ntrlj] \propto g^2_{_{Z \ntrli \ntrlj}} =
\frac{g^2_2}{4 \cos^2\theta_W} (N_{i3}N_{j3} - N_{i4}N_{j4})^2 \, .
\eea
By using eqs.~(\ref{eqn:N13-N15relation}),~(\ref{eqn:N14-N15relation}) and the 
involved unitarity condition ($N_{j3}^2+N_{j4}^2+N_{j5}^2=1$, for the
higgsino-singlino system), individual elements on the right-hand side of 
eq.~(\ref{eqn:gamma-chi-z}) can be expressed as
\bea
\label{eqn:nj3-value}
N^2_{j3} &=& \Big({\lambda v \over \mueff}\Big)^2{\Big[\Big({\mntrlj \over \mueff}\Big) \sin\beta - \cos\beta\Big]^2 \over 
\Big[1 - \Big({\mntrlj \over \mueff}\Big)^2\Big]^2 
+ \Big[ 1 + \Big({\mntrlj \over \mueff}\Big)^2 - 2{\mntrlj \over \mueff}\sin2\beta\Big] \big({\lambda v \over \mueff}\big)^2} \, , \quad \\ \nonumber \\
\label{eqn:nj4-value}
N^2_{j4} &=& \Big({\lambda v \over \mueff}\Big)^2{\Big[\Big({\mntrlj \over \mueff}\Big) \cos\beta - \sin\beta\Big]^2 \over 
\Big[1 - \Big({\mntrlj \over \mueff}\Big)^2\Big]^2
+ \Big[ 1 + \Big({\mntrlj \over \mueff}\Big)^2 - 2{\mntrlj \over \mueff}\sin2\beta\Big] \big({\lambda v \over \mueff}\big)^2} \, , \quad
\eea
along with
\bea
\label{eqn:nj5-value}
\hskip -49pt
N^2_{j5} &=& {\Big[1 - \Big({\mntrlj \over \mueff}\Big)^2\Big]^2 \over 
\Big[ 1 - \Big({\mntrlj \over \mueff}\Big)^2\Big]^2
+ \Big[ 1 + \Big({\mntrlj \over \mueff}\Big)^2 - 2{\mntrlj \over \mueff}\sin2\beta\Big] \big({\lambda v \over \mueff}\big)^2} \, .
\eea
Clearly, $N_{j3}^2$, $N_{j4}^2$ and $N_{j5}^2$ are explicit functions of `$\lambda$'. These also have some implicit dependence on `$\lambda$' via the neutralino mass, $\mntrlj$. It can be checked from eqs.~(\ref{eqn:nj3-value}),~(\ref{eqn:nj4-value}) and~(\ref{eqn:nj5-value}) that when $\mntrlj=\pm \mueff$
implying the $j$-th neutralino is a pure higgsino, both $N_{j3}^2$ and $N_{j4}^2$ equal to 1/2 while $N_{j5}^2=0$. With increasing `$\lambda$', higgsino component in the otherwise singlino-dominated neutralino increases at the expense of the higgsino content of the higgsino-dominated state(s).

Substituting the expressions for the elements in eqs.~(\ref{eqn:nj3-value}) 
and~(\ref{eqn:nj4-value}) into the right-hand side of eq.~(\ref{eqn:gamma-chi-z}), we find
\bea
\label{eqn:chij-chi-z-coupling}
g^2_{_{Z \ntrli \ntrlj }} = \frac{g_2^2}{4\cos^2\theta_W} {\Big(1 - {\mntrli\mntrlj \over \mueff^2}\Big)^2 \cos^22\beta \over \prod\limits_{k = i,j}\Big[ 1 + \Big({\mntrlk \over \mueff}\Big)^2 - 2{\mntrlk \over \mueff}\sin2\beta + \Big\{1 - \Big({\mntrlk \over \mueff}\Big)^2 \Big\}^2 \big({\mueff \over \lambda v}\big)^2\Big]} \,. \quad
\eea
For our discussion, $\mntrli \gg \mntrlj$ and $\ntrli \, (\ntrlj)$ is 
taken to be higgsino(singlino)-like such that $\Bigl| {\mntrli \over \mueff} 
\Bigr| \sim 1$ and $\Bigl| \frac{\mntrlj}{\mueff} \Bigr| \ll 1$ when the
terms $\Big({\mntrlj \over \mueff}\Big)^2$ can be ignored. In this limit, 
expression~(\ref{eqn:chij-chi-z-coupling}) reduces to
\bea
\label{eqn:chij-chi-z-coupling-approx}
g^2_{_{Z \ntrli \ntrlj}} \approx \frac{g_2^2}{4\cos^2\theta_W} {\Big(1 - {\mntrli\mntrlj \over \mueff^2}\Big)^2 \cos^22\beta \over 2\Big[1 - {\mntrli \over \mueff}\sin2\beta\Big]\Big[1 - 2{\mntrlj \over \mueff}\sin2\beta + \big({\mueff \over \lambda v}\big)^2\Big]} \, . \quad
\eea
Expression~(\ref{eqn:chij-chi-z-coupling-approx}) reveals that the coupling of such a pair of neutralinos to the $Z$-boson increases with increasing `$\lambda$' when the participating masses could be held fixed. On the other hand, the variation of this coupling with $\mueff$ is more subtle and depends upon relative signs among the mass parameters.

Similarly, the decay width of a neutralino ($\ntrli$) to another neutralino ($\ntrlj$) and the SM-like Higgs boson is given by
\bea
\label{eqn:eqn:gamma-chi-hsm}
\Gamma[\ntrli \rightarrow \hsm \, \ntrlj] &\propto & g^2_{_{\hsm \ntrli \ntrlj}} 
\simeq
 {\lambda^2 \over 2} \Big[ \big[ N_{i5} (N_{j3} \sin\beta + N_{j4} \cos\beta) \big] +  \big[i \longleftrightarrow j  \big]  \Big]^2 .
\quad
\eea
Again, substituting the expressions for the elements in eqs.~(\ref{eqn:nj3-value}),~(\ref{eqn:nj4-value}) and~(\ref{eqn:nj5-value}) into the right-hand side of eq.~(\ref{eqn:eqn:gamma-chi-hsm}), we find
\bea
\label{eqn:coupling-hsmnjnk}
g^2_{_{\hsm \ntrli \ntrlj}} &\simeq& {\lambda^2 \over 2}\Big({\lambda v \over \mueff}\Big)^2
{\Big[ \Big( {\mntrli \over \mueff} - \sin 2\beta \Big) \Big\{ 1 - \Big({\mntrlj \over \mueff}\Big)^2 \Big\} + \Big( {\mntrlj \over \mueff} - \sin 2\beta \Big) \Big\{1 - \Big({\mntrli \over \mueff}\Big)^2 \Big\} \Big]^2  \over \prod\limits_{k = i,j}\Big[ \Big\{ 1 + \Big({\mntrlk \over \mueff}\Big)^2 - 2{\mntrlk \over \mueff}\sin2\beta \Big\} \big({\lambda v \over \mueff}\big)^2 + \Big\{ 1 - \Big({\mntrlk \over \mueff}\Big)^2\Big\}^2\Big]} \nonumber \\ \nonumber \\
&=& {1 \over 2}\Big({\mueff \over v}\Big)^2 {\Big[\Big( {\mntrli \over \mueff} - \sin 2\beta \Big) \Big\{1 - \Big({\mntrlj \over \mueff}\Big)^2 \Big\}
+ \Big({\mntrlj \over \mueff} - \sin 2\beta \Big) \Big\{1 - \Big({\mntrli \over \mueff}\Big)^2 \Big\} \Big]^2  \over \prod\limits_{k = i,j} \Big[ 1 + \Big({\mntrlk \over \mueff}\Big)^2 - 2{\mntrlk \over \mueff}\sin2\beta  + \Big\{ 1 - \Big({\mntrlk \over \mueff}\Big)^2\Big\}^2\big({\mueff \over \lambda v}\big)^2\Big]} \,.
\quad \nonumber \\
\eea
Using similar approximations as has been done after eq.~(\ref{eqn:chij-chi-z-coupling}), eq.~(\ref{eqn:coupling-hsmnjnk}) simplifies to
\bea
\label{eqn:chij-chi-hsm-coupling-approx}
g^2_{_{\hsm \ntrli \ntrlj}} \approx {1 \over 2}\Big({\mueff \over v}\Big)^2{ \Big({\mntrli \over \mueff} - \sin 2\beta \Big)^2 \over 2\Big(1 - {\mntrli \over \mueff}\sin2\beta\Big) \Big\{ \big(1 - 2{\mntrlj \over \mueff}\sin2\beta\big) + \big({ \mueff \over \lambda v}\big)^2\Big\}} \, .\quad
\eea
Again, just like in the case of $g^2_{_{Z \ntrli \ntrlj}}$, here also one finds
$g^2_{_{\hsm \ntrli \ntrlj}}$ increases with increasing `$\lambda$' under similar circumstances.

Finally, the partial decay width of a chargino to a lighter neutralino and
a $W^\pm$-boson would also go as the involved coupling-squared and is given by
\bea
\label{eqn:gamma-cha-nj-w}
\Gamma[\chi_i^\pm \rightarrow W^\pm \ntrlj] \; \propto g^2_{_{W^\pm \chi_i^\pm \ntrlj}}
&\equiv& \left| -i g_2 \big(U^*_{i1}N_{j2} + {1 \over \sqrt{2}}U^*_{i2}N_{j3}\big)\Big(\gamma_{\mu} . {1 - \gamma_{5} \over 2}\Big) \right|^2   \nonumber \\
 &+&  \left| i g_2 \big({1 \over \sqrt{2}}V^*_{i2}N^*_{j4} - V^*_{i1}N^*_{j2}   \big)\Big(\gamma_{\mu} . {1 + \gamma_{5} \over 2}\Big) \right|^2  \,.     
\quad
\eea
With a decoupled wino that we consider, the lighter chargino is higgsino-like.
Thus, the wino components in the same, $U_{11}, V_{11} \sim 0$ while the higgsino admixtures $U_{12}, V_{12} \sim 1$. As before, $N_{j2} \sim 0$.
Thus, $g^2_{_{W^\pm \charonepm \ntrlj}}$ will depend on the two higgsino components, $N^2_{j3}$ and $N^2_{j4}$ of the only neutralino state taking
part in the interaction. Again, it is straightforward to expect from expressions~(\ref{eqn:nj3-value}) and~(\ref{eqn:nj4-value}) that this coupling also increases with `$\lambda$' when the $j$-th state is singlino-dominated to start with.

\end{document}